\documentclass{jpp}

\usepackage{graphicx}
\usepackage{natbib}
\usepackage{amsmath}
\usepackage{color}

\ifCUPmtlplainloaded \else
  \checkfont{eurm10}
  \iffontfound
    \IfFileExists{upmath.sty}
      {\typeout{^^JFound AMS Eulerq Roman fonts on the system,
                   using the 'upmath' package.^^J}%
       \usepackage{upmath}}
      {\typeout{^^JFound AMS Euler Roman fonts on the system, but you
                   dont seem to have the}%
       \typeout{'upmath' package installed. JPP.cls can take advantage
                 of these fonts, if you use 'upmath' package.^^J}%
      }
  \else
  \fi
\fi

\ifCUPmtlplainloaded \else
  \checkfont{msam10}
  \iffontfound
    \IfFileExists{amssymb.sty}
      {\typeout{^^JFound AMS Symbol fonts on the system, using the
                'amssymb' package.^^J}%
       \usepackage{amssymb}%
       \let\le=\leqslant  
       \let\ge=\geqslant  
      }{}
  \fi
\fi

\ifCUPmtlplainloaded \else
  \IfFileExists{amsbsy.sty}
    {\typeout{^^JFound the 'amsbsy' package on the system, using it.^^J}%
     \usepackage{amsbsy}}
    {}
\fi

\newsavebox{\astrutbox}
\sbox{\astrutbox}{\rule[-5pt]{0pt}{20pt}}

\newcommand{\V}[1]{\mathbf{#1}} 
\newcommand{\T}[1]{\texttt{#1}} 
\newcommand\Alfven{Alfv\'en }
\newcommand\Alfvenic{Alfv\'enic }
\newcommand{\figref}[1]{Fig.~\ref{#1}}
\newcommand{\secref}[1]{\S\ref{#1}}
\newcommand{\appref}[1]{Appendix~\ref{#1}}
\newcommand{\xhat}{\mbox{$\hat{\mathbf{x}}$}} 
\newcommand{\yhat}{\mbox{$\hat{\mathbf{y}}$}} 
\newcommand{\zhat}{\mbox{$\hat{\mathbf{z}}$}}

\title[Particle Energization in \Alfvenic Current Sheets] {Spatially Localized Particle Energization by Landau Damping in Current Sheets Produced by Strong \Alfven Wave Collisions}

\author[ G.~G.~Howes, A.~J.~McCubbin, and K.~G.~Klein]%
       {G\ls R\ls E\ls G\ls O\ls R\ls Y\ns G.\ns H\ls O\ls W\ls E\ls S$^1$\thanks{Email address for correspondence: gregory-howes@uiowa.edu},\ns
         A\ls N\ls D\ls R\ls E\ls W\ns 
         J.\ns M\ls C\ls C\ls U\ls B\ls B\ls I\ls N$^{1}$, and
         K\ls R\ls I\ls S\ls T\ls O\ls P\ls H\ls E\ls R\ns 
G.\ns K\ls L\ls E\ls I\ls N$^{2}$}

\affiliation{$^1$Department of Physics and Astronomy, 
University of Iowa,
Iowa City, IA 52242, USA\\[\affilskip]
$^2$Department of Climate and Space Sciences and Engineering, University of Michigan, Ann Arbor, MI 48109, USA}

\pubyear{2017}
\volume{?}
\pagerange{?}
\date{?; revised ?; accepted ?.}

\begin{document}

\maketitle

\begin{abstract}
Understanding the removal of energy from turbulent fluctuations in a
magnetized plasma and the consequent energization of the constituent
plasma particles is a major goal of heliophysics and astrophysics.
Previous work has shown that nonlinear interactions among
counterpropagating Alfven waves---or Alfven wave collisions---are the
fundamental building block of astrophysical plasma turbulence and
naturally generate current sheets in the strongly nonlinear limit. A
nonlinear gyrokinetic simulation of a strong Alfven wave collision is
used to examine the damping of the electromagnetic fluctuations and
the associated energization of particles that occurs in
self-consistently generated current sheets.  A simple model explains
the flow of energy due to the collisionless damping and the associated
particle energization, as well as the subsequent thermalization of the
particle energy by collisions.  The net particle energization by the
parallel electric field is shown to be spatially intermittent, and the 
nonlinear evolution is essential in enabling that spatial
non-uniformity.  Using the recently developed field-particle
correlation technique, we show that particles resonant with the Alfven
waves in the simulation dominate the energy transfer, demonstrating
conclusively that Landau damping plays a key role in the spatially
intermittent damping of the electromagnetic fluctuations and
consequent energization of the particles in this strongly nonlinear
simulation.
\end{abstract}

\section{Introduction}
\label{sec:intro}
The space and astrophysical plasmas that fill the heliosphere, and
other more remote astrophysical environments, are found generally to
be both magnetized and turbulent.  Understanding the removal of energy
from turbulent fluctuations in a magnetized plasma and the consequent
energization of the constituent plasma particles is a major goal of
heliophysics and astrophysics. Although plasma heating and particle
energization are governed by microscopic processes typically occurring
at kinetic length scales in the plasma, these important energy
transport mechanisms can have a significant impact on the macroscopic
evolution of the systems. For example, the diffuse plasma of the solar
corona is found to be nearly three orders of magnitude hotter than the
solar photosphere. The dissipation of turbulent fluctuations, through
a physical mechanism that is poorly understood at present, is believed
to be responsible for this dramatic heating of the coronal plasma.
This very high coronal temperature leads to the supersonic solar wind
that pervades the entire heliosphere \citep{Parker:1958}, so the
kinetic plasma physics governing the heating of the coronal plasma at
small scales indeed impacts the global structure of the heliosphere.

The low density and high temperature conditions of the plasma in many
astrophysical systems lead to a mean free path for collisions among
the constituent charged particles that is often much longer than the
length scales of the turbulent fluctuations.  Under such weakly
collisional plasma conditions, the dynamics of the turbulence and its
dissipation is governed by kinetic plasma physics. Unlike in the more
well-known case of fluid systems (which corresponds to the strongly
collisional regime), in weakly collisional plasmas, the dissipation of
turbulent energy into plasma heat is inherently a two-step process
\citep{Howes:2017c}. First, energy is removed from the turbulent
electromagnetic fluctuations through collisionless interactions
between the fields and particles, transferring that energy to
non-thermal fluctuations in the particle velocity distribution
functions, a process that is reversible. Subsequently, arbitrarily
weak collisions can smooth out the small fluctuations in velocity
space, leading to entropy increase and irreversible heating of the
plasma \citep{Howes:2006,Howes:2008c,Schekochihin:2009}. In this
two-step process, the removal of energy from turbulent fluctuations
and the subsequent conversion of that energy into plasma heat may even
occur at different locations \citep{Navarro:2016}.

In fluid simulations of plasma turbulence using the
magnetohydrodynamic (MHD) approximation---a strongly collisional limit
of the large-scale dynamics (relative to the characteristic kinetic
plasma length scales)---the nonlinear evolution leads to the
development of intermittent current sheets
\citep{Matthaeus:1980,Meneguzzi:1981}. Furthermore, it has been found
that the dissipation of turbulent energy is largely concentrated in
these intermittent current sheets
\citep{Uritsky:2010,Osman:2011,Zhdankin:2013}.  Numerous studies have
recently sought evidence for the spatial localization of plasma
heating by the dissipation of turbulence in current sheets through
statistical analyses of solar wind observations
\citep{Osman:2011,Borovsky:2011,Osman:2012a,Perri:2012a,Wang:2013,Wu:2013,Osman:2014b}
and numerical simulations
\citep{Wan:2012,Karimabadi:2013,TenBarge:2013a,Wu:2013,Zhdankin:2013}.

The mechanisms of the spatially localized dissipation found in MHD
simulations are resistive (Ohmic) heating and viscous heating
\citep{Zhdankin:2013,Brandenburg:2014,Zhdankin:2015b}.  But,
resistivity and viscosity arise from microscopic collisions in the
strongly collisional (or small mean free path) limit, a limit that is
not applicable to the dynamics of dissipation in many space and
astrophysical environments \citep{Howes:2017c}.  Under the weakly
collisional conditions appropriate for most space and astrophysical
plasmas, which physical mechanisms are responsible for the damping of the
turbulent fluctuations and the consequent energization of the plasma
particles remains an open question.  Our aim here is to identify the
mechanisms governing the damping of the turbulent fluctuations and the
particle energization using a kinetic simulation code that follows the
three-dimensional evolution of a weakly collisional plasma in which
current sheets develop self-consistently.

Early research on incompressible MHD turbulence in the 1960s
\citep{Iroshnikov:1963,Kraichnan:1965} emphasized the wave-like nature
of turbulent plasma motions, suggesting that nonlinear interactions
between counterpropagating \Alfven waves---or simply \emph{\Alfven
  wave collisions}---mediate the turbulent cascade of energy from
large to small scales. In fact, the physics of the nonlinear
interactions among \Alfven waves provides the foundation for modern
scaling theories of plasma turbulence that explain the anisotropic
nature of the turbulent cascade \citep{Goldreich:1995} and the dynamic
alignment of velocity and magnetic field fluctuations
\citep{Boldyrev:2006}.

Following a number of previous investigations of weak incompressible
MHD turbulence \citep{Sridhar:1994,Ng:1996,Galtier:2000}, the
nonlinear energy transfer in \Alfven wave collisions in the weakly
nonlinear limit has been solved analytically \citep{Howes:2013a},
confirmed numerically with gyrokinetic numerical simulations
\citep{Nielson:2013a}, and verified experimentally in the laboratory
\citep{Howes:2012b,Howes:2013b,Drake:2013}, establishing \Alfven wave
collisions as the fundamental building block of astrophysical plasma
turbulence. More recent research has found that \Alfven wave
collisions in the strongly nonlinear limit naturally generate current
sheets \citep{Howes:2016b}, providing a first-principles explanation
for the ubiquitous development of intermittent current sheets in
plasma turbulence. This self-consistent generation of current sheets is
found to persist even in the more realistic case of strong collisions
between localized \Alfven wave packets \citep{Verniero:2017a}.

Here we explore the damping of the electromagnetic fluctuations and
the associated energization of particles that occurs in current sheets
that are generated self-consistently by strong \Alfven wave
collisions. Previous work using a simulation of kinetic \Alfven wave
turbulence has shown that, although enhanced plasma heating rates are
well correlated with the presence of current sheets, the rate of
heating as a function of wavenumber is well predicted assuming that
linear Landau damping is entirely responsible for the removal of
energy from the turbulence \citep{TenBarge:2013a}.  This result
suggests that the physical mechanism governing the removal of energy
from turbulent fluctuations, even in spatially intermittent current
sheets, is Landau damping. Using nonlinear gyrokinetic simulations of
strong \Alfven wave collisions, we aim to answer two questions:
\begin{enumerate}
  \item[(i)] Is the
dissipation associated with current sheets that are generated by
strong Alfven wave collisions spatially intermittent?
\item[(ii)] What is the physical mechanism governing the removal of
  energy from the turbulence and the consequent energization of the
  particles?
\end{enumerate}


\section{Simulation}
\label{sec:sim}
Similar to a previous study showing the development of current sheets
in strong \Alfven wave collisions \citep{Howes:2016b}, we employ the
Astrophysical Gyrokinetics code \T{AstroGK} \citep{Numata:2010} to
perform a gyrokinetic simulation of the nonlinear interaction between
two counterpropagating \Alfven waves in the strongly nonlinear limit.
\T{AstroGK} evolves the perturbed gyroaveraged distribution function
$h_s(x,y,z,\lambda,\varepsilon)$ for each species $s$, the scalar
potential $\varphi$, the parallel vector potential $A_\parallel$, and
the parallel magnetic field perturbation $\delta B_\parallel$
according to the gyrokinetic equation and the gyroaveraged Maxwell's
equations \citep{Frieman:1982,Howes:2006}. Velocity space coordinates
are $\lambda=v_\perp^2/v^2$ and $\varepsilon=v^2/2$. The domain is a
periodic box of size $L_{\perp }^2 \times L_{\parallel }$, elongated
along the straight, uniform mean magnetic field $\V{B}_0=B_0 \zhat$,
where all quantities may be rescaled to any parallel dimension
satisfying $L_{\parallel } /L_{\perp } \gg 1$. Uniform Maxwellian
equilibria for ions (protons) and electrons are chosen, with a reduced
mass ratio $m_i/m_e=36$ such that, even with the modest spatial
resolution of this simulation, the collisionless damping by ions and
electrons is sufficiently strong within the resolved range of length
scales to terminate the nonlinear transfer of energy to small scales.
Spatial dimensions $(x,y)$ perpendicular to the mean field are treated
pseudospectrally; an upwind finite-difference scheme is used in the
parallel direction, $z$. Collisions employ a fully conservative,
linearized collision operator with energy diffusion and pitch-angle
scattering \citep{Abel:2008,Barnes:2009}.

To set up the simulation of an \Alfven wave collision, following
\citet{Nielson:2013a}, we initialize two perpendicularly polarized,
counterpropagating plane \Alfven waves, $\V{z}^{+} = z_+ \cos (k_\perp
x -k_\parallel z -\omega_0 t) \yhat$ and $\V{z}^{-} =z_- \cos (k_\perp
y +k_\parallel z -\omega_0 t)\xhat$, where $\omega_0=k_\parallel v_A$,
$k_\perp=2 \pi/L_\perp$, and $k_\parallel=2 \pi/L_\parallel$.  Here
$\V{z}^{\pm} = \V{u} \pm \delta \V{B}/\sqrt{4 \pi (n_i m_i + n_e
  m_e)}$ are the Elsasser fields \citep{Elsasser:1950} which represent
\Alfven waves that propagate up or down the mean magnetic field at the
\Alfven velocity $v_A= B_0/\sqrt{4 \pi (n_i m_i + n_e m_e)}$ in the
MHD limit, $k_\perp \rho_i \ll 1$.  We specify a balanced collision
with equal counterpropagating wave amplitudes, $z_+=z_-$, such that
the nonlinearity parameter is $\chi=k_\perp z_\pm/(k_\parallel
v_A)=1$, relevant to the regime of strong turbulence
\citep{Goldreich:1995}. To study the nonlinear evolution in the limit
$k_\perp \rho_i\ll 1$, we choose a perpendicular simulation domain
size $L_{\perp}=8 \pi \rho_i$ with simulation resolution
$(n_x,n_y,n_z,n_\lambda,n_\varepsilon,n_s)= (64,64,32,128,32,2)$.  The
fully resolved perpendicular range in this dealiased pseudospectral
method covers $0.25 \le k_\perp \rho_i \le 5.25$, or $0.042 \le
k_\perp \rho_e \le 0.875$ given the chosen mass ratio $m_i/m_e=36$ and
temperature ratio $T_i/T_e=1$. Here the ion thermal Larmor radius is
$\rho_i= v_{ti}/\Omega_i$, the ion thermal velocity is $v_{ti}^2 =
2T_i/m_i$, the ion cyclotron frequency is $\Omega_i= q_i B_0/(m_i c)$,
and the temperature is given in energy units. The plasma parameters of
the simulation are $\beta_i=1$ and $T_i/T_e=1$, typical of near-Earth
solar wind conditions. The linearized Landau collision operator
\citep{Abel:2008,Barnes:2009} is employed with collisional
coefficients $\nu_i=\nu_e=6 \times 10^{-4} k_\parallel v_A$, yielding
weakly collisional dynamics with $ \nu_s /\omega \ll 1$.

To prepare the simulation, the two initial \Alfven wave modes are
evolved linearly for five periods with enhanced collision frequencies
$\nu_i=\nu_e=0.01 k_\parallel v_A$ to eliminate any transient behavior
arising from the initialization that does not agree with the
properties of the \Alfven wave mode \citep{Nielson:2013a}. The
simulation is then restarted with the nonlinear terms enabled,
beginning the nonlinear evolution of the strong \Alfven wave
collision. Note that the two \Alfven waves are already overlapping at
the beginning of this simulation before the nonlinear evolution
begins, an idealized case which facilitates the comparison to an
asymptotic analytical solution in the weakly nonlinear limit
\citep{Howes:2013a,Howes:2016b}. The nonlinear evolution of
development of current sheets is found to persist in the more
realistic case of collisions between two initially separated \Alfven
wavepackets of finite parallel extent
\citep{Verniero:2017a,Verniero:2017b}.

\begin{figure}
  \centerline{\resizebox{4.0in}{!}{\includegraphics*[0.35in,2.in][8.0in,7.5in]{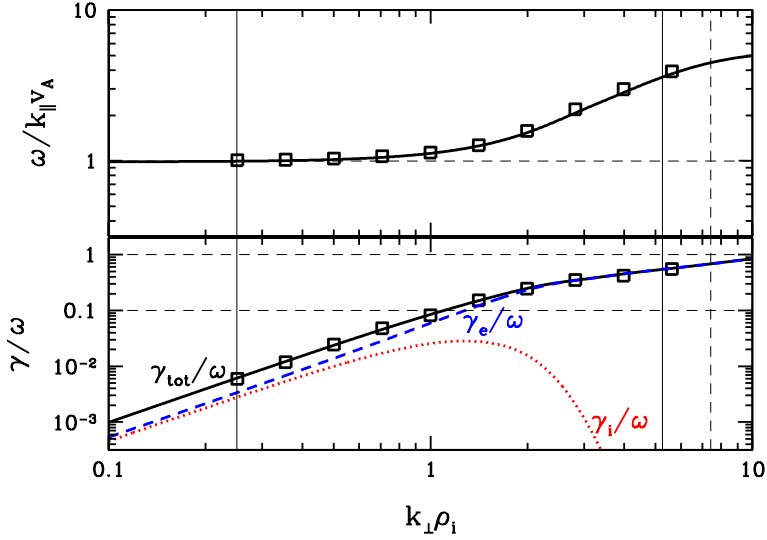}}
  }
  \caption{(a) The normalized frequency $\omega/k_\parallel v_A$ and
    (b) total collisionless damping rate $\gamma_{tot}/\omega$ (black
    solid) vs.~ $k_\perp \rho_i$ for \Alfven and kinetic \Alfven waves
    with $m_i/m_e=36$ from the linear collisionless gyrokinetic
    dispersion relation, including the separate contributions to the
    linear collisionless damping rate from the ions $\gamma_i/\omega$
    (red dotted) and the electrons $\gamma_e/\omega$ (blue
    dashed). Squares indicate values computed from linear runs of
    \T{AstroGK}. Solid vertical lines indicate the limits of the fully
    resolved perpendicular scales of the nonlinear simulation at
    $k_\perp \rho_i=0.25$ and $k_\perp \rho_i=5.25$.  The vertical
    dashed line indicates the highest $k_\perp \rho_i$ value, $k_\perp
    \rho_i =5.25\sqrt{2} \simeq 7.42$, of the modes in the corner of
    Fourier space.}
\label{fig:fpac22_gk_disp}
\end{figure}

For the plasma parameters of this gyrokinetic simulation, we solve the
linear collisionless gyrokinetic dispersion relation
\citep{Howes:2006} for the Alfv\'en/kinetic \Alfven wave mode to
determine the linear frequency and collisionless damping rate for this
mode as a function of perpendicular wavenumber.  Note that the
collisionless damping of this mode is due to the Landau resonances
with the ions and electrons. In the upper panel of
\figref{fig:fpac22_gk_disp} is plotted the normalized real frequency
$\omega/k_\parallel v_A$ vs.~the normalized perpendicular wavenumber
$k_\perp \rho_i$. In the lower panel is plotted the total
collisionless damping rate normalized to the wave frequency
$\gamma/\omega$ (solid black), as well as the separate contributions
to this linear collisionless damping rate from the ions (red dotted)
and electrons (blue dashed).  These gyrokinetic results have been
verified by comparison with the solutions of the full Vlasov-Maxwell
linear dispersion relation using the PLUME solver
\citep{Klein:2015a}. Since gyrokinetic theory resolves the Landau
resonances but not the cyclotron resonances, this agreement between
the gyrokinetic and the Vlasov-Maxwell results confirms that the linear
collisionless damping is due to the Landau resonance.

\figref{fig:fpac22_gk_disp} shows that the collisional damping by the
ions (red dotted) has a relatively broad peak over the range $0.5
\lesssim k_\perp \rho_i \lesssim 2.0$.  The range of resonant parallel
phase velocities $\omega / k_\parallel$ associated with this broad
peak in damping, normalized in terms of the ion thermal velocity, is
$1.0 \lesssim \omega / k_\parallel v_{ti} \lesssim 1.5$. Therefore, if
Landau damping with the ions is active, the energy transfer should be
dominated by resonant ions with parallel velocities in the range $1.0
\lesssim v_\parallel/ v_{ti} \lesssim 1.5$.  The collisionless damping
by the electrons, on the other hand, increases monotonically with
perpendicular wavenumber, becoming sufficiently strong with
$\gamma_e/\omega \gtrsim 0.1$ at $k_\perp \rho_i \gtrsim 1.2$. From
this point, up to the maximum fully resolved perpendicular scale of
$k_\perp \rho_i =5.25$, the range of resonant parallel phase
velocities $\omega / k_\parallel$ in terms of the electron thermal
velocity is $0.17 \lesssim \omega / k_\parallel v_{te} \lesssim
0.6$. Therefore, if the collisionless energy transfer from the
turbulent electromagnetic fields to the plasma particles is governed
by a Landau resonant mechanism, we would expect to see the transfer of
energy localized at parallel velocities within this range of resonant
values.

\section{Evolution of Energy}
\label{sec:energy}
Under weakly collisional plasma conditions typical of many
heliospheric and astrophysical plasmas, the removal of energy from
turbulent fluctuations and the eventual conversion of that energy into
plasma heat, unlike in the more familiar fluid limit, is a two-step
process \citep{Howes:2017c}.  Specifically, the turbulent fluctuations
are first damped through reversible, collisionless interactions
between the electromagnetic fields and the plasma particles, leading
to energization of the particles.  This non-thermal energization of
the particle velocity distributions is subsequently thermalized by
arbitrarily weak collisions, thereby accomplishing the ultimate
conversion of the turbulent energy into particle heat. An analysis of
the flow of energy in this \Alfven wave collision simulation
illustrates these two distinct steps of the turbulent dissipation.

In a gyrokinetic system, the \emph{total fluctuating energy} $\delta
W$ \citep{Howes:2006,Brizard:2007,Schekochihin:2009} is given
by\footnote{Note that in the gyrokinetic approximation, the electric
  field energy is relativistically small relative to the magnetic
  field energy \citep{Howes:2006}.}
\begin{equation}
\delta W =
\int\!d^3\V{r} \left[ \frac{|\delta \V{B}|^2+|\delta \V{E}|^2}{8\pi}+\sum\limits_s 
\int d^3 \V{v}\ \frac{T_{0s}\delta f_s^2}{2F_{0s}} \right], 
\label{eq:deltaW_GK}
\end{equation}
where the index $s$ indicates the plasma species and $T_{0s}$ is the
temperature of each species' Maxwellian equilibrium. The left term
represents the electromagnetic energy and the right term
represents that microscopic fluctuating kinetic energy of the
particles of each plasma species $s$.  Note that the elimination of
the parallel nonlinearity in the standard form of gyrokinetic theory
means that the appropriate conserved quadratic quantity in
gyrokinetics is the Kruskal-Obermann energy, $E^{(\delta f)}_s\equiv
\int d^3 \V{r} \int d^3 \V{v}\ T_{0s}\delta f_s^2/2F_{0s}$
\citep{Kruskal:1958,Morrison:1994}, in contrast to the usual kinetic
theory definition of microscopic kinetic energy, $ \int d^3 \V{r} \int
d^3 \V{v} \ (m_s v^2/2) f_s$.  Note also that $\delta W$ includes
neither the equilibrium thermal energy, $\int d^3 \V{r} \frac{3}{2}
n_{0s} T_{0s} = \int d^3 \V{r} \int d^3 \V{v} \frac{1}{2} m_sv^2
F_{0s}$, nor the equilibrium magnetic field energy, $\int d^3 \V{r}
\ B_0^2/8\pi$. Thus, the terms of $\delta W$ in \eqref{eq:deltaW_GK}
represent the perturbed electromagnetic field energies and the
microscopic kinetic energy of the deviations from the Maxwellian
velocity distribution for each species.

A more intuitive form of the total fluctuating energy $\delta W$ can
be obtained by separating out the kinetic energy of the bulk motion of
the plasma species from the non-thermal energy in the distribution
function that is not associated with bulk flows \citep{TCLi:2016},
\begin{equation}
\delta W =
\int\!d^3\V{r} \left[\frac{|\delta \V{B}|^2+|\delta \V{E}|^2}{8\pi}+\sum\limits_s
  \left( \frac{1}{2}n_{0s}m_s|\delta
  \V{u_s}|^2 + \frac{3}{2} \delta P_s  \right) \right]
\label{eq:deltaW}
\end{equation}
where $n_{0s}$ is the equilibrium density, $m_s$ is mass, and
$\delta\V{u_s}$ is the fluctuating bulk flow velocity. The
\emph{non-thermal energy} in the distribution function (not including
the bulk kinetic energy) is defined by $E^{(nt)}_s\equiv \int d^3
\V{r}\frac{3}{2} \delta P_s \equiv \int d^3 \V{r} (\int d^3
\V{v}\ T_{0s}\delta f_s^2/2F_{0s} - \frac{1}{2}n_{0s}m_s|\delta
\V{u_s}|^2)$ \citep{TenBarge:2014b}.  The \emph{turbulent energy} is
defined as the sum of the electromagnetic field and the bulk flow
kinetic energies \citep{Howes:2015b,TCLi:2016}, $E^{(turb)}\equiv \int
d^3 \V{r} [ (|\delta \V{B}|^2+|\delta \V{E}|^2)/8\pi + \sum_s
  \frac{1}{2}n_{0s}m_s|\delta \V{u_s}|^2 ]$.  Therefore the total
fluctuating energy is simply the sum of the turbulent energy and
species non-thermal energies, $\delta W=E^{(turb)} + E^{(nt)}_i +
E^{(nt)}_e$. 

\subsection{Evolution of Turbulent and Non-Thermal Energies}

\begin{figure}
  \centerline{\resizebox{5.5in}{!}{\includegraphics{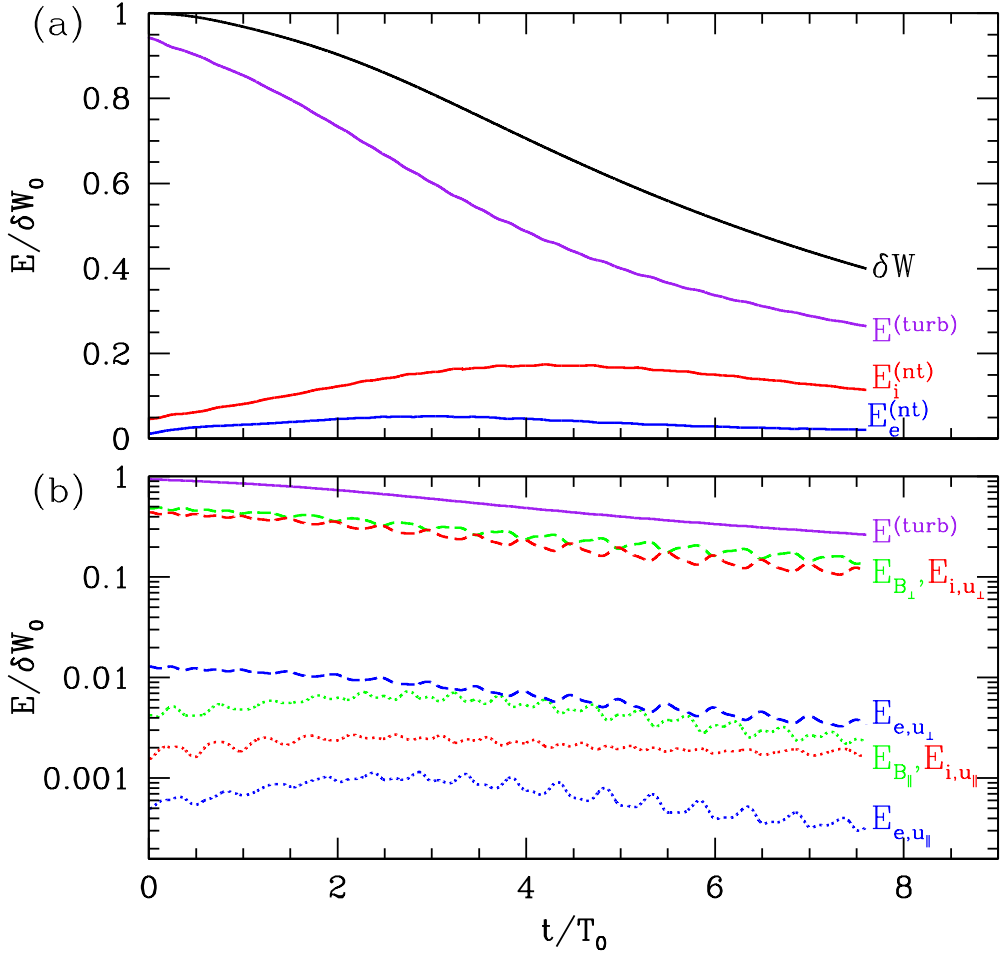}}
  }
  \caption{(a) Evolution of the normalized energy $E/\delta W_0$ as a
    function of time $t/T_0$ for the total fluctuating energy $\delta
    W$ (black), the turbulent energy $E^{(turb)}$ (purple), the
    ion non-thermal energy $E^{(nt)}_i$ (red) and the electron
    non-thermal energy $E^{(nt)}_e$ (blue). (b) Evolution of the
    different components of the turbulent energy $E^{(turb)}$
    (purple), dominated by the perpendicular magnetic field energy
    $E_{B_\perp}$ (green dashed) and the perpendicular ion bulk flow
    kinetic energy $E_{i, u_\perp}$ (red dashed), with successively
    smaller contributions by the perpendicular electron bulk kinetic
    energy $E_{e, u_\perp}$ (blue dashed), the parallel magnetic field
    energy $E_{B_\parallel}$ (green dotted), the parallel ion bulk
    flow kinetic energy $E_{i, u_\parallel}$ (red dotted), and the
    parallel electron bulk flow kinetic energy $E_{e, u_\parallel}$ (blue
    dotted).}
    \label{fig:fpac22_energy}
\end{figure}

In \figref{fig:fpac22_energy}, we plot the evolution of these three
different contributions to the total fluctuating energy normalized to
the total initial fluctuating energy $\delta W_0\equiv \delta W
(t=0)$. In \figref{fig:fpac22_energy}(a), we plot the total
fluctuating energy $\delta W/\delta W_0$ (black), the turbulent energy
$E^{(turb)}/\delta W_0$ (purple), the ion non-thermal energy
$E^{(nt)}_i /\delta W_0$ (red), and the electron non-thermal energy
$E^{(nt)}_e /\delta W_0$ (blue). Note that collisions in \T{AstroGK},
as well as in real plasma systems, convert non-thermal to thermal
energy, representing irreversible plasma heating with an associated
increase of entropy. The energy lost from $\delta W$ by collisions is
tracked by \T{AstroGK} and represents thermal heating of the plasma
species, but this energy is not fed back into the code to evolve the
equilibrium thermal temperature, $T_{0s}$
\citep{Howes:2006,Numata:2010,TCLi:2016}. The evolution in
\figref{fig:fpac22_energy}(a) makes clear that, over 7.5 periods of
the initial \Alfven waves, more than 60\% of the initial fluctuating
energy in the simulation is lost to collisional heating.

In \figref{fig:fpac22_energy}(b), we plot the different components
that contribute to the turbulent energy $E^{(turb)}$. In order of
decreasing magnitude, these contributions are the perpendicular
magnetic energy $E_{B_\perp}$ (green dashed), perpendicular ion
kinetic energy $E_{u_{i,\perp}}$ (red dashed), perpendicular electron
kinetic energy $E_{u_{e,\perp}}$ (blue dashed), parallel magnetic
energy $E_{B_\parallel}$ (green dotted), parallel ion kinetic energy
$E_{u_{i,\parallel}}$ (red dotted), and parallel electron kinetic
energy $E_{u_{e,\parallel}}$ (blue dotted).  The turbulent energy is
dominated by the perpendicular magnetic energy and perpendicular ion
kinetic energy. This is expected for \Alfvenic fluctuations at
$k_\perp \rho_i \ll 1$: transverse motion of the plasma dominated by
ion kinetic energy is first arrested by magnetic tension, followed by
the acceleration of the plasma back toward the equilibrium point by
magnetic tension, thereby leading to the oscillatory transfer of
energy back and forth between perpendicular magnetic energy and
perpendicular ion kinetic energy, as evident in
\figref{fig:fpac22_energy}(b). Note that this energy is integrated
over the entire simulation domain, so neither of these energies is
expected to drop to zero, as would occur for the energy density at a
single point in space as an \Alfven wave passes through that point.
In the MHD limit $k_\perp \rho_i \ll 1$, \Alfvenic fluctuations also
have very little parallel motion, $u_\parallel \ll u_\perp$ and a very
small parallel magnetic field fluctuation, $\delta B_\parallel \ll
\delta B_\perp$. Finally, the electron kinetic energies are down from
the respective ion kinetic energies approximately by a factor of the
mass ratio, $m_e/m_i = 1/36$, so electrons make a subdominant
contribution to the turbulent energy. Finally, note that although the
volume integrated energy of each component of $E^{(turb)}$ shows
oscillations with the period $T_0$, their sum varies smoothly in time,
suggesting that this definition of turbulent energy is physically well
motivated.

\subsection{Evolution of Collisional Heating}

\begin{figure}
  \centerline{\resizebox{5.50in}{!}{\includegraphics*[0.25in,2.in][8.0in,5.5in]{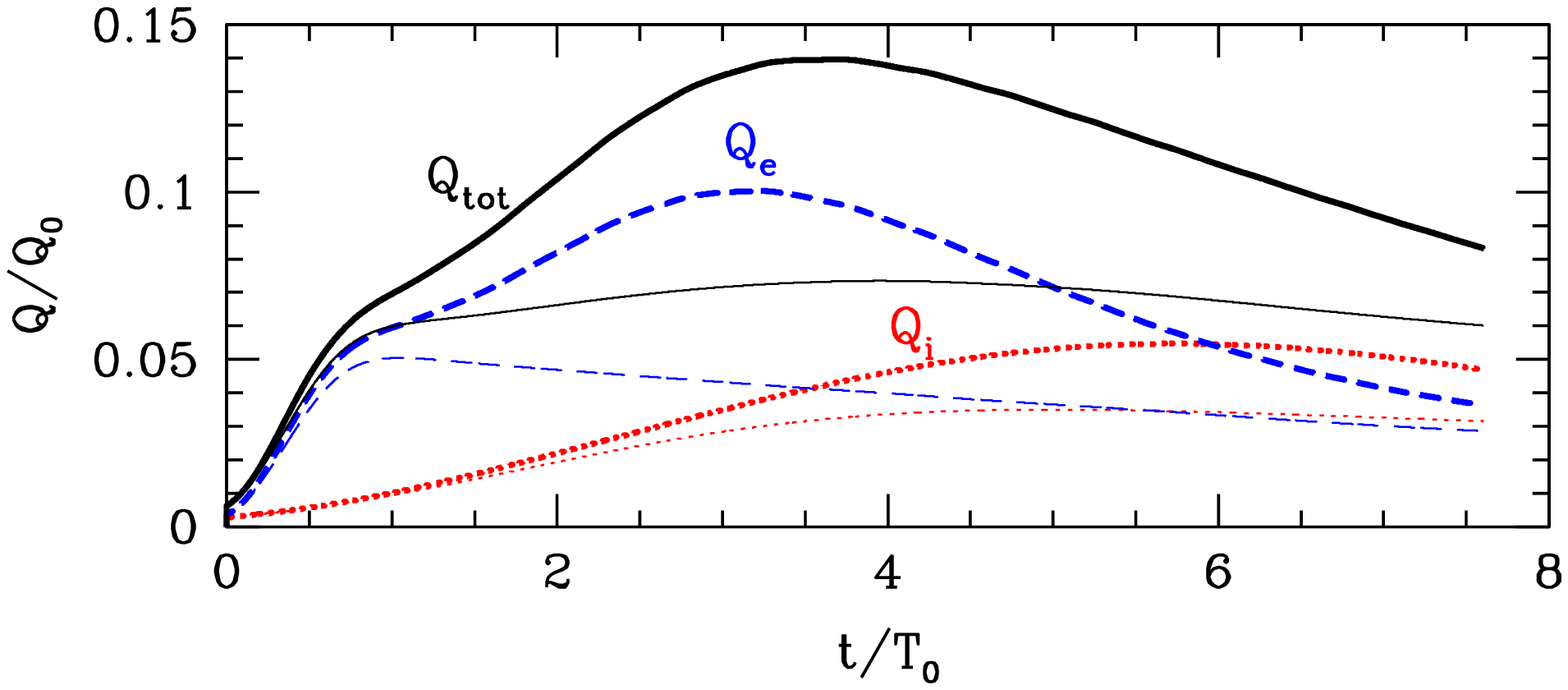}}
  }
  \caption{Ion collisional heating rate $Q_i/Q_0$ (red), electron
    collisional heating rate (blue) and total collisional heating rate
    $Q_{tot}=Q_i+Q_e$ (black) and as a function of time $t/T_0$ for
    the nonlinear simulation (thick lines). Also plotted (thin lines)
    is the linear evolution from the same initial conditions.
\label{fig:fpac22_heat}}
\end{figure}

In \figref{fig:fpac22_heat}, we present the evolution of the
collisional heating rate per unit volume of ions $Q_i$ (red) and
electrons $Q_e$ (blue) as well as the total collisional heating rate
$Q_{tot}=Q_i+Q_e$ (black) for this nonlinear \Alfven wave collision
simulation (thick lines). The heating rates are normalized by a
characteristic heating rate per unit volume, $Q_0=(n_{0i} T_{0i}
v_{ti}/L_\parallel)(\pi/8)(L_\perp/L_\parallel)^2$. The total
fluctuating energy $\delta W$ in \figref{fig:fpac22_energy}(a)
diminishes in time due to thermalization by collisions.  This
collisional energy loss from $\delta W$ is tracked in \T{AstroGK} by
this collisional heating rate, enabling energy conservation to be
measured in the simulation.

Note that the rapid initial rise in the collisional damping rate for
the electrons $Q_e$ at $t/T_0 \lesssim 0.5$ in
\figref{fig:fpac22_heat} is due to the fact that the linear
initialization uses higher collision coefficients, $\nu_s = 0.01
k_\parallel v_A$, than the subsequent nonlinear evolution, $\nu_s = 6
\times 10^{-4} k_\parallel v_A$.  When the collisional coefficients
are reduced, smaller velocity scale structures in the velocity
distribution must develop (through the kinetic evolution) before the
collisional heating is able to effectively thermalize the non-thermal
energy contained in those fluctuations.

Also plotted in \figref{fig:fpac22_heat} is the evolution of the
collisional heating rates in a linear simulation (thin lines), where
the simulation is started from the same initial conditions but the
nonlinear terms are turned off.  In this linear simulation, there is
no nonlinear transfer of energy to other Fourier modes---meaning that
there is no nonlinear turbulent cascade of energy to small scales---so
the evolution of the energy is solely due to linear Landau damping of the
initial \Alfven waves and the subsequent collisional thermalization of
the fluctuations in the velocity distribution functions that were
generated by this linear Landau damping. It is important to note that
the nonlinear evolution eventually leads to a higher collisional
heating rate, presumably through the nonlinear transfer of energy to
smaller scale fluctuations that have higher collisionless damping
rates than the initial \Alfven waves, although we do not directly
analyze that nonlinear cascade of energy in this study.

\subsection{Model of Energy Flow}

\begin{figure}
  \centerline{\resizebox{5.20in}{!}{\includegraphics{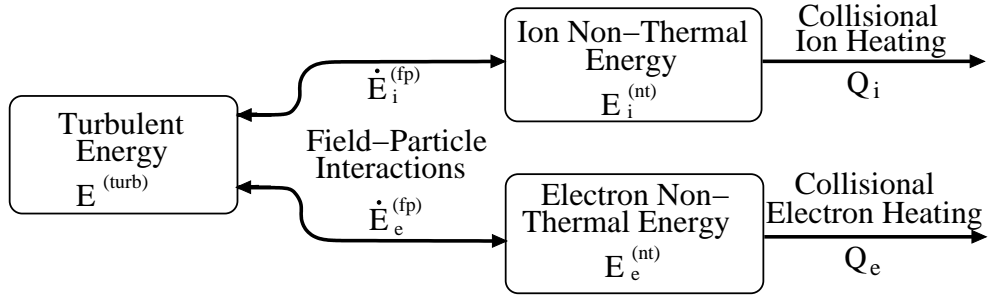}}
  }
  \caption{Diagram of the energy flow in weakly collisional turbulent
    plasmas, showing that interactions between the electromagnetic
    fields and plasma particles $\dot{E}^{(fp)}_s$ can reversibly
    transfer energy between the turbulent energy $E^{(turb)}$ and the
    non-thermal energy in the velocity distribution function of each
    species $E^{(nt)}_s$. Collisional heating $Q_s$ then can
    irreversibly convert this non-thermal energy, represented by
    fluctuations in velocity space of each species, into heat of each
    plasma species $s$. This is the two-step process of reversible
    particle energization and subsequent irreversible thermalization
    of that particle energy.}
\label{fig:energyflow}
\end{figure}

A physical interpretation of the two-step energy flow in this strong
\Alfven wave collision simulation is illustrated by the diagram in
\figref{fig:energyflow}. The energy of turbulent fluctuations
$E^{(turb)}$, consisting of the sum of the electromagnetic field
fluctuations and the kinetic energy of the bulk flows (first velocity
moment) of each plasma species \citep{Howes:2015b,Howes:2017c}, can be
removed by collisionless interactions $\dot{E}^{(fp)}_s$ between the
electromagnetic fields and the plasma particles. This energy is
converted to non-thermal energy of the ions and electrons,
$E^{(nt)}_s$.  This non-thermal energy is represented by fluctuations
in the particle velocity distribution functions that have no
associated bulk flow (first moment), and therefore do not contribute
to the turbulent motions. A key property of this collisionless energy
transfer $\dot{E}^{(fp)}_s$ is that it is reversible (two-headed
arrows in \figref{fig:energyflow}), representing the electromagnetic
work done on the particles by the fields, which can be positive or
negative.

The non-thermal energy $E^{(nt)}_s$ is contained in fluctuations in
velocity-space of the particle velocity distribution functions for
each species, $\delta f_s(\V{v})$.  If these fluctuations reach
sufficiently small scales in velocity space, arbitrarily weak
collisions can smooth out those fluctuations, thermalizing their
energy and thereby realizing irreversible plasma heating, $Q_s$. The
kinetic equation for each species governs two mechanisms that
facilitate the transfer of energy to ever smaller scales in velocity
space: linear phase mixing and nonlinear phase mixing.

The first mechanism is linear phase mixing governed by the ballistic
term in the kinetic equation, which couples spatial variations with
velocity-space fluctuations and can lead to the transfer of energy to
small scales in velocity space.\footnote{It has been recently
  suggested that, under particular conditions in a turbulent plasma of
  sufficiently low collisionality, a turbulent anti-phase-mixing
  process can prevent velocity space fluctuations from reaching
  sufficiently small-scales to enable thermalization by collisions
  \citep{Schekochihin:2016,Parker:2016}.}  In linear Landau damping,
for example, the energy of a damped wave is first transferred
collisionlessly into non-thermal velocity space fluctuations, which
subsequently phase mix linearly to small enough scales in velocity
space that weak collisions can irreversibly convert the non-thermal
energy into plasma heat.  Boltzmann's $H$ theorem proves that the
entropy increase associated with irreversible plasma heating is
ultimately collisional \citep{Howes:2006}.

In addition to this linear phase-mixing process, at perpendicular
spatial scales comparable to the particle thermal Larmor radii,
$k_\perp \rho_s \gtrsim 1$, a nonlinear phase-mixing process
\citep{Dorland:1993}, also known as the entropy cascade
\citep{Schekochihin:2009,Tatsuno:2009,Plunk:2010,Plunk:2011,Kawamori:2013},
can be very effective at transferring energy to ever smaller scales in
velocity space. Ultimately, when the non-thermal particle energy in
the velocity distribution functions $\delta f_s(\V{v})$ has reached
sufficiently small scales in velocity, due to some combination of
linear and nonlinear phase mixing, collisions may thermalize that
particle energy, completing the final step in the conversion of
turbulent energy into plasma heat. In \T{AstroGK}, this collisional
heating removes energy from fluctuating energy in the plasma, $\delta
W$.

It is worthwhile to contrast this two-step mechanism in weakly
collisional plasmas---collisionless particle energization followed by
collisional thermalization---with the more familiar picture of
turbulent dissipation in the fluid (strongly collisional) limit.  A
dimensionless measure of the collisionality is the ratio of the
thermal collision rate to the frequency of typical fluctuations in the
plasma, $\nu/\omega$.  In the strongly collisional limit, $\nu/\omega
\gg 1$, collisions can directly remove energy from both the bulk
plasma flows through viscosity and the plasma currents through
resistivity.  Because both viscosity and resistivity are collisional,
entropy increases through these mechanisms, and the energy from the
turbulent electromagnetic field and plasma flow fluctuations is
immediately thermalized to plasma heat. Thus, the dissipation of
turbulence in the strongly collisional, fluid limit is a single-step
process. Consider the example of resistive MHD, where Ohm's Law gives
the electric field in terms of the plasma fluid velocity, magnetic
field, and current density, $\V{E} + \V{U}/c \times \V{B} = \eta
\V{j}$ \citep{Spitzer:1962,Kulsrud:1983}. The work done by the
electric field is $\V{j} \cdot \V{E} = - \V{j} \cdot (\V{U}/c \times
\V{B}) + \eta \V{j}^2$, where second term is the non-negative Ohmic
heating due to resistive dissipation of the current, showing that the
resistivity leads directly to plasma heating.

The strong \Alfven wave collision simulation presented here has
$\nu/\omega \sim 6 \times 10^{-4} \ll 1$, firmly in the weakly
collisional limit. Unlike in the MHD Ohm's Law above, where the
current density $\V{j}$ and electric field $\V{E}$ due to the
resistive term are in phase, and thereby yield a zero or positive
change in energy, in the weakly collisional case the current density
$\V{j}$ and electric field $\V{E}$ need not be in phase, enabling the
work done by collisionless interactions between the fields and
particles to give energy to or take energy from the particles. In
fact, if the current and electric field are exactly 90 degrees out of
phase, there is zero net energy transfer between fields and particles
over one complete oscillation, corresponding to undamped wave motion.
The bottom line, a point that cannot be overstated, is that in a
weakly collisional plasma, the electromagnetic work $\V{j} \cdot
\V{E}$ does not correspond to irreversible plasma heating, but rather
to reversible work done on the particles by the fields, or vice versa.

Developing a detailed understanding of particle energization and
plasma heating in heliospheric plasmas is grand challenge problem in
heliophysics, and this simple model of the energy flow provides
important constraints to focus efforts in that endeavor.  Note that
the final step of the process in \figref{fig:energyflow}, the
thermalization of the particle energy, is fundamentally collisional,
independent of what mechanism (which we have not specified here)
removed energy from the turbulent fluctuations initially.  The key
question in understanding particle energization and plasma heating in
heliospheric plasmas is therefore to understand the first step: what
collisionless and reversible mechanism is responsible for the removal
of energy from the turbulent fluctuations and conversion of that
energy into non-thermal energy of the plasma species?

\subsection{Rate of Energy Transfer}
\label{sec:rate}
Now we use the strong \Alfven wave collision simulation presented
here to analyze the channels of energy transfer shown in
\figref{fig:energyflow}.
For each species, the rate of change of non-thermal energy is given by
\begin{equation}
  \dot{E}^{(nt)}_s =  \dot{E}^{(fp)}_s - Q_s,
  \label{eq:edotnt}
\end{equation}
where the irreversible collisional heating $Q_s\ge 0$ but the
reversible collisionless field-particle energy transfer
$\dot{E}^{(fp)}_s$ can be either positive or negative.  In addition,
the rate of change of turbulent energy must be the sum of the
collisionless field-particle energy transfer for each species,
\begin{equation}
 -\dot{E}^{(turb)} = \dot{E}^{(fp)}_i + \dot{E}^{(fp)}_e.
\end{equation}
Note that we have not specified the physical mechanism governing the
field-particle energy transfer, but we are simply showing that the
transfers of energy indeed follow the diagram in
\figref{fig:energyflow}. The rate of field-particle energy transfer
presented below is calculated from \eqref{eq:edotnt} as the difference
between the rate of change of non-thermal particle energy and the
collisional heating rate for each species.

\begin{figure}
  \centerline{\resizebox{5.50in}{!}{\includegraphics{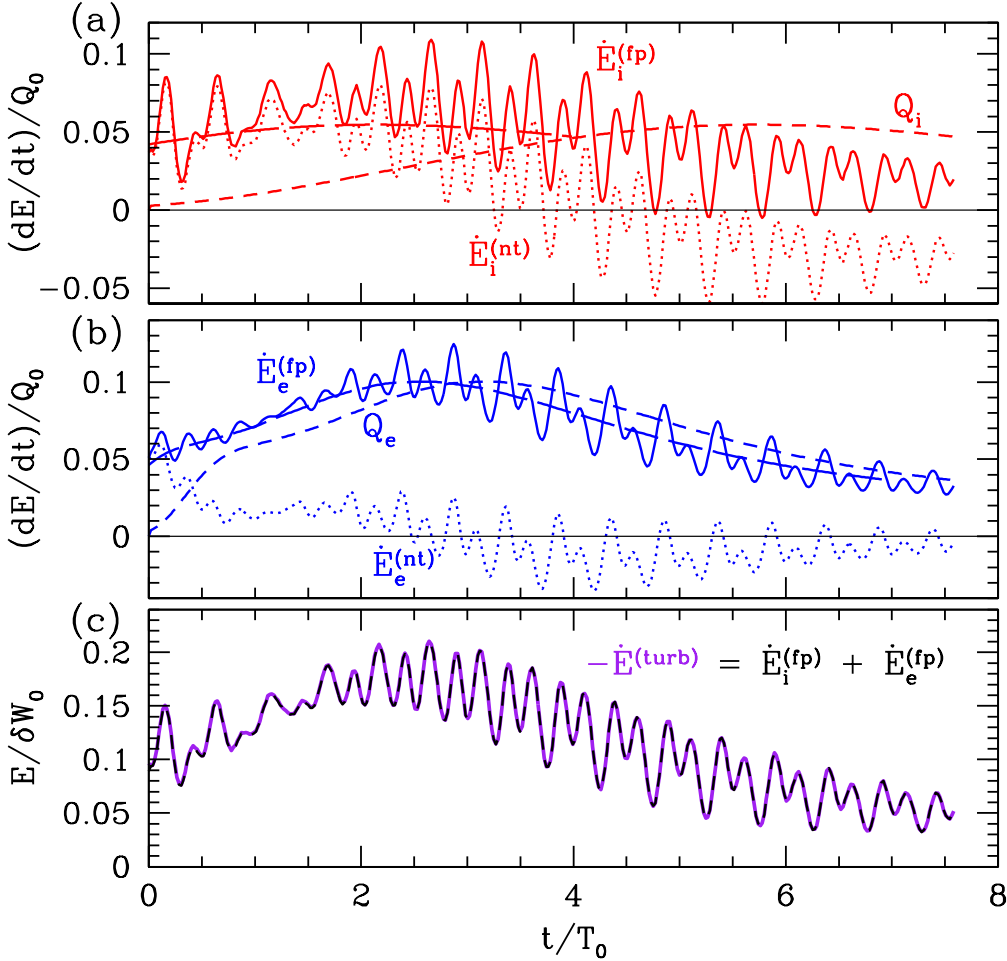}}
  }
  \caption{The rate of energy transfer by field-particle interactions
    $\dot{E}^{(fp)}_s$ (solid), the rate of change of non-thermal
    energy $\dot{E}^{(nt)}_s$ (dotted), and the collisional heating
    rate $Q_s$ (dashed) for (a) ions (red) and (b) electrons (blue).  (c) The
    energy balance between the loss of turbulent energy $
    -\dot{E}^{(turb)}$ (purple solid) and the summed transfer of energy to both ions
    and electrons, $\dot{E}^{(fp)}_i + \dot{E}^{(fp)}_e$ (black dashed).}
\label{fig:fpac22_dedt}
\end{figure}

In \figref{fig:fpac22_dedt}, we present the terms of these energy
transfer relations for the (a) ions and (b) electrons, as well as (c)
the balance between the loss of turbulent energy and the
field-particle energy transfer to each species.  A few very
interesting aspects of \figref{fig:fpac22_dedt} are worth
highlighting. First, although the change of turbulent energy
$E^{(turb)}$ and non-thermal energies $E^{(nt)}_s$ in
\figref{fig:fpac22_energy} appears to be smooth, the time derivative, which 
gives the rate of change, indeed varies rapidly, including a
significant fluctuation with period $T_0/2$.

Second, in \figref{fig:fpac22_dedt}(b), the energy
transferred into electron non-thermal energy at the rate
$\dot{E}^{(fp)}_e$ (solid) is very quickly thermalized by collisions
into electron heat (dashed); the time lag between these two curves is
$\Delta t = 0.6 T_0$ (not shown), suggesting that non-thermal energy
transferred into the electron velocity distribution is rapidly
transferred by phase mixing to sufficiently small velocity-space
scales to be thermalized by the weak collisions.  For the ions in
\figref{fig:fpac22_dedt}(a), on the other hand, the time lag between
the energy transferred into non-thermal ion energy $\dot{E}^{(fp)}_i$
and the thermalization of that ion energy is approximately $\Delta t =
3.6 T_0$, a factor of $\sqrt{m_i/m_e}=6$ longer, suggesting that the
phase-mixing occurs more slowly for ions by the ratio of the
electron-to-ion thermal velocity.  Note also that the collisionless
field-particle energy transfer to ions indeed becomes negative at a
few points in time, as allowed for a reversible process.

Furthermore, note that the magnitudes of $\dot{E}^{(fp)}_i$ and
$\dot{E}^{(fp)}_e$ are fairly similar, as expected because the linear
damping rates, shown in \figref{fig:fpac22_gk_disp}, are fairly
similar for ions and electrons, $\gamma_i \simeq \gamma_e$, over the
range of spatial scales $k_\perp \rho_i < 1$ that contain most of the
energy in the simulation. Finally, in \figref{fig:fpac22_dedt}(c), we
see that the energy lost by the turbulence $-\dot{E}^{(turb)}$ (purple
solid) is indeed balanced by the sum of the field-particle energy
transfer to ions and electrons (black dashed).

\subsection{Evolution of the Total Energy Budget}
Plots of the total energy budget as a function of time in the
simulation nicely summarize the flow of energy in the simulation.
First, we account for the energy lost from $\delta W$ in the
\T{AstroGK} simulation to collisional plasma heating by accumulating
the thermalized energy in each species over time, $E^{(coll)}_s (t) =
\int_0^t dt' Q_S(t')$.

\begin{figure}
  \centerline{\resizebox{4.25in}{!}{\includegraphics*[0.45in,2.5in][7.5in,8.25in]{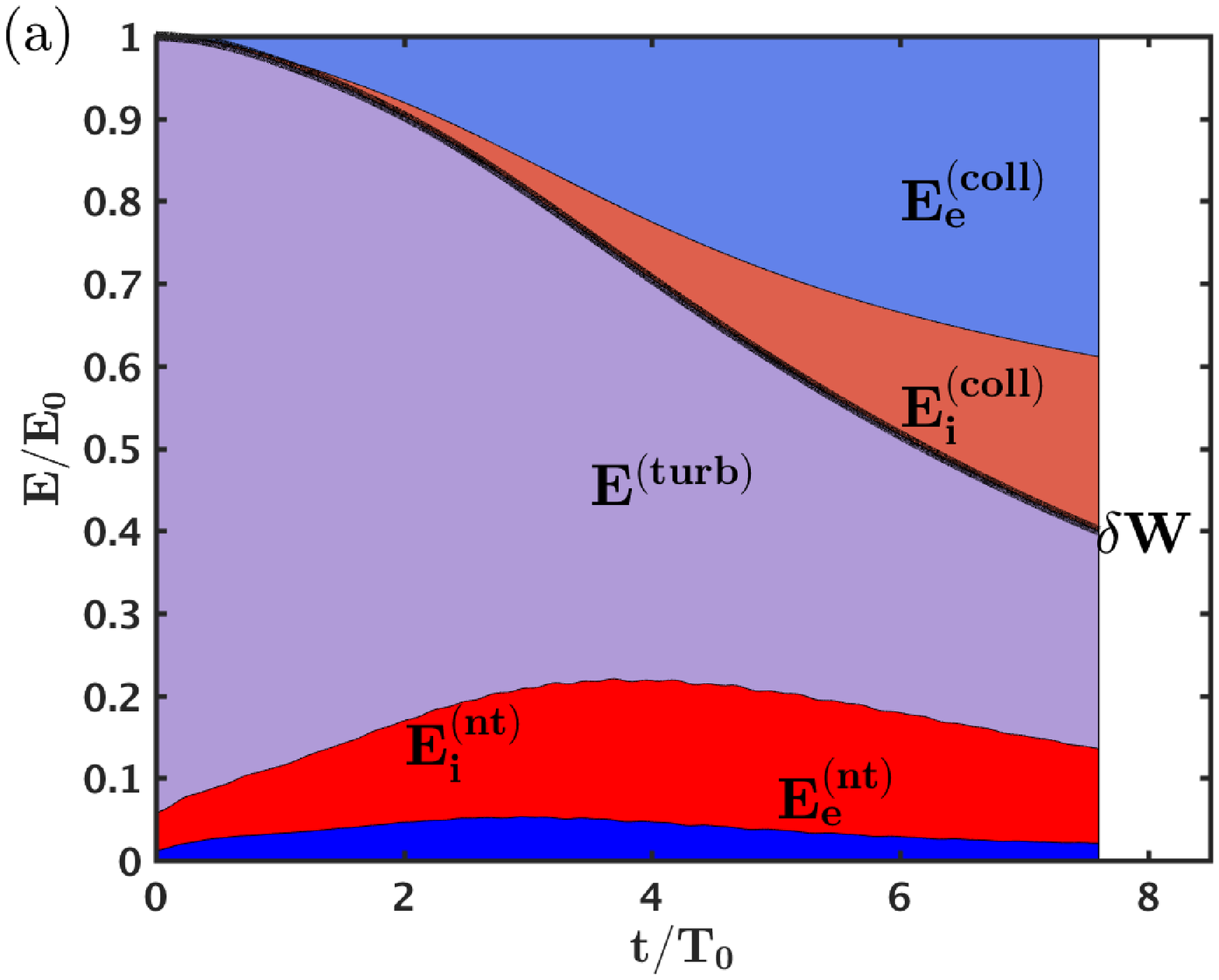}}
  }
  \centerline{\resizebox{4.25in}{!}{\includegraphics*[0.45in,2.5in][7.5in,8.25in]{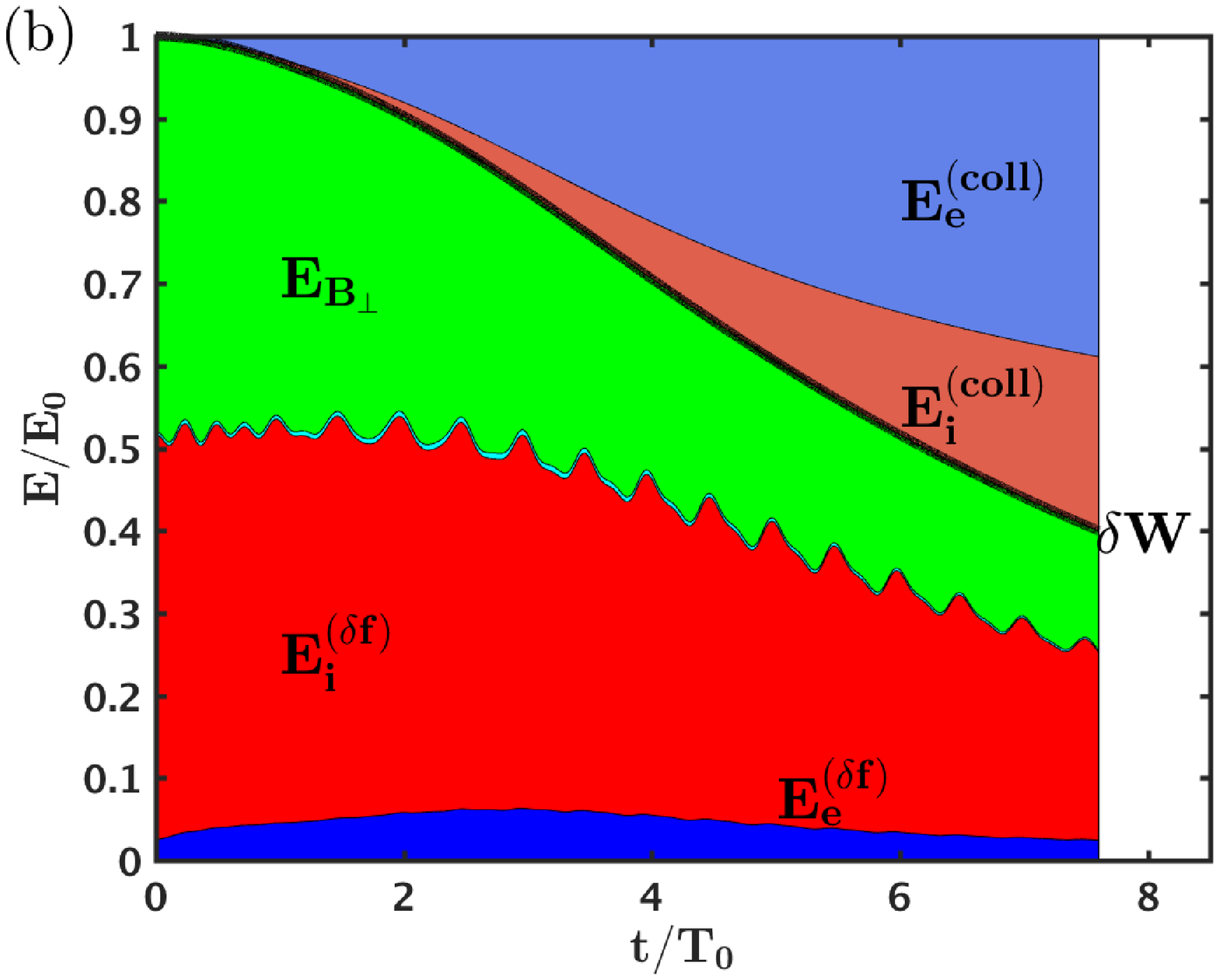}}
  }
  \caption{(a) The energy budget of the simulation vs.~time, showing
    the turbulent energy $E^{(turb)}$, non-thermal ion energy
    $E^{(nt)}_i$, non-thermal electron energy $E^{(nt)}_e$, ion heat
    $E^{(coll)}_i$ and electron heat $E^{(coll)}_e$. (b) The same
    energy budget decomposed according to \eqref{eq:deltaW_GK},
    showing the perpendicular magnetic field energy $E_{B_\perp}$,
    parallel magnetic field energy $E_{B_\parallel}$ (cyan, not
    labeled, appearing between $E_{B_\perp}$ and $E^{(\delta f)}_i$,),
    total fluctuating ion kinetic energy $E^{(\delta f)}_i$, total
    fluctuating electron kinetic energy $E^{(\delta f)}_e$, ion heat
    $E^{(coll)}_i$ and electron heat $E^{(coll)}_e$. The total
    fluctuating energy $\delta W$ is shown in both panels (thick black
    line).}
\label{fig:fpac22_area}
\end{figure}

In \figref{fig:fpac22_area}(a), we plot the evolution of the energy
budget over the course of the simulation, showing that turbulent
energy $E^{(turb)}$, which dominates at the beginning of the
simulation, is largely converted to ion heat $E^{(coll)}_i$ and
electron heat $E^{(coll)}_e$ by the end of the simulation, with a
smaller fraction of the lost  turbulent energy persisting as
non-thermal ion energy $E^{(nt)}_i$ and electron energy $E^{(nt)}_e$.
Also indicated in \figref{fig:fpac22_area}(a) is the evolution of the
total fluctuating energy $\delta W$ (thick black line), showing that
60\% of this energy has been lost to plasma heating over 7.5 periods
of the initial \Alfven waves. Another interesting point is that,
although electrons are heated twice as much as ions, the non-thermal
electron energy content of the simulation always remains very
small. This point is consistent with the idea, introduced in
\secref{sec:rate} above, that non-thermal energy transferred into the
electron velocity distribution function by collisionless damping of
the turbulence is very rapidly thermalized into electron heat.  This
analysis of the evolution of the total energy budget shows that energy
is conserved to within 0.1\% over the course of the simulation.

One can alternatively divide the contributions to the energy budget in
terms of \eqref{eq:deltaW_GK}, as shown in
\figref{fig:fpac22_area}(b), showing the perpendicular magnetic field
energy $E_{B_\perp}$ (green), the parallel magnetic field energy
$E_{B_\parallel}$ (cyan), the total fluctuating ion kinetic energy
$E^{(\delta f)}_i$ (red), and the total fluctuating electron kinetic
energy $E^{(\delta f)}_e$ (blue).  Note that, as anticipated from the
contributions to the turbulent energy in
\figref{fig:fpac22_energy}(b), the turbulent energy in
\figref{fig:fpac22_area}(a) is largely composed of perpendicular
magnetic energy $E_{B_\perp}$ and kinetic energy of the perpendicular
ion bulk flows $E_{i, u_\perp}$. The wiggly boundary between
$E_{B_\perp}$ and $E_{i, u_\perp}$ is a consequence of the \Alfvenic
fluctuations, and their nonlinear interactions, in the simulation.

One final point is that, although one may choose to decompose the
different contributions to the energy using \eqref{eq:deltaW_GK} in
\figref{fig:fpac22_area}(b), by organizing the energies instead
according to the turbulent energy $E^{(turb)}= \int d^3 \V{r} [
  (|\delta \V{B}|^2+|\delta \V{E}|^2)/8\pi + \sum_s
  \frac{1}{2}n_{0s}m_s|\delta \V{u_s}|^2 ]$ and the species
non-thermal energies $E^{(nt)}_s$, the interpretation of the energy
flow is much more physically motivated, as illustrated by
\figref{fig:energyflow}. By simply plotting $E^{(\delta f)}_i$ as a
function of time, one does not see the important split between the
large fraction of the total fluctuating ion kinetic energy $E^{(\delta
  f)}_i$ that is associated with turbulent fluctuations and the
remainder that corresponds to non-thermal energy not associated with
turbulent fluctuations.



\section{Development of Current Sheets and Intermittent Particle Energization}
\label{sec:cs}
In the limit of strong nonlinearity, $\chi \sim 1$---corresponding to
the important case of critically balanced, strong MHD turbulence
\citep{Goldreich:1995}---recent work has shown that \Alfven wave
collisions self-consistently develop intermittent current sheets
\citep{Howes:2016b}.  This finding may indeed explain the ubiquitous
current sheets found to develop in simulations of plasma turbulence
\citep{Wan:2012,Karimabadi:2013,TenBarge:2013a,Wu:2013,Zhdankin:2013}
and inferred from spacecraft observations of the solar wind
\citep{Osman:2011,Borovsky:2011,Osman:2012a,Perri:2012a,Wang:2013,Wu:2013,Osman:2014b}.
Yet how this self-consistent development of current sheets influences
the physical mechanisms that remove energy from plasma turbulence
remains unanswered. We show in this section that the simulation
reported here indeed develops intermittent currents sheets
(intermittent in both time and space), and in section
\secref{sec:fpcorr} we employ the field-particle correlation technique
to examine the physical mechanism that removes energy from the
turbulent fluctuations.

\begin{figure}
  \centerline{\resizebox{2.6in}{!}{\includegraphics{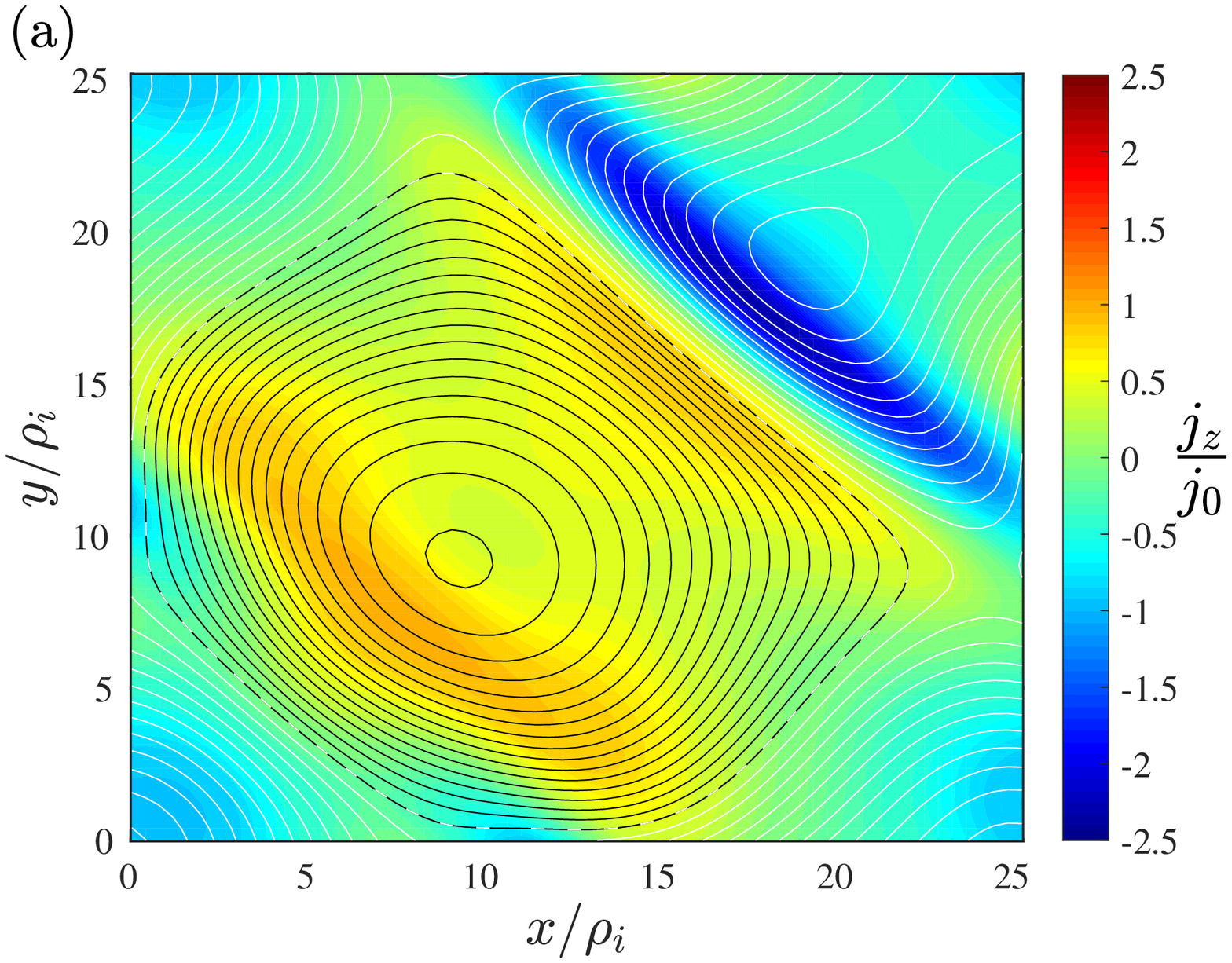}}
\resizebox{2.6in}{!}{\includegraphics{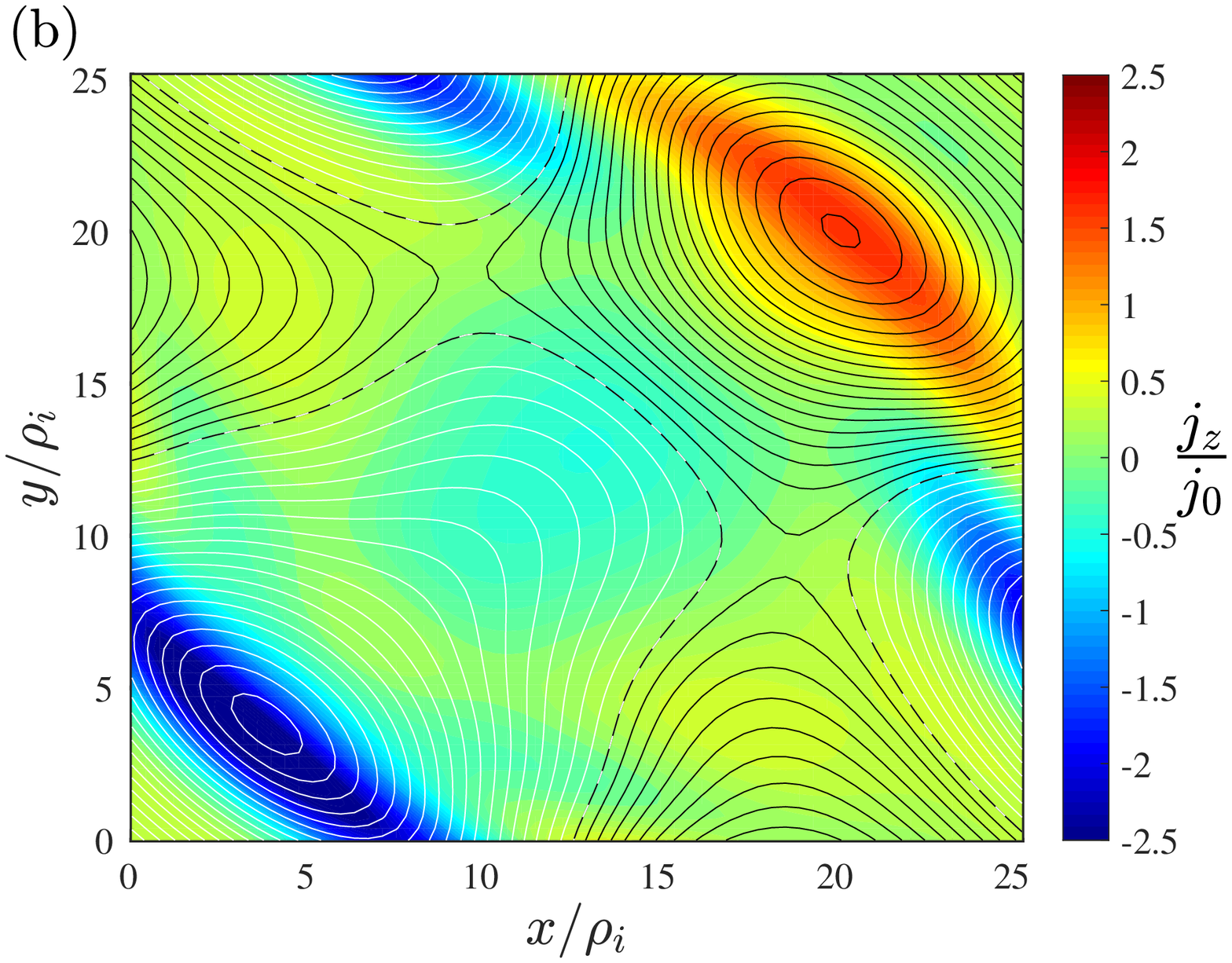} }   }
  \centerline{\resizebox{2.6in}{!}{\includegraphics{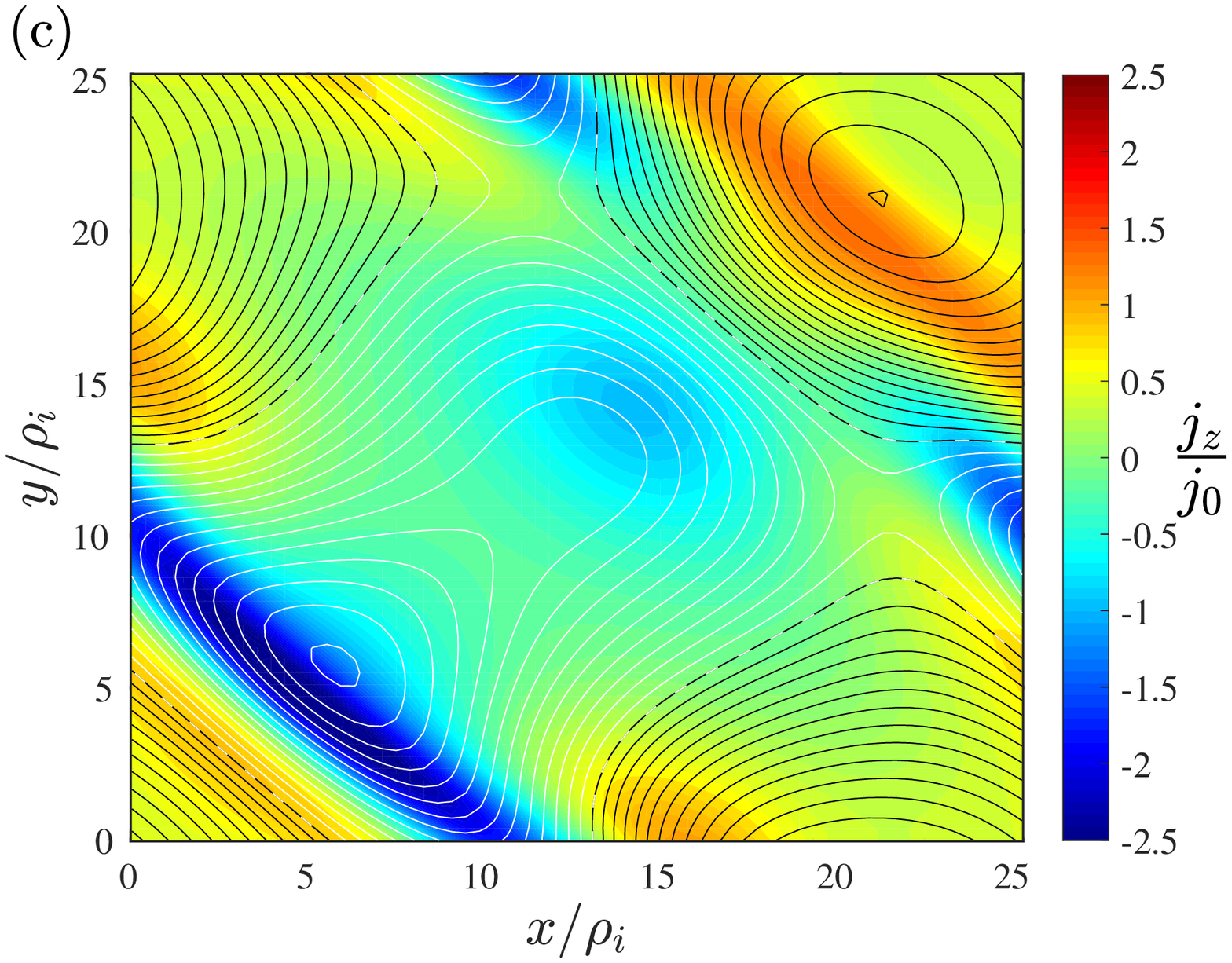}}
\resizebox{2.6in}{!}{\includegraphics{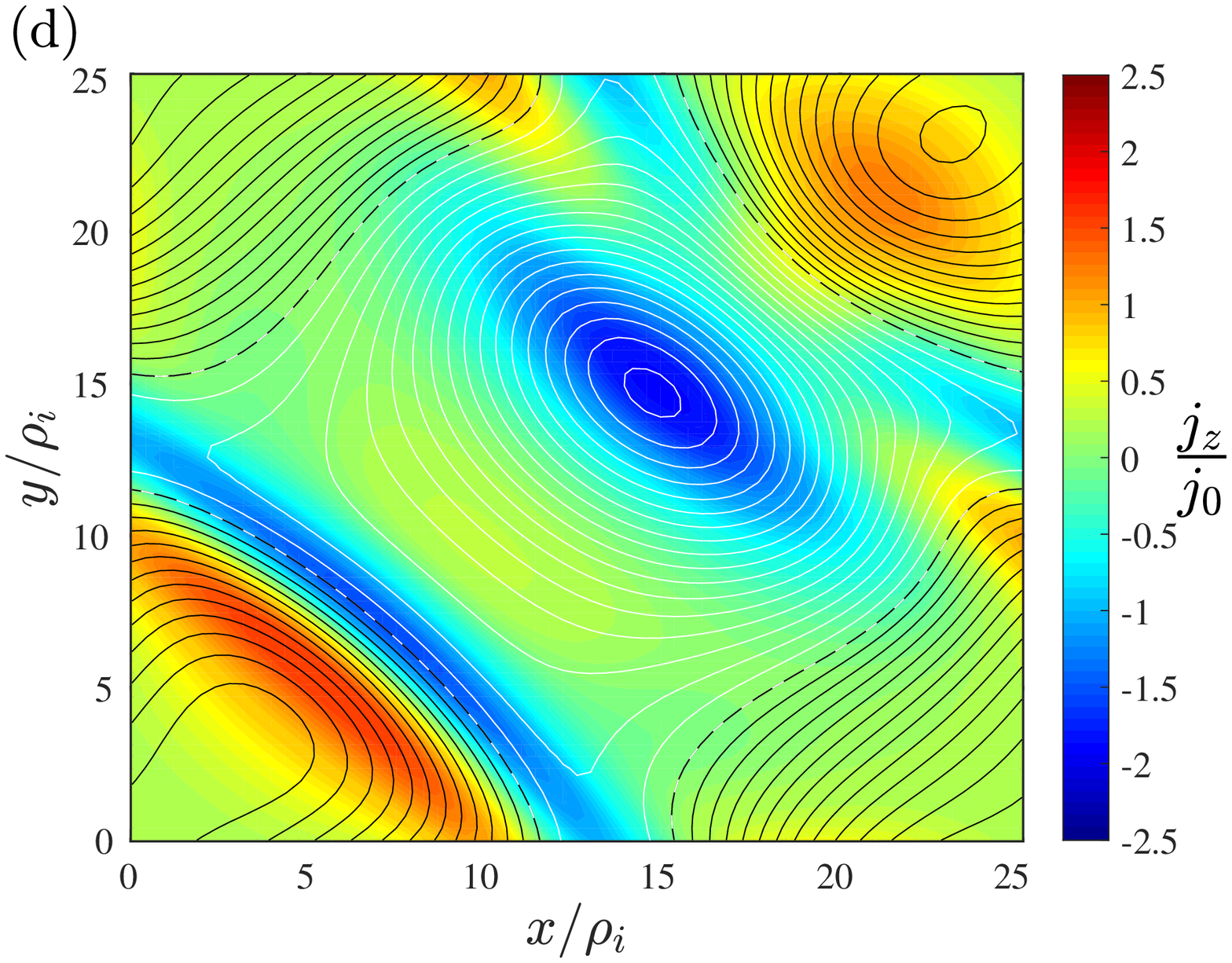} }   }
\caption{Plots of parallel current $j_\parallel/j_0$ (colorbar) and contours of
  the parallel vector potential $A_\parallel$ (contours, positive
  black, negative white) at times $t/T_0=$ (a) 1.38, (b) 1.75, (c)
  1.86, and (d) 2.03.}
\label{fig:fpac22_jz}
\end{figure}

In \figref{fig:fpac22_jz} we plot the current density parallel to the
mean magnetic field $j_\parallel/j_0$ (colorbar) and contours of parallel
vector potential $A_\parallel$ (positive black, negative white) in the
plane $z/L_\parallel=-0.25$, where the simulation domain spans
$-L_\parallel/2 \le z \le L_\parallel/2$ and $j_0=n_0q_i v_{ti}
L_\perp/L_\parallel$.  We plot evolution of the current in this plane
at four different times in the evolution of the strong \Alfven wave
collision, $t/T_0=$ (a) 1.38, (b) 1.62, (c) 1.86, and (d) 2.10. Here
$T_0=2 \pi /\omega$ is the period of the initial \Alfven waves, where
the gyrokinetic linear dispersion relation gives $\omega/k_\parallel
v_A = 0.995$ and $\gamma/k_\parallel v_A = -6.10 \times
10^{-3}$. These plots show the presence of intermittent, elongated
sheets of localized current density. Over a single initial \Alfven
wave period $T_0$, two current sheets form at slightly different
times, become thinner and more intense, and then disappear. One of
these current sheets appears in the upper right quadrant of the plane
$z/L_\parallel=-0.25$, and the other in the lower left quadrant, as
shown in \figref{fig:fpac22_jz}.  During this time, their cross
sections in the plane plotted in \figref{fig:fpac22_jz} moves slowly
across the quadrant of the domain in which each appears (but these
intermittent current sheets do not cross the entire domain, as would
be expected from a strictly linear fluctuation).  The general picture
of current sheet development and evolution in a strong \Alfven wave
collision is described in more quantitative detail by
\citet{Howes:2016b}; although the parameters of this simulation are
slightly different, the evolution of the current sheets is
qualitatively similar here.

\subsection{Spatial Distribution of Parallel Electromagnetic Work, $j_\parallel E_\parallel$ }
As shown in \secref{sec:energy}, over the full time of the simulation,
$7.5 T_0$, 60\% of the fluctuating energy $\delta W$ of the initial
\Alfven waves is removed from the fluctuations in the
plasma. \figref{fig:fpac22_heat} shows that this energy is ultimately
irreversibly converted into electron and ion heat through the weak but
finite collisionality in the plasma. As the model of energy flow
illustrated in \figref{fig:energyflow} shows, this energy is initially
removed from the turbulent electromagnetic fluctuations
\citep{Howes:2015b,Howes:2017c} through collisionless interactions
between the electromagnetic fields and the individual plasma
particles.  In a kinetic plasma, the rate of electromagnetic work done
on the particles by the fields is given by $d W /dt=\int d^3\V{r}
\ \V{j} \cdot \V{E}$ \citep{Howes:2017a,Klein:2017a}. Therefore,
plotting the rate of electromagnetic work $\V{j} \cdot \V{E}$ as a
function of position provides useful insights into the particle
energization in the plasma.

\begin{figure}
  \centerline{\resizebox{2.6in}{!}{\includegraphics{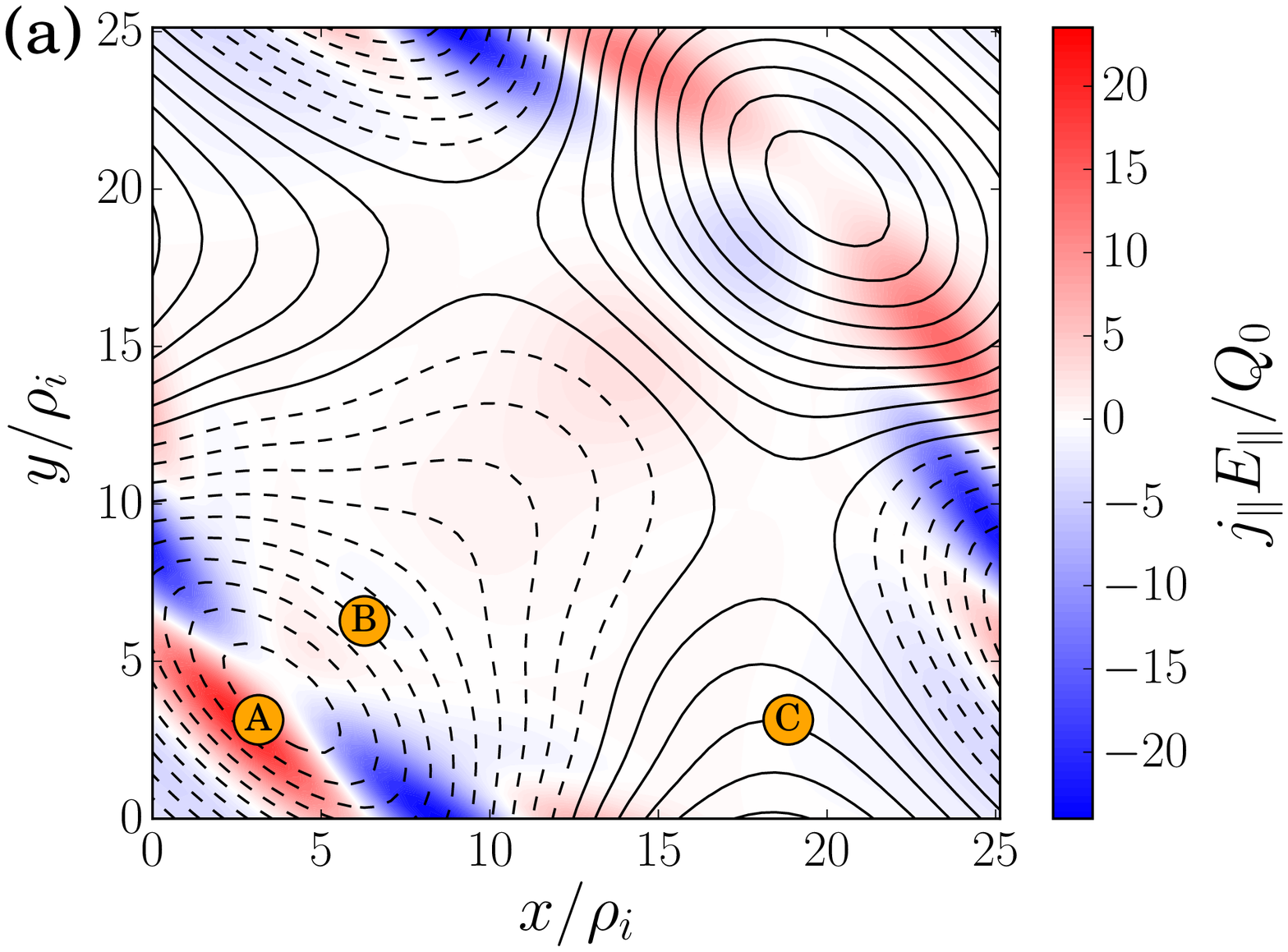}}
\resizebox{2.6in}{!}{\includegraphics{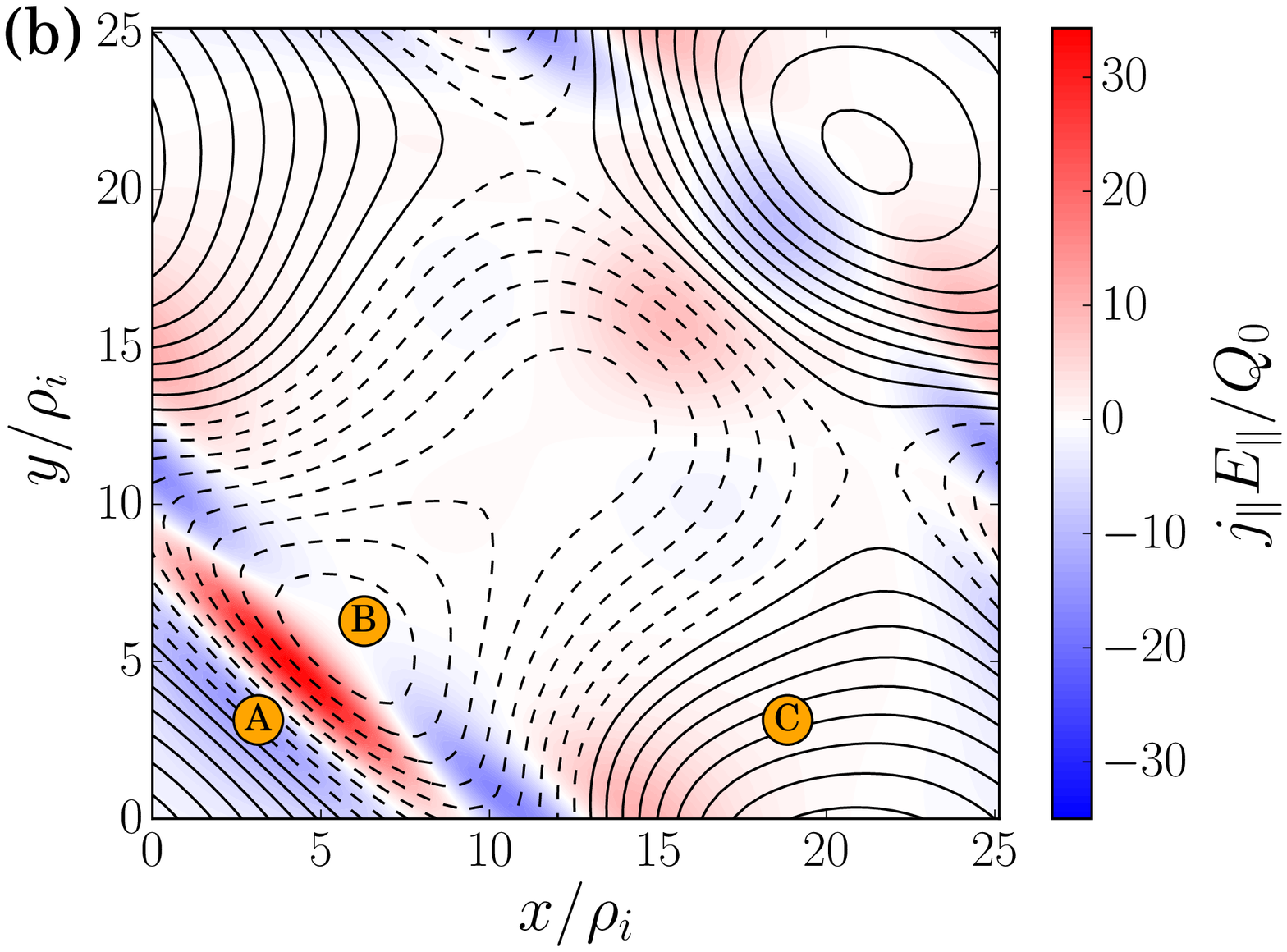} }   }
  \centerline{\resizebox{2.6in}{!}{\includegraphics{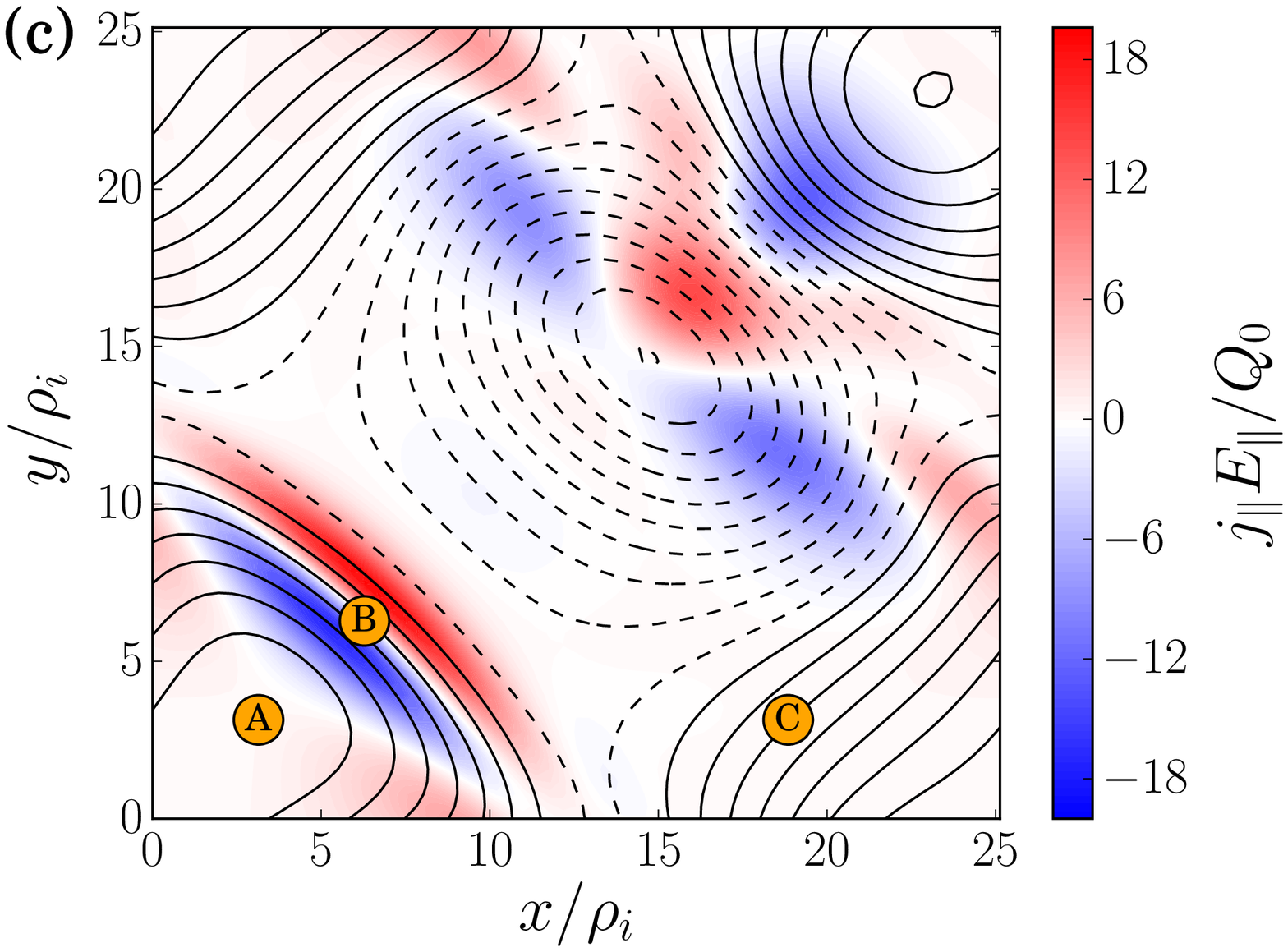}}
\resizebox{2.6in}{!}{\includegraphics{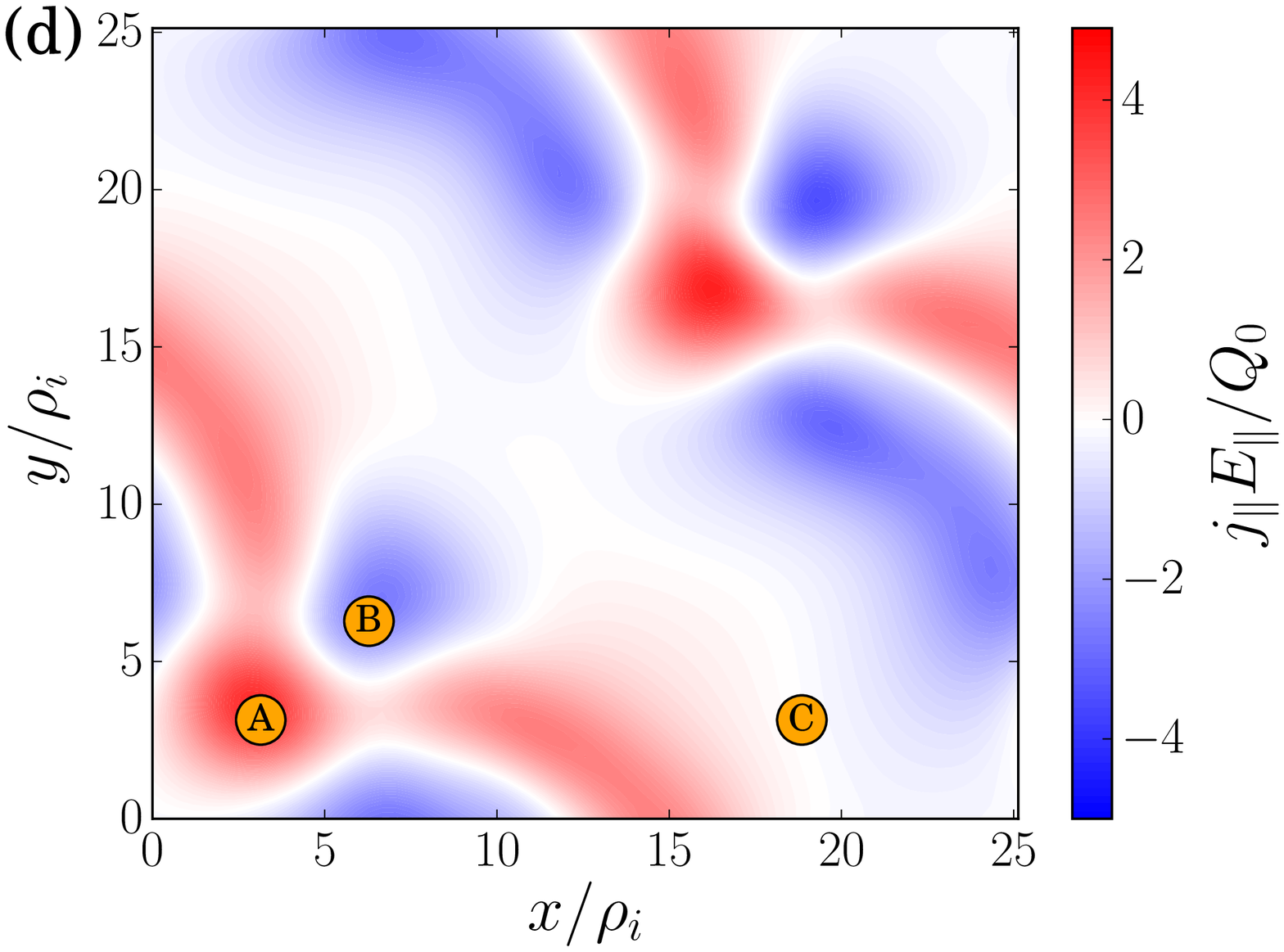} }   }
\caption{Plots of $j_\parallel E_\parallel$ (colorbar)  and contours of
  the parallel vector potential $A_\parallel$ (contours, positive
  solid, negative dashed) at times $t/T_0=$ (a)
  1.75, (b) 1.86, and (c) 2.03, as well as (d) $\langle j_\parallel
  E_\parallel\rangle_\tau$, the rate of electromagnetic work per unit
  volume averaged over approximately one full wave period, $\tau=0.992
  T_0$, centered at time $t/T_0= 1.86$.}
\label{fig:fpac22_jzez}
\end{figure}

As shown in \appref{app:je}, in this simulation the dominant
electromagnetic work is done by the component of the electric field
parallel to the magnetic field, $E_\parallel$, so in
\figref{fig:fpac22_jzez} we plot the instantaneous value of
dimensionless rate of work per unit volume $j_\parallel
E_\parallel/Q_0$ as a function of position in the plane
$z/L_\parallel=-0.25$ at three different times during the simulation
$t/T_0=$ (a) 1.75, (b) 1.86, and (c) 2.03.  Note that the value of
$j_\parallel E_\parallel$ is physically interpreted as the rate of
transfer of spatial energy density between the parallel electric field
$E_\parallel$ and the plasma ions and electrons. Since this
electromagnetic work is reversible, its value can be positive or
negative, where positive means work done on the particles by the
field, and negative means work done on the field by the particles.

As emphasized in \citet{Howes:2017a}, the instantaneous energy
transfer between fields and particles has two components: (i) an
\emph{oscillating energy transfer} back and forth between fields and
particles that is typical of undamped linear wave motion in a kinetic
plasma, and (ii) a \emph{secular energy transfer} that represents the
energy lost from the electromagnetic fluctuations to the plasma
particles through collisionless damping. To determine the particle
energization, it is the secular energy transfer that is of interest,
but the challenge is that the oscillating energy transfer often has a
much larger amplitude than the secular energy transfer.  However, if a
time-average is taken over a suitably chosen averaging interval, the
oscillating energy transfer will largely cancel out, exposing the
smaller secular energy transfer that is sought.  In this strong
\Alfven wave collision simulation, the linear period $T_0$ of the initial
\Alfven waves is an appropriate choice for this time-averaging, and we
plot in \figref{fig:fpac22_jzez}(d) the time-average of $\langle
j_\parallel E_\parallel \rangle_\tau$ over an interval $\tau=0.992 T_0$ centered at
time $t/T_0= 1.86$. 

The plots shown in \figref{fig:fpac22_jzez} convey a number of
valuable insights into the particle energization in this simulation.
First, the plots in \figref{fig:fpac22_jzez}(a)-(c) show clearly that
the instantaneous rate of energy transfer is both spatially and
temporally intermittent, with the energy transfer localized in
sheet-like structures reminiscent of the current sheets plotted in
\figref{fig:fpac22_jz}. An example of the temporal variation is
illustrated by observing the changes in the instantaneous energy
transfer rate at point A marked on each plot. At (a) $t/T_0=1.75$, the
plasma is energized by $E_\parallel$, but later at (b) $t/T_0=1.86$
the plasma is losing energy to $E_\parallel$, and finally at (c)
$t/T_0=2.03$ there is very little energy transfer either
direction. Averaged over one period, \figref{fig:fpac22_jzez}(d) shows
that the plasma gains energy from the parallel electric field at point
A. Curiously, the instantaneous energy transfer from fields to
particles at $t/T_0=1.86$ in \figref{fig:fpac22_jzez}(b) is negative
at point A, but the single-period average, centered at that same time
$t/T_0=1.86$ in \figref{fig:fpac22_jzez}(d) shows a positive transfer
of energy to the particles at the same position. This plot stresses
the importance of appropriate time-averaging to properly understand
the net particle energization in a turbulent plasma.

Second, the net plasma energization over one period in
\figref{fig:fpac22_jzez}(d) is also spatially intermittent, with
plasma energization at point A, a net loss of energy at point B, and
little energy change at point C. It is also worthwhile pointing out
that the magnitude of the time-averaged energy transfer is smaller in
magnitude than the instantaneous energy transfer, as expected if some
fraction of this energy transfer is oscillatory and largely cancels
out when averaged over one period $T_0$.  Together, the four panels
demonstrate the key point that that the particle energization is
spatially non-uniform, both instantaneously as well as when averaged
over one period $T_0$ of the initial \Alfven waves.

Third, something that cannot be appreciated by the single time slice
in \figref{fig:fpac22_jzez}(d), is the surprising result that the
single-period-averaged plasma energization has very little temporal
variation as the center of the time-average window is advanced over one
period. In fact, one observes only a very slow evolution of this
particle energization pattern over a number of periods, probably due
to the long-term evolution and accumulating loss of fluctuating energy
$\delta W$ over the course of the simulation.

A final point is that the plasma energization---the sum of the energy
transfer to both ions and electrons---has a net positive value when
integrated over the entire simulation domain, as demonstrated by the
sum $ \dot{E}^{(fp)}_i + \dot{E}^{(fp)}_e$ plotted in
\figref{fig:fpac22_dedt}(c).  Therefore, although there is a loss of
plasma energy in some regions of the domain, the net effect is that
plasma species gain energy at the expense of the turbulent
electromagnetic field and bulk plasma flow fluctuations, as depicted
in the energy flow diagram  in \figref{fig:energyflow}.

\begin{figure}
  \centerline{\resizebox{2.6in}{!}{\includegraphics{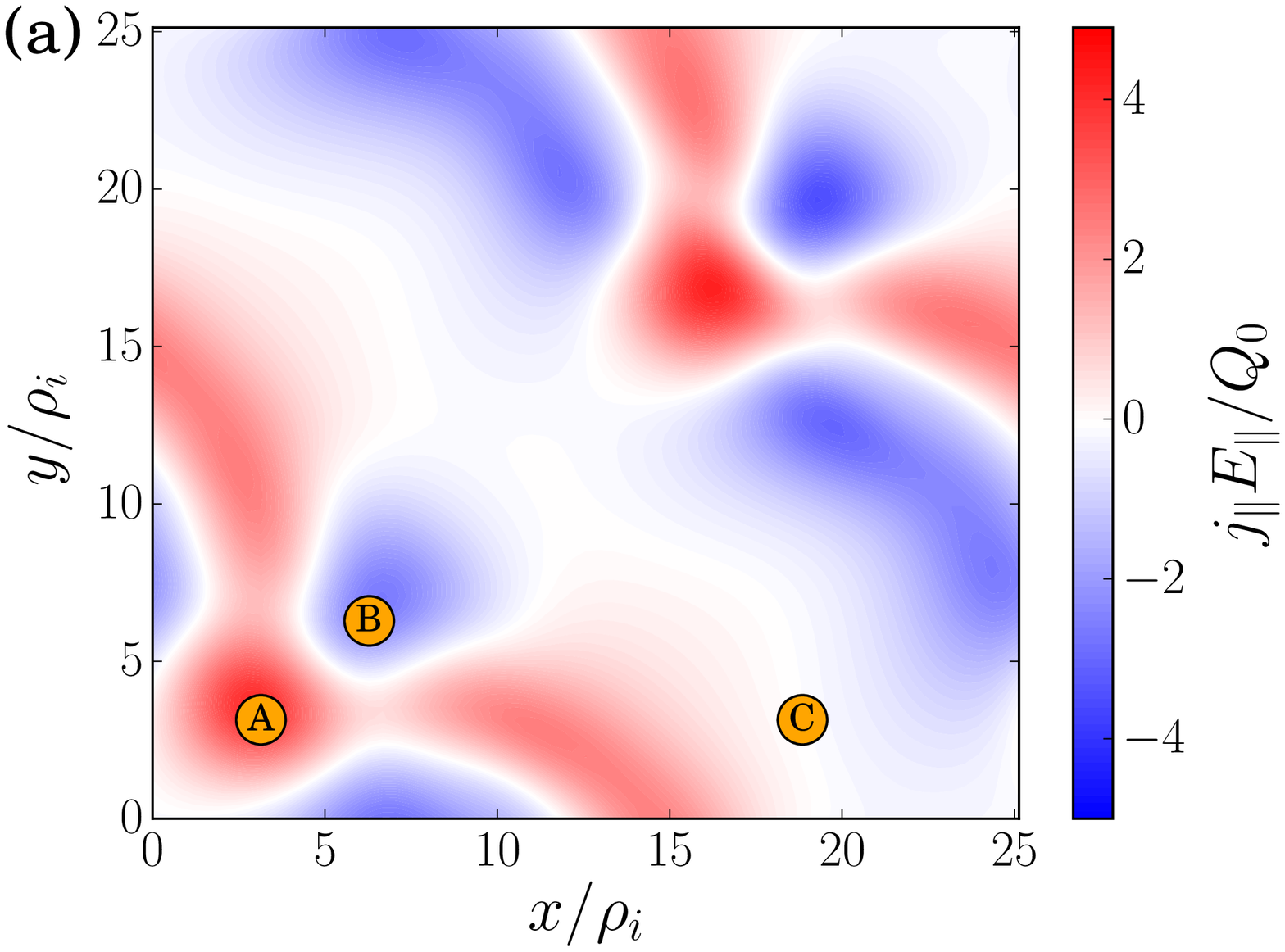} }
    \resizebox{2.6in}{!}{\includegraphics{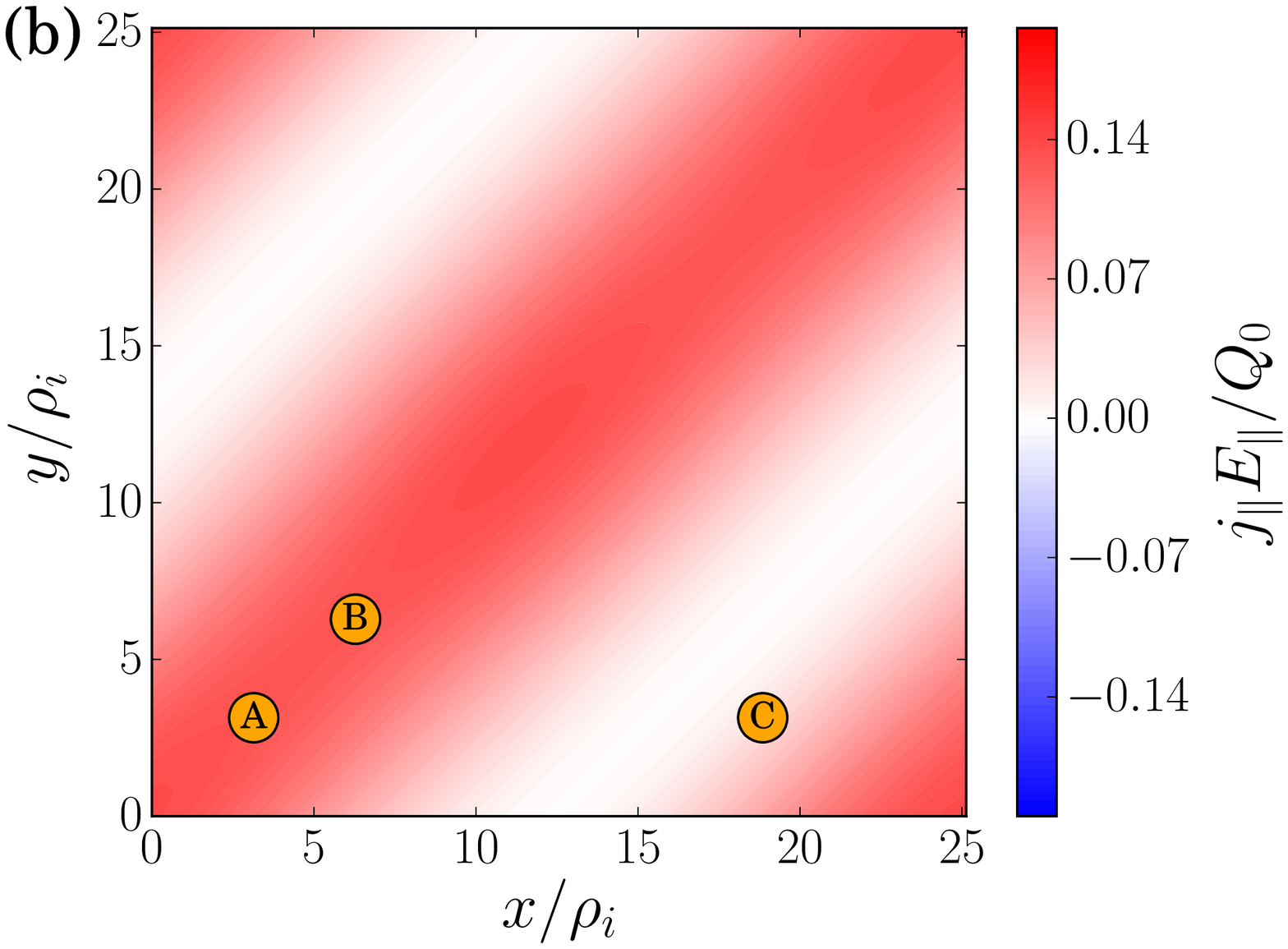}}}
\caption{ Comparison of time-averaged $\langle j_\parallel E_\parallel \rangle_\tau$
  over an interval $\tau=0.992 T_0$ centered at time $t/T_0= 1.86$ for
  both  (a) a nonlinear run and  (b) a  linear run, starting from identical
  initial conditions, showing a much more spatially intermittent
  distribution of plasma energization in the nonlinear case.}
\label{fig:fpac22_jzez_lin_nl}
\end{figure}

Further insight into the effect of the nonlinear evolution on the
resulting plasma energization can be gained by comparing a linear
simulation starting from identical initial conditions. In the linear
case, no energy is transferred to other Fourier modes, and all of the
particle energization is due to linear collisionless damping via the
Landau resonance.  In \figref{fig:fpac22_jzez_lin_nl}, we plot the
time-averaged $\langle j_\parallel E_\parallel \rangle_\tau$ over an
interval $\tau=0.992 T_0$ centered at time $t/T_0= 1.86$ for both (a)
the nonlinear run and (b) the linear run.  This figure directly
demonstrates the striking fact that the spatial non-uniformity of
particle energization arises due to the nonlinear transfer of energy to
other Fourier modes.  This is fully consistent with the picture of
current sheet generation by constructive interference among the
initial \Alfven wave modes and the nonlinearly generated fluctuations
\citep{Howes:2016b}. In \secref{sec:fpcorr}, the field-particle
correlation technique will be used to identify the nature of the
collisionless energy transfer that yields this spatially non-uniform
particle energization.

The rate of plasma energization is the sum of the rates of ion and
electron energization, $j_\parallel E_\parallel = j_{\parallel i}
E_\parallel + j_{\parallel e} E_\parallel$, and, in \appref{app:je}, we
plot in Figs.~\ref{fig:fpac22_jzez_s1} and~\ref{fig:fpac22_jzez_s2}
the separate ion and electron energization contributing to
\figref{fig:fpac22_jzez}.  In this simulation, the single-period
averaged particle energization in the plane $z/L_\parallel=-0.25$
shown in \figref{fig:fpac22_jzez} yield about twice the energy
transfer to electrons relative to the ions at $t/T_0= 1.86$.

\section{Analysis of Energy Transfer Mechanism Using Field-Particle Correlations}
\label{sec:fpcorr}
The rates of total energy transfer as a function of time between the
turbulent fluctuations and the ions and electrons, plotted in
\figref{fig:fpac22_dedt}, give the desired information about the net
collisionless particle energization over the entire simulation domain.
But this simple approach cannot be applied to the analysis of
spacecraft measurements to understand heating in heliospheric plasmas,
because spacecraft measure  the particle velocity distributions and
electromagnetic fields at only one or a few points in space, so it is not
possible to integrate the plasma heating over the entire plasma
volume. In addition, such an energy flow analysis alone, such as that given
in the diagram in \figref{fig:energyflow}, tells us nothing of the
mechanism leading to the particle energization.

The electromagnetic work, $\V{j} \cdot \V{E}$, can be computed with
single point measurements, providing more insight into the nature of
the particle energization mechanism and the spatial distribution of
energy transfer than an energy analysis alone, but the newly developed
field-particle correlation technique \citep{Klein:2016a,Howes:2017a},
which yields the distribution of the energy transfer as a function of
particle velocity, gives far greater detail about the nature of the
energy transfer mechanism. This technique requires only a single-point
time series of both field and velocity distribution function
measurements, which can be obtained using modern spacecraft
instrumentation.

The field-particle correlation technique has been successfully applied
to examine the electron energization due to the damping of
electrostatic fluctuations in a 1D-1V Vlasov-Poisson plasma
\citep{Klein:2016a,Howes:2017a}, to determine the transfer of free
energy in kinetic instabilities from unstable particle velocity
distributions to electrostatic fluctuations \citep{Klein:2017a}, and
to explore the particle energization caused by the collisionless
damping of strong, broadband, gyrokinetic plasma turbulence
\citep{Klein:2017b}. Here we apply the technique to discover the
nature of the physical mechanism responsible for the spatially
intermittent transfer of energy from the turbulent fluctuations to the
non-thermal energy of the ions and electrons in the plasma.

Specifically, since we know from \figref{fig:fpac22_jxyzfull} in
\appref{app:je} that the net energy transfer is dominated by the
parallel electric field, we will evaluate the correlation of the ion
and electron fluctuations with the parallel electric field.  The
correlation of the parallel electric field, $E_\parallel$, with a species $s$
is defined by
\begin{equation}
  C_{E_\parallel,s} (\V{v},t,\tau)= C\left(- q_s\frac{v_\parallel^2}{2}
  \frac{\partial f_s(\V{r}_0,\V{v},t)}{\partial
    v_\parallel},E_\parallel(\V{r}_0,t)\right).
   \label{eq:cepar}
\end{equation}
This unnormalized correlation is taken over an appropriately chosen
correlation interval $\tau$ to suppress the signal of the oscillatory
energy transfer relative to the secular energy transfer
\citep{Howes:2017a}.  Defining the \emph{phase-space energy density}
by $w_s(\V{r},\V{v},t) = m_s v^2 f_s(\V{r},\V{v},t)/2$, this
unnormalized correlation yields the phase-space energy transfer rate
between the parallel electric field $E_\parallel$ and species $s$
given by the Lorentz term in the Vlasov equation
\citep{Howes:2017c,Klein:2017b}. A key aspect of this novel analysis
method is that it retains the dependence of the energy transfer on
velocity space.  Note that integrating this correlation over velocity
space simply yields the parallel contribution to the electromagnetic
work, $j_\parallel E_\parallel = \int d\V{v} C_{E_\parallel}
(\V{v},t,\tau)$ \citep{Howes:2017a,Klein:2017b}.

For the application of this technique to data from our gyrokinetic
simulation using \T{AstroGK}, we note that the gyrokinetic
distribution function $h_s(x,y,z,v_\perp,v_\parallel)$ is related to
the total distribution function $f_s$ via \citep{Howes:2006}
\begin{equation}
f_s(\V{r}, \V{v}, t) = 
F_{0s}(v)\left( 1 - \frac{q_s \phi(\V{r},t)}{T_{0s}} \right)
+ {h_s}(\V{r},  v_\parallel, v_\perp,t).
\label{eqn:fullF}
\end{equation}
As a technical step, we transform from the gyrokinetic distribution
function $h_s$ to the complementary perturbed distribution function 
\begin{equation}
{g_s}(\V{r}, v_\parallel,v_\perp) = {h_s}(\V{r}, v_\parallel,v_\perp) 
- \frac{q_s F_{0s}}{T_{0s}} 
\left\langle 
\phi 
- \frac{\V{v}_\perp \cdot \V{A}_\perp}{c}
\right\rangle_{\V{R}_s},
\label{eqn:g+h}
\end{equation}
where $\langle ...\rangle$ is the gyroaveraging operator
\citep{Schekochihin:2009}. The complementary distribution function
$g_s$ describes perturbations to the background distribution in the
frame of reference moving with the transverse oscillations of an
\Alfven wave.  Field-particle correlations calculated using $h_s$ or
$f_s$ yield qualitatively and quantitatively similar results to those
computed with $g_s$ \citep{Klein:2017b}.

Below, we present the correlations between the complementary perturbed 
distribution function and the parallel electric field $E_\parallel$ at a
single-point $\V{r}_0$
\begin{equation}
  C_{E_\parallel,s} (v_\parallel,v_\perp,t,\tau)= C\left(- q_s\frac{v_\parallel^2}{2}
  \frac{\partial g_s(\V{r}_0,v_\parallel,v_\perp,t)}{\partial
    v_\parallel},E_\parallel(\V{r}_0,t)\right).
   \label{eq:cepar_gs}
\end{equation}
To explore the particle energization over time, we can also integrate
over the perpendicular velocity $v_\perp$ to obtain a reduced parallel correlation
\begin{equation}
  C_{E_\parallel,s} (v_\parallel,t,\tau)= \int v_\perp d v_\perp
  C_{E_\parallel,s} (v_\parallel,v_\perp,t,\tau).
   \label{eq:cepar_reduced}
\end{equation}
This reduced parallel correlation $C_{E_\parallel,s}
(v_\parallel,t,\tau)$ can be plotted as a function of
$(v_\parallel,t)$ to illustrate the time evolution particle
energization using a timestack plot of the energy transfer as a
function of the parallel velocity of the particles.

\subsection{Timestack Plots of Field-Particle Correlations}
\label{sec:timestack}

\begin{figure}
  \centerline{\resizebox{4.25in}{!}{\includegraphics{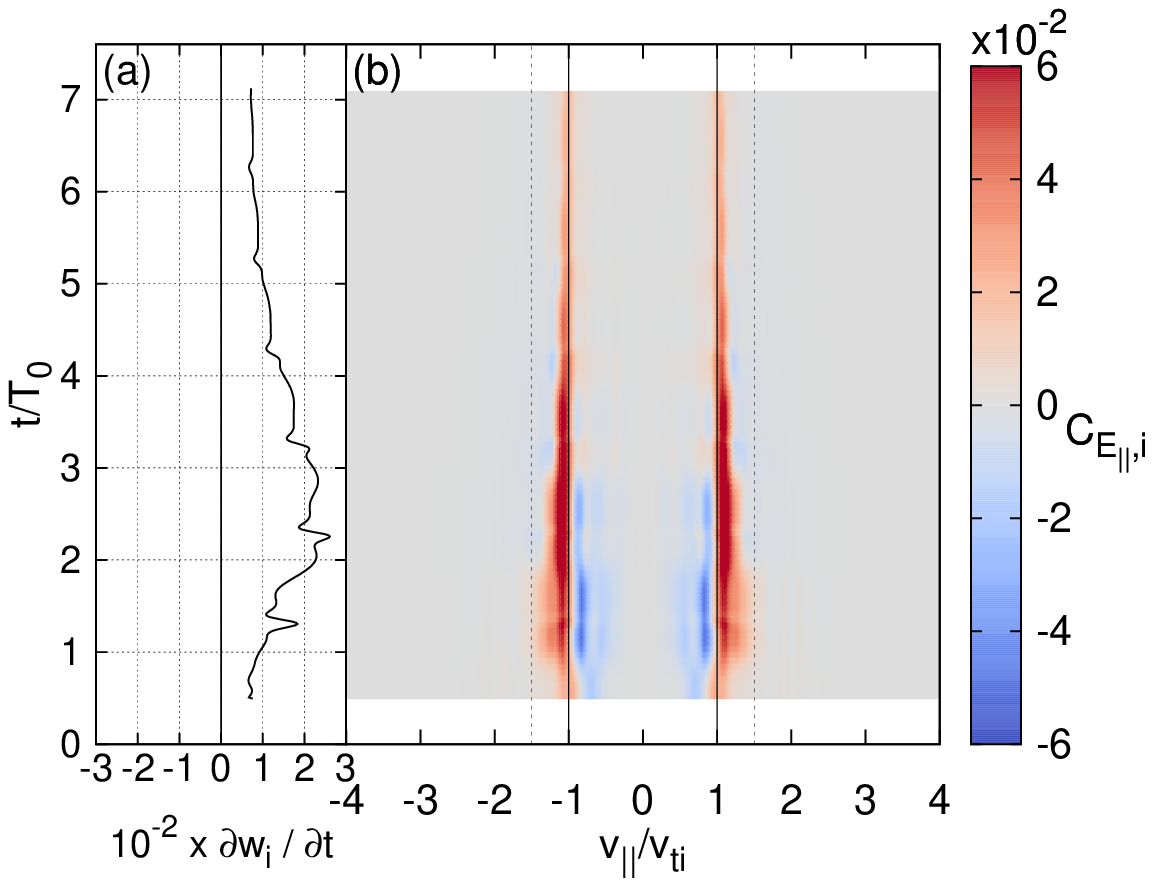}}}
  \centerline{\resizebox{4.25in}{!}{\includegraphics{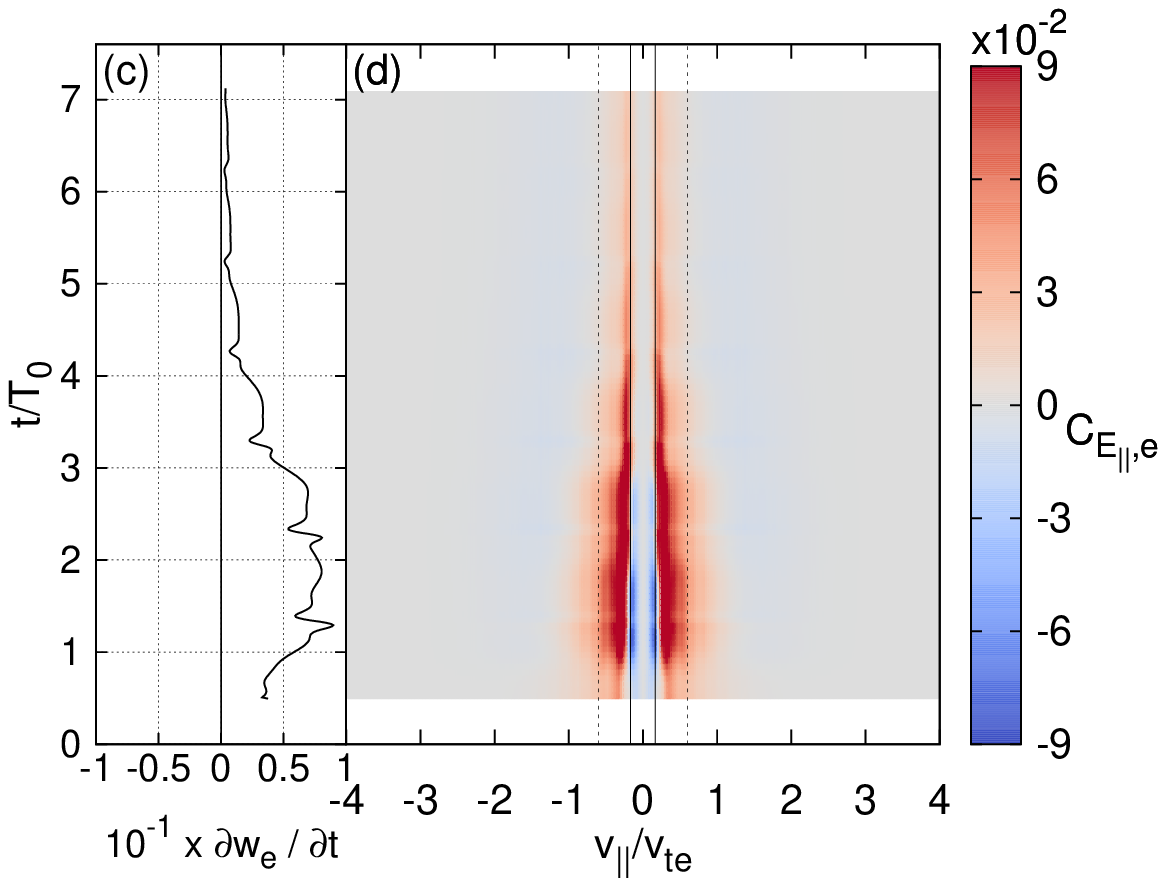}}}
  \caption{Timestack plots of ion and electron energization at
    position A for a correlation interval $\tau/T_0=0.992$. (a)
    Velocity-space integrated correlation, giving  the rate of ion
    energization $\partial w_i/\partial t$ due to ion interactions
    with $E_\parallel$. (b) The reduced parallel field-particle
    correlation for the ions $C_{E_\parallel,i}
    (v_\parallel,t,\tau)$. (c) Velocity-space integrated correlation,
    giving  the rate of electron energization $\partial w_e/\partial t$
    due to electron interactions with $E_\parallel$.(d) The reduced
    parallel field-particle correlation for the electrons
    $C_{E_\parallel,e} (v_\parallel,t,\tau)$.  Vertical solid black
    indicate resonant velocities for a parallel phase velocity at the
    \Alfven speed $\omega/(k_\parallel v_{ts})= v_A/v_{ts}.$ Vertical
    dashed lines indicate the highest parallel phase velocities for
    modes with significant collisionless damping in the simulation.}
\label{fig:cvpar_a}
\end{figure}

Here we present the results of the field-particle technique applied at
three points in the simulation domain, labeled A, B, and C in
\figref{fig:fpac22_jzez}.  From \figref{fig:fpac22_jzez}(d), we see
that, averaged over one period $\tau/T_0=0.992$ centered at
$t/T_0=1.86$, there is a net gain of energy by the plasma at point A,
a net loss of energy by the plasma at point B, and little net change
in the plasma energy at point C. Note that the reduced parallel
correlation $C_{E_\parallel,s} (v_\parallel,t,\tau)$ in the plots
presented in this section is normalized by the energy transfer rate
per unit volume per unit velocity, $Q_0/v_{ti}$.

In \figref{fig:cvpar_a}(b), we present a timestack plot of the reduced
parallel field-particle correlation for the ions $C_{E_\parallel,i}
(v_\parallel,t,\tau)$ at position A with a correlation interval
$\tau/T_0=0.992$, showing the distribution of the energy transfer to
the ions as a function of the parallel velocity $v_\parallel/v_{ti}$
vs.~normalized time $t/T_0$. Vertical solid and dashed black lines
indicate the limits of resonant parallel phase velocities from
\figref{fig:fpac22_gk_disp}, $1.0 \lesssim |\omega / k_\parallel
v_{ti}| \lesssim 1.5$ for ions; there are both positive and negative
ranges of parallel phase velocities, corresponding to \Alfven waves
traveling up or down the mean magnetic field.  Also plotted in
\figref{fig:cvpar_a}(a) is the velocity-space integrated correlation,
$\partial w_i/\partial t = \int d v_\parallel C_{E_\parallel i}
(v_\parallel,t,\tau)$, equivalent to the parallel ion contribution of
the electromagnetic work $j_{\parallel i} E_\parallel$ at position A,
showing a net energization of the ions over the course of the
simulation.

The distribution of the energy transfer as a function of
$v_\parallel/v_{ti}$ in \figref{fig:cvpar_a}(b) is the \emph{velocity
  space signature} of the energy transfer mechanism.  The localization
of the energy transfer in the marked range of resonant parallel phase
velocities for kinetic \Alfven waves clearly indicates that the energy
transfer is resonant.  The specific distribution of this energy
transfer, with a transfer of energy from $E_\parallel$ to the ions
(red) at $|v_\parallel/v_{ti}|>|v_{res}/v_{ti}|$ and a loss of energy
from the ions (blue) at $|v_\parallel/v_{ti}|<|v_{res}/v_{ti}|$ is the
characteristic signature of the Landau damping of kinetic \Alfven
waves \citep{Howes:2017c,Klein:2017b}. The change of sign in the
energy transfer occurs at the resonant phase velocity for the
collisionlessly damped wave. Here the change of sign occurs at
$v_\parallel/v_{ti} = \omega / k_\parallel v_{ti} \simeq v_A/v_{ti} =
\pm 1$, indicating that larger scale \Alfven waves with $k_\perp
\rho_i \ll 1$, which have a parallel phase velocity $\omega /
k_\parallel = v_A$, appear to dominate the energy transfer at point
A. This is consistent with the fact the energy in the electromagnetic
and plasma bulk flow fluctuations in this simulation is dominated by
the low $k_\perp \rho_i \ll 1$ modes. This novel field-particle
correlation analysis shows definitively the key result that ion Landau
damping contributes to the energization of ions at position A in this
strong \Alfven wave collision simulation.

In \figref{fig:cvpar_a}(d), we plot the the reduced parallel
field-particle correlation for the electrons $C_{E_\parallel,e}
(v_\parallel,t,\tau)$ at position A with the same correlation interval
$\tau/T_0=0.992$, where vertical solid and dashed black lines indicate
the range of resonant parallel velocities from
\figref{fig:fpac22_gk_disp}, $0.17 \lesssim \omega / k_\parallel
v_{te} \lesssim 0.6$ for electrons. Also plotted in
\figref{fig:cvpar_a}(c) is the the velocity-space integrated
correlation, $\partial w_e/\partial t = \int d v_\parallel
C_{E_\parallel e} (v_\parallel,t,\tau)$, equivalent to the parallel
electron contribution of the electromagnetic work $j_{\parallel e}
E_\parallel$, showing a net energization of the electrons at position
A.

The velocity space signature of the electron energization in
\figref{fig:cvpar_a}(d) also shows the typical characteristics of
electron Landau damping, with a slight difference from the ion case.
Because kinetic \Alfven waves are dispersive, with a parallel phase
velocity that increases for $k_\perp \rho_i \gtrsim 1$ given
approximately by $\omega=k_\parallel v_A \sqrt{1 + (k_\perp
  \rho_i)^2/[\beta_i + 2/(1+T_e/T_i)]}$ \citep{Howes:2014a}, the
parallel resonant velocity will increase for kinetic \Alfven waves
with larger $k_\perp \rho_i$. The velocity space signature of linear
Landau damping typically shows the change of sign of the energy
transfer at the resonant velocity.  In \figref{fig:cvpar_a}(d), that
change of sign for $1\le t/T_0\le 3$ occurs at a resonant velocity
slightly larger than $v_A/v_{te}$ (vertical black line), suggesting
that the kinetic \Alfven wave involved in the electron Landau damping
has a value of $k_\perp \rho_i \gtrsim 1$ leading to a higher resonant
parallel velocity.  Despite this minor detail, the energy transfer
still shows that the electron energization is mediated by resonant
electrons, with a velocity space signature typical of electron Landau
damping \citep{Howes:2017c}. Therefore, this analysis definitively
yields a second key result, that electron Landau damping contributes
to the energization of electrons at position A in this strong \Alfven
wave collision simulation.

\begin{figure}
  \centerline{\resizebox{4.25in}{!}{\includegraphics{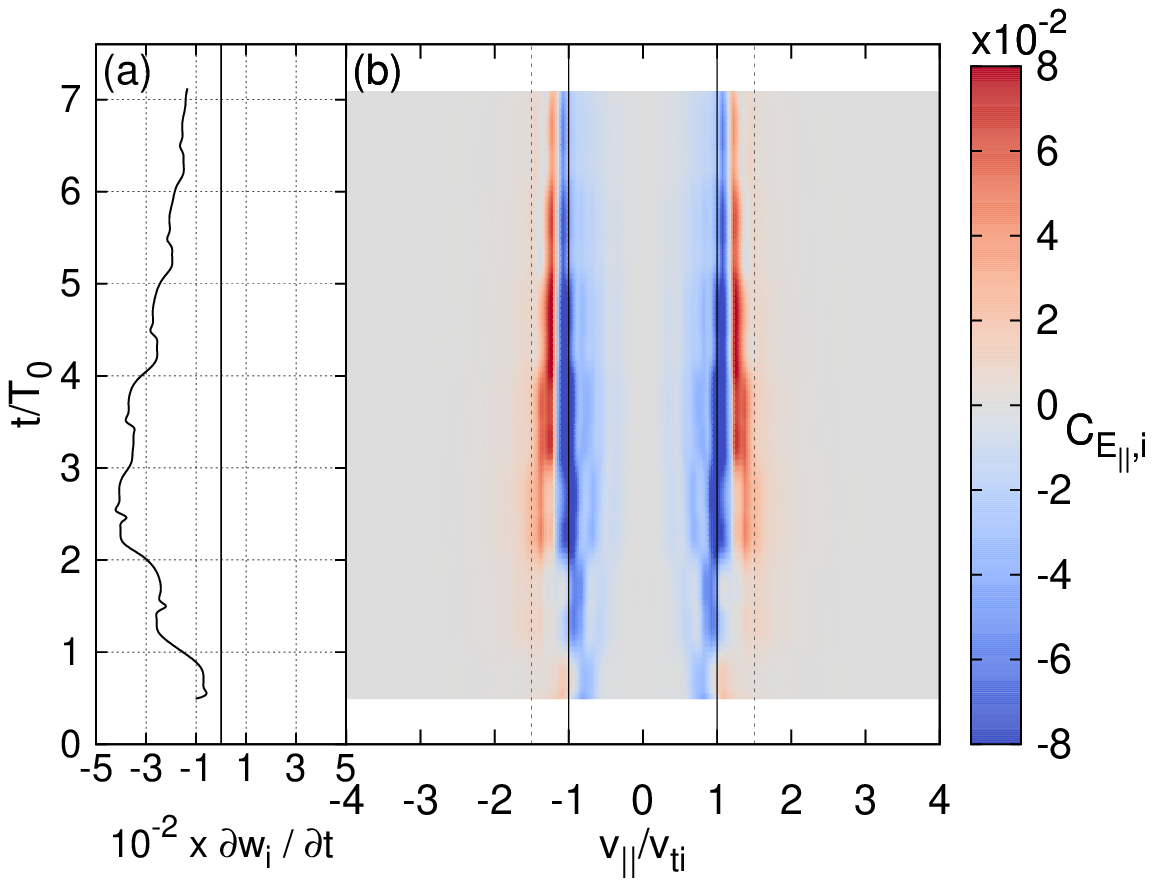}}}
  \centerline{\resizebox{4.25in}{!}{\includegraphics{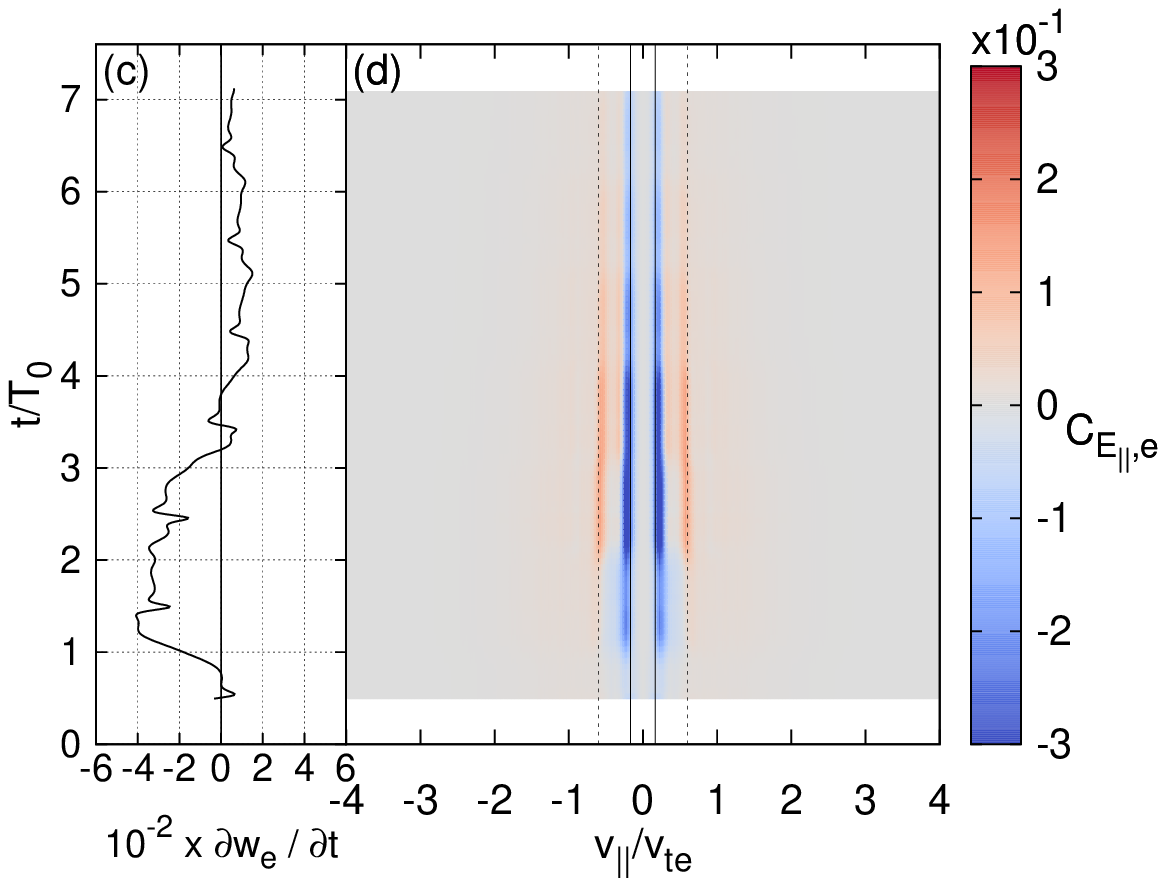}}}
\caption{Plots of the same field-particle correlation analysis as
  \figref{fig:cvpar_a}, but taken at point B.}
\label{fig:cvpar_b}
\end{figure}

We can also investigate the regions in the simulation domain where the
plasma loses energy to the parallel electric field at point B, plotted
in \figref{fig:cvpar_b}. The (a) velocity-integrated ion energization
$\partial w_i/\partial t $ due to $E_\parallel$ and (c)
velocity-integrated electron energization $\partial w_e/\partial t $
due to $E_\parallel$ both show that the net energy transfer to ions
and electrons at point B is negative.  Here, as in
\figref{fig:cvpar_a}, we see that the energy transfer for both ions in
panel (b) and electrons in panel (d) is dominated by particles with
velocities that fall within the range of parallel velocities expected
to be resonant with the parallel phase velocity of \Alfven waves,
demonstrating directly that energy transfer between the particles and
the parallel electric field $E_\parallel$ is governed by Landau
resonant interactions.

\begin{figure}
  \centerline{\resizebox{4.25in}{!}{\includegraphics{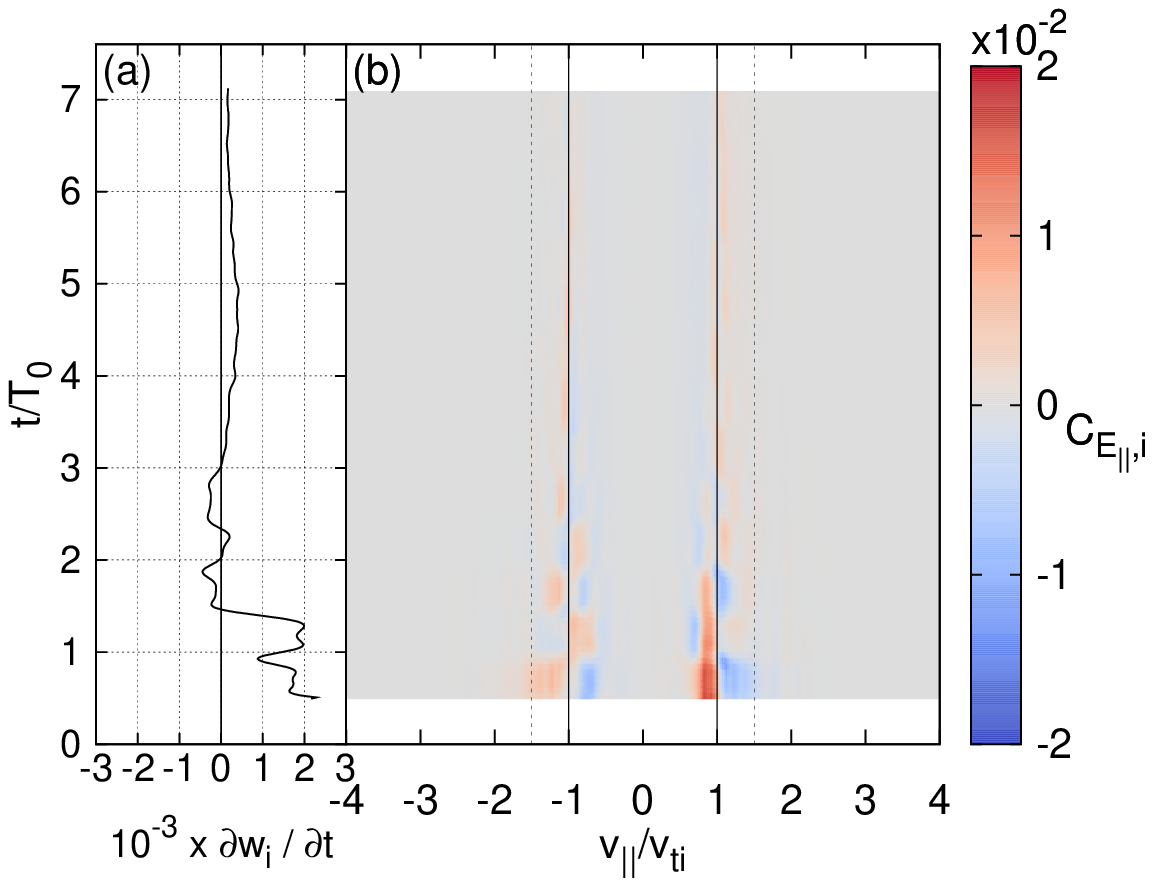}}}
  \centerline{\resizebox{4.25in}{!}{\includegraphics{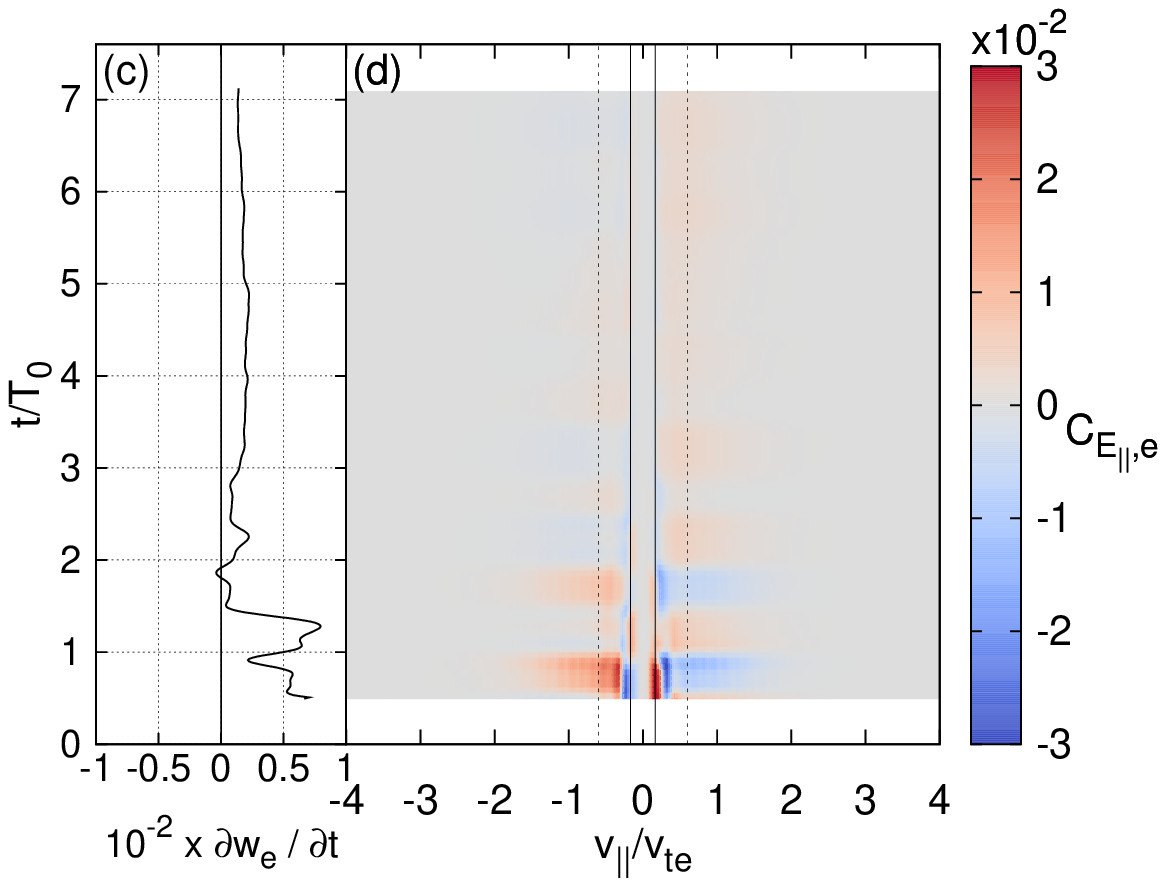}}}
\caption{Plots of the same field-particle correlation analysis as
  \figref{fig:cvpar_a}, but taken at point C.}
\label{fig:cvpar_c}
\end{figure}

At point C in the simulation domain, there is very little net energy
transfer from the parallel electric field to the plasma particles.
The same field-particle correlation analysis at point C, presented in
\figref{fig:cvpar_c}, shows that the velocity-integrated (a) ion
energization $\partial w_i/\partial t $ and (c) electron energization
$\partial w_e/\partial t $ due to $E_\parallel$ yields a very small
positive transfer of energy to the particles over the first couple of
periods $T_0$, with an amplitude about an order of magnitude smaller
than the energy transfer at points A and B.  The reduced parallel
field-particle correlation $C_{E_\parallel,s} (v_\parallel,t,\tau)$
for (b) the ions and (d) the electrons shows that the majority of
this very small amount of energy transfer is still dominated by
resonant particles.

But there is a very significant difference between the reduced parallel
field-particle correlation $C_{E_\parallel,s}$ at point C for both
ions and electrons compared to the same correlation at points A and B:
the pattern of energy transfer at point C is dominantly odd in
$v_\parallel$, whereas the patterns at points A and B are dominantly
even in $v_\parallel$.  When integrated over the parallel velocity to
obtain the net change of energy of a species, an odd pattern largely
cancels out, whereas an even pattern does not.  Therefore, there is
little net particle energization at point C, even though individual
particles do gain and lose energy through resonant interactions with
$E_\parallel$. Particles with $v_\parallel > 0 $ gain nearly the same
amount of energy as that lost by particles with $v_\parallel< 0 $,
yielding little net change of particle energy.

The important point that the field-particle correlation analysis here
demonstrates is that collisionless interactions of the Landau
resonance between $E_\parallel$ and the ions and electrons contribute
to the spatially intermittent pattern of time-averaged particle
energization, shown in \figref{fig:fpac22_jzez}(d). This result
disproves by counterexample the commonly stated belief that Landau
damping can only lead to spatially uniform particle energization.
Rather, we see clearly here that collisionless damping via the Landau
resonance can indeed be responsible for spatially localized particle
energization, as previously suggested in the literature
\citep{TenBarge:2013a,Howes:2015b,Howes:2016b}.  Furthermore, the
nonlinear energy transfer by collisionless damping via the Landau
resonance is not inhibited by the strong nonlinear interactions that
play an important role in this strong \Alfven wave collision
simulation.  Indeed, nonlinear gyrokinetic simulations of strong,
broadband plasma turbulence have indeed shown that the collisionless
transfer of energy between fields and ions is dominated by particles
approximately in Landau resonance with the parallel phase velocity of
\Alfvenic fluctuations \citep{Klein:2017b}.

\subsection{Particle Energization in Gyrotropic Velocity Space}
\label{sec:gyro}

\begin{figure}
  \centerline{\resizebox{2.65in}{!}{\includegraphics*[1.05in,1.3in][5.0in,3.9in]{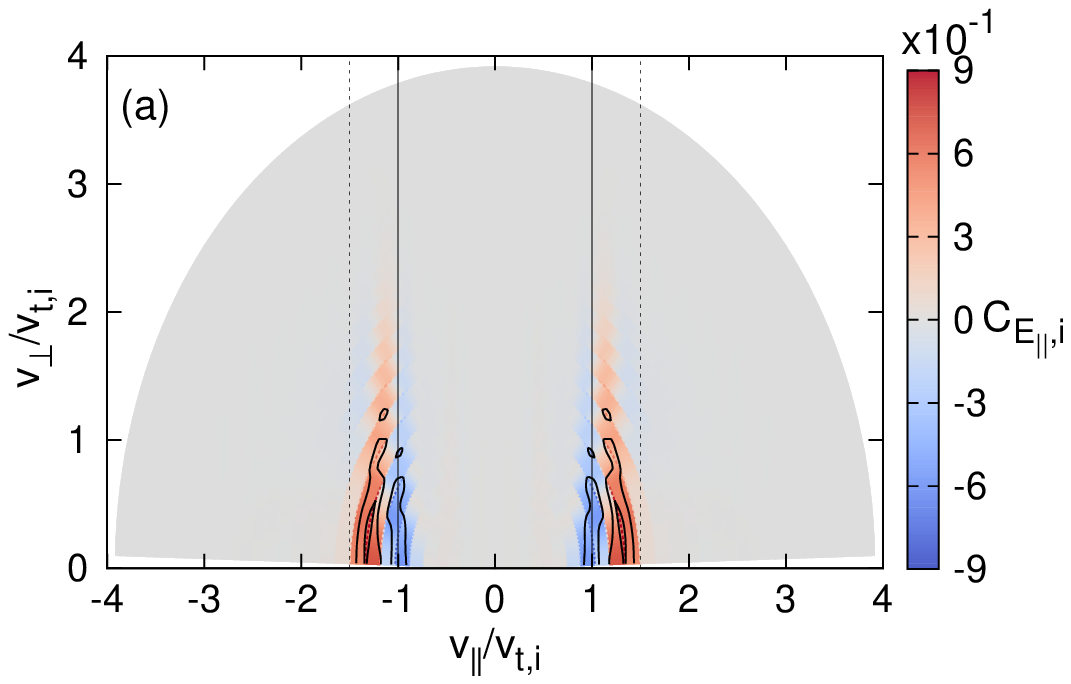}}
\resizebox{2.65in}{!}{\includegraphics[1.05in,1.3in][5.0in,3.8in]{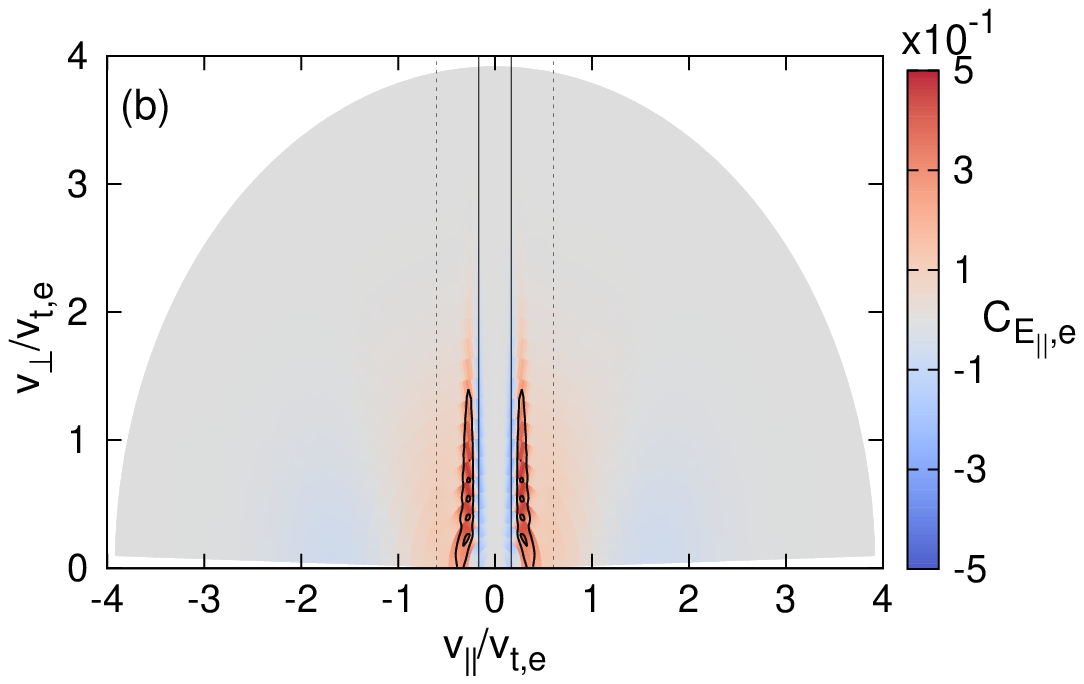} }   }
  \centerline{\resizebox{2.65in}{!}{\includegraphics[1.05in,1.3in][5.0in,3.9in]{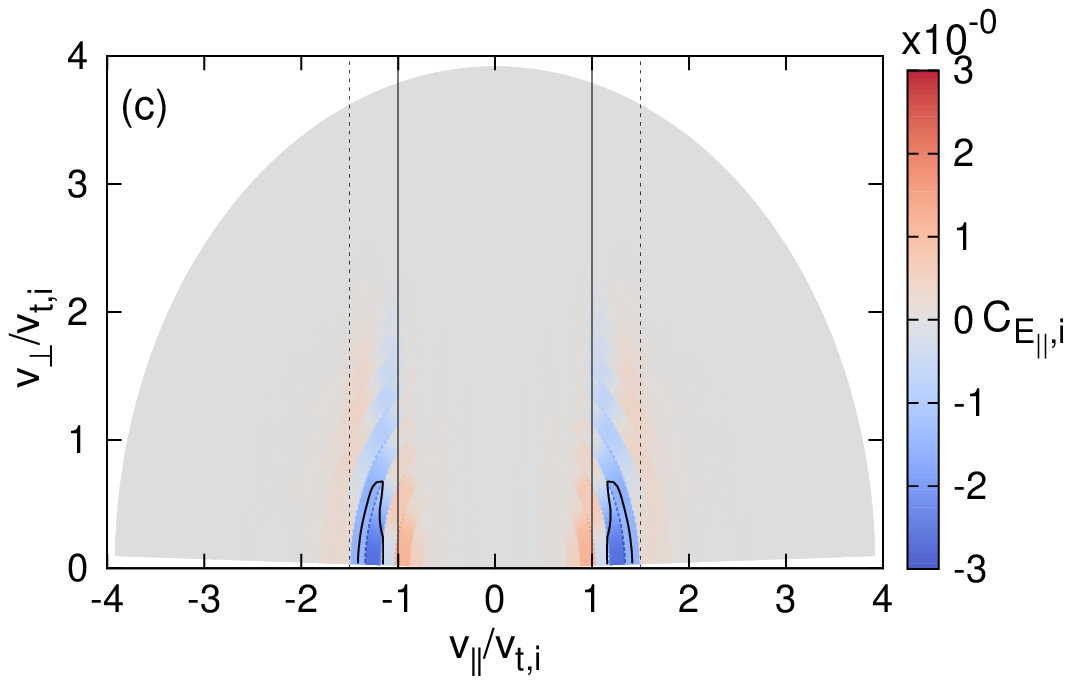}}
\resizebox{2.65in}{!}{\includegraphics[1.05in,1.3in][5.0in,3.8in]{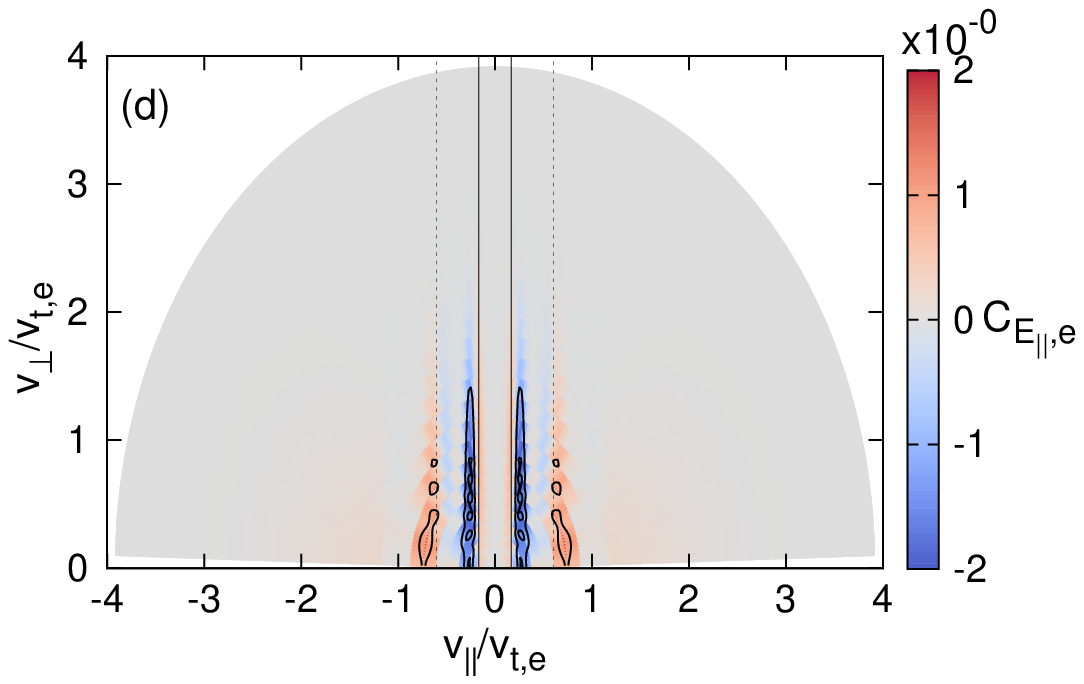} }   }
  \centerline{\resizebox{2.65in}{!}{\includegraphics[1.05in,1.3in][5.0in,3.9in]{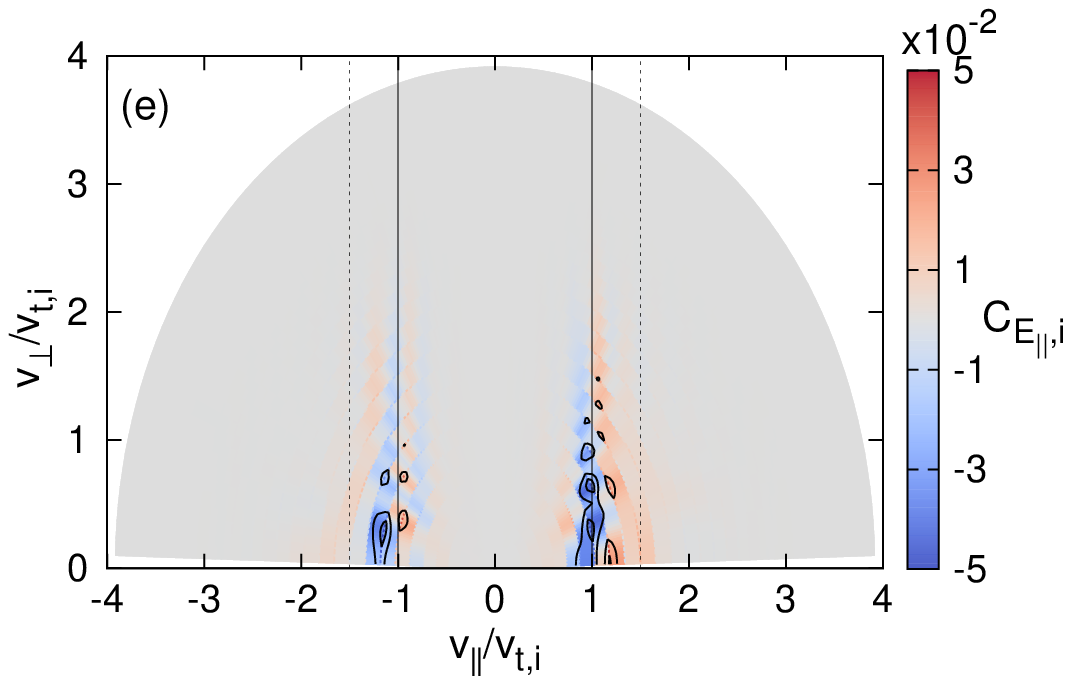}}
\resizebox{2.65in}{!}{\includegraphics[1.05in,1.3in][5.0in,3.9in]{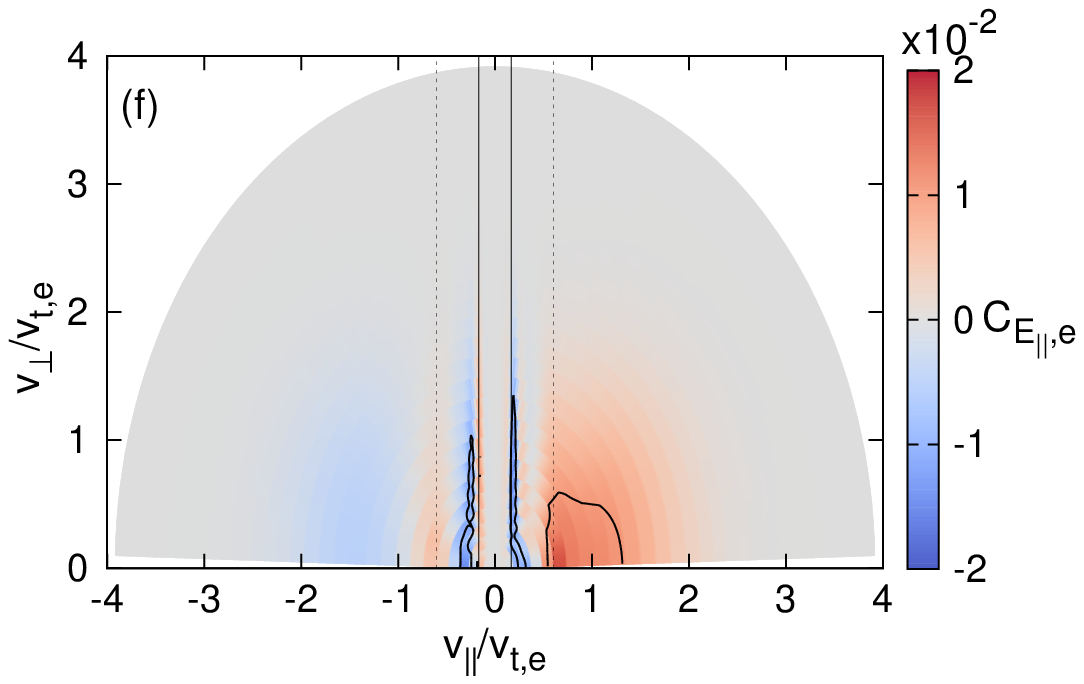} }   }
  \caption{Plots of the field-particle correlation $ C_{E_\parallel}
    (v_\parallel,v_\perp,t,\tau)$ on gyrotropic velocity space
    $(v_\parallel, v_\perp)$ for a correlation interval
    $\tau/T_0=0.992$  centered at time $t/T_0=2.10$: (a) ion and (b)
    electron energization at point A, (c) ion and (d) electron
    energization at point B, and (e) ion and (f) electron energization
    at point C. Vertical solid lines denote the resonant parallel
    velocities for a parallel phase velocity at the \Alfven speed
    $\omega/(k_\parallel v_{ts})= v_A/v_{ts}$.}
\label{fig:fpac22_gyro_abc}
\end{figure}

Finally, we can examine the distribution of particle energization in
gyrotropic velocity space $(v_\parallel,v_\perp)$ \citep{Howes:2017c}
using the field-particle correlation $ C_{E_\parallel,s}
(v_\parallel,v_\perp,t,\tau)$. Although plots of this analysis are
limited to the correlation centered at just a single point in time, by
not integrating over perpendicular velocity $v_\perp$ one obtains
complete information about which particles in gyrotropic velocity
space $(v_\parallel, v_\perp)$ participate in the collisionless
transfer of energy.  Note that the parallel correlation in gyrotropic
velcotiy space, $C_{E_\parallel,s} (v_\parallel,v_\perp,t,\tau)$ in
the plots presented in this section is normalized by the energy
transfer rate per unit volume per unit velocity squared,
$Q_0/v_{ti}^2$.

In \figref{fig:fpac22_gyro_abc}, we plot $ C_{E_\parallel,s}
(v_\parallel,v_\perp,t,\tau)$ for the same correlation interval
$\tau/T_0=0.992$ centered at time $t/T_0=2.10$: (a) ion and (b)
electron energization at point A, (c) ion and (d) electron
energization at point B, and (e) ion and (f) electron energization at
point C.  As before, vertical solid and dashed black lines denote the
range of resonant parallel velocities for \Alfven waves. Three
important points can be inferred from these gyrotropic velocity space
plots. First, other than a steady decrease of the amplitude of the
signal with increasing $v_\perp$---as expected because the equilibrium
Maxwellian distribution drops off exponentially as
$\exp(-v^2/v_{ts}^2)$, so the amplitude of fluctuations $\delta
f(\V{v})$ would be expected to have a similar drop off in
amplitude---the energy transfer shows very little variation with
$v_\perp$. The variation in the energy transfer is organized almost
entirely by $v_\parallel$, as expected for a Landau resonant energy
transfer process. Second, this energy transfer is dominated by
particles with parallel velocities resonant with \Alfvenic
fluctuations, $v_\parallel \simeq v_A$, demonstrating that the energy
transfer is governed by the Landau resonance.  Third, the odd or even
character in $v_\parallel$ at the different points A, B, and C seen in
the timestack plots is also clearly apparent here in these gyrotropic
velocity space plots.

Summarizing, the gyrotropic velocity space $(v_\parallel,v_\perp)$
plots in \figref{fig:fpac22_gyro_abc} demonstrate how the
field-particle correlation technique maximizes the use of the full
particle velocity distribution function information, enabling the
physical mechanism responsible for the removal of energy from
turbulent fluctuations and consequent particle energization to be
identified definitively.  In this case, the velocity-space signature
of the field-particle correlation is unmistakably that of Landau
damping of a kinetic \Alfven wave \citep{Howes:2017c}, proving that
Landau damping indeed plays a role in the spatially intermittent
removal of the energy of electromagnetic and bulk plasma flow
fluctuations, even in the presence of strong nonlinearity.

\section{Conclusion}

Using a nonlinear gyrokinetic simulation of a strong \Alfven wave
collision, we examine here the damping of the electromagnetic
fluctuations and the associated energization of particles that occurs
in current sheets that are generated self-consistently during the
nonlinear evolution.

The flow of energy due to the collisionless damping and the associated
particle energization, as well as the subsequent thermalization of the
particle energy by collisions, provides an important framework for
interpreting the nonlinear dynamics and dissipation.
\figref{fig:energyflow} presents a simple model of the energy flow
from turbulent energy to plasma heat in the simulation, with the
following two key stages: (i) the turbulent fluctuation energy is
removed by collisionless field-particle interactions, transferring
that energy reversibly into non-thermal energy of the plasma species;
and (ii) the non-thermal energy, represented by fluctuations in the
particle velocity distribution functions, is driven to sufficiently
small velocity-space scales that weak collisions can thermalize that
energy, irreversibly heating the plasma species. In the strong \Alfven
wave collision simulated here, this two-step processes ultimately
leads to more than 60\% of the original fluctuation energy being
dissipated collisionally as thermal ion and electron energy.

It has long been appreciated that the nonlinear evolution of plasma
turbulence leads to the development of intermittent current sheets
\citep{Matthaeus:1980,Meneguzzi:1981}, and a recent study has shown
that strong \Alfven wave collisions---nonlinear interactions between
counterpropagating \Alfven waves---self-consistently develop
intermittent current sheets through the constructive interference of
the original \Alfven waves and nonlinearly generated fluctuations
\citep{Howes:2016b}. MHD turbulence simulations have shown that the
dissipation of turbulent energy is largely concentrated in these
intermittent current sheets
\citep{Uritsky:2010,Osman:2011,Zhdankin:2013}, so a natural question
is whether the collisionless damping of current sheets generated by
strong \Alfven wave collisions leads to such spatially intermittent
particle energization.  Plotting the spatial distribution of the
electromagnetic work done by the parallel electric field
$E_\parallel$, shown in \figref{fig:fpac22_jzez}(a)--(c), shows that the
instantaneous particle energization is indeed spatially intermittent
with a sheet-like morphology.

A key point, however, is that much of this reversible electromagnetic
work leads to an oscillatory transfer of energy to and from the
particles associated with undamped wave motion. Only by averaging over
a suitable time interval, in this case an averaging interval that is
approximately a single wave period $\tau \simeq T_0$, can we determine
the secular particle energization $\langle j_\parallel E_\parallel
\rangle_\tau$ associated with the net removal of energy from
the turbulent fluctuations.  \figref{fig:fpac22_jzez}(d) shows that
the secular particle energization $\langle j_\parallel E_\parallel
\rangle_\tau$ in this strong \Alfven wave collision indeed remains
spatially intermittent, although less localized than the instantaneous
rates of energy transfer in \figref{fig:fpac22_jzez}(a)--(c). The
bottom line is that the current sheets arising in strong \Alfven wave
collisions indeed generate spatially localized particle energization,
consistent with that found in simulations of plasma turbulence
\citep{Wan:2012,Karimabadi:2013,TenBarge:2013a,Wu:2013,Zhdankin:2013}
and inferred from spacecraft observations of the solar wind
\citep{Osman:2011,Osman:2012a,Perri:2012a,Wang:2013,Wu:2013,Osman:2014b}.

The next obvious question is what is the physical mechanism governing
the removal of energy from the turbulence and the consequent spatially
intermittent energization of the particles? Using the recently
developed field-particle correlation technique
\citep{Klein:2016a,Howes:2017a}, we examine how the energy transfer to
ions and electrons by the parallel electric field $E_\parallel$ is distributed
in velocity space. In other words, which particles in velocity space
receive the energy transferred collisionlessly from the
electromagnetic fields? The results, exemplified by
\figref{fig:cvpar_a}, show that the particles that are resonant with the
parallel velocity of the \Alfven waves in the simulation dominate the
energy transfer, demonstrating conclusively that Landau damping plays
a role in the damping of the electromagnetic fluctuations and
consequent energization of the particles in this strongly nonlinear
simulation.

Based on the plane-wave decomposition typically used to derive linear
Landau damping analytically, one may naively expect that Landau
damping leads to spatially uniform energization.  Together, the
results presented here definitively show instead that Landau damping
can indeed lead to spatially intermittent particle energization. The
comparison to a strictly linear simulation from identical initial
conditions, presented in \figref{fig:fpac22_jzez_lin_nl}, shows that
the nonlinear energy transfer to other Fourier modes is essential for
the localization of the particle energization. This is consistent with
the model for current sheet generation in \Alfven wave collisions in
which nonlinearly generated modes constructively interfere with the
initial \Alfven waves to create spatially intermittent current sheets;
linear Landau damping of each of these modes, which occurs spatially
locally, leads to the intermittent spatial pattern of the
energization, as previously suggested \citep{Howes:2015b,Howes:2016b}.

Our result here, that Landau damping is effective even in a plasma
where strong nonlinear interactions are playing an important role,
also addresses the important question of whether Landau damping is
effective in a strongly turbulent plasma
\citep{Plunk:2013,Schekochihin:2016}.  Our results here complement a recent
field-particle correlation analysis of gyrokinetic turbulence
simulations showing that Landau damping indeed persists as an
effective physical mechanism for ion energization in broadband, strong
plasma turbulence \citep{Klein:2017b}.

We emphasize here that we have not shown that Landau damping is the
\emph{only} damping mechanism, but we have provided conclusive
evidence that Landau damping does play a role in the collisionless damping
of turbulence in intermittent current sheets that arise from strong
\Alfven wave collisions. As mentioned in \appref{app:je}, transit-time
damping \citep{Barnes:1966,Quataert:1998}---which does work on
particles via their magnetic moment through the magnetic mirror force
arising from fluctuations in the magnetic field magnitude---is another
effective physical mechanism for collisionless damping and
energization of particles via the Landau resonance in
gyrokinetics. Here we have focused only on the contribution to the
particle energization by the parallel electric field $E_\parallel$;
future work will address the additional contribution by the magnetic
mirror force arising from $\nabla_\parallel |\V{B}|$.

The comparison of the time evolution of the field-particle energy
transfer $\dot{E}^{(fp)}_s$ and the collisional heating $Q_s$ for each
species in \figref{fig:fpac22_dedt} also raises important questions
about the relative rates of linear and nonlinear phase-mixing
processes that enable the non-thermal energy, represented by
fluctuations in the particle velocity distribution function, to reach
sufficiently small velocity-space scales to be thermalized by
arbitrarily weak collisions. These questions, and more, about the flow
of energy in weakly collisional heliospheric plasmas, lie at the
forefront of kinetic heliophysics \citep{Howes:2017c}, and will drive
research efforts by the next generation of space plasma physicists.

 This work was supported by NSF CAREER Award AGS-1054061, NASA HSR
 grant NNX16AM23G, DOE grant DE-SC0014599, and the University of Iowa
 Mathematical and Physical Sciences Funding Program.  This work used
 the Extreme Science and Engineering Discovery Environment (XSEDE),
 which is supported by National Science Foundation grant number
 ACI-1053575, through NSF XSEDE Award PHY090084.

\appendix
\section{Collisionless Damping Rate as a Function of Mass Ratio}
\label{app:implement}
\begin{figure}
  \centerline{\resizebox{4.0in}{!}{\includegraphics*[0.35in,2.in][8.0in,7.5in]{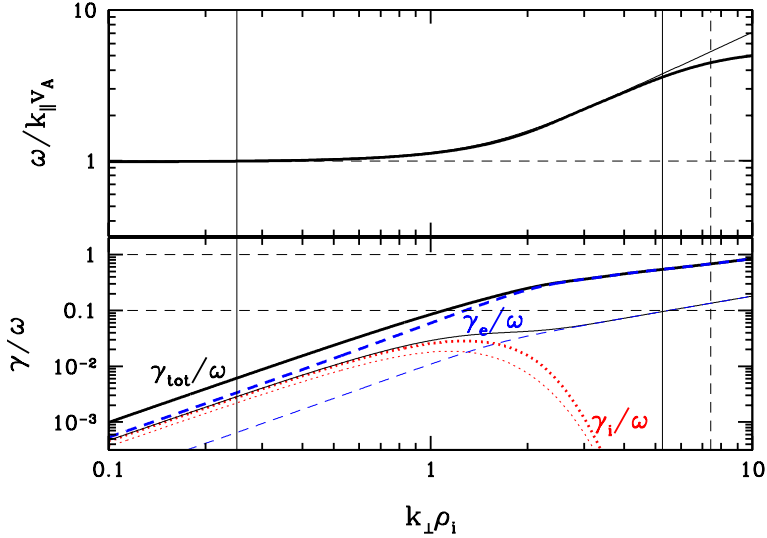}}
  }
  \caption{Comparison of the results from the linear collisionless
    gyrokinetic dispersion relation for the reduced mass ratio
    $m_i/m_e=36$ (thick) and a realistic proton-to-electron mass ratio
    $m_i/m_e=1836$ (thin): (a) normalized frequency
    $\omega/k_\parallel v_A$ and (b) total collisionless damping rate
    $\gamma_{tot}/\omega$ (black solid), ion collisionless damping
    rate $\gamma_i/\omega$ (red dotted), and electron collisionless
    damping rate $\gamma_e/\omega$ (blue dashed).
 Solid and dashed vertical lines are the same as in \figref{fig:fpac22_gk_disp}.
  }
\label{fig:fpac22_gk_disp_mucomp}
\end{figure}

Here we present in \figref{fig:fpac22_gk_disp_mucomp} a comparison of
the linear physics of the Alfv\'en/kinetic \Alfven wave mode for the
reduced mass ratio used here $m_i/m_e=36$ (thick lines) to that for a
realistic proton-to-electron mass ratio $m_i/m_e=1836$ (thin lines).
Specifically, for a plasma with $\beta_i=1$ and $T_i/T_e=1$, we solve
the linear collisionless gyrokinetic dispersion relation
\citep{Howes:2006} for the (a) normalized wave frequency
$\omega/k_\parallel v_A$ and (b) total collisionless damping rate
$\gamma_{tot}/\omega$ (black solid) as a function of the perpendicular
wavenumber $k_\perp \rho_i$. In addition, in panel (b) we also show
the separate contributions of the ions (red dotted) and electrons
(blue dashed) to the collisionless damping rate.

The comparison shows that the linear wave frequency
$\omega/k_\parallel v_A$ begins to differ only slightly between the
two cases at $k_\perp \rho_i \gtrsim 5$. Note that the fully resolved
perpendicular range of the dealiased pseudospectral method for the
strong \Alfven wave collision simulation covers $0.25 \le k_\perp
\rho_i \le 5.25$, denoted by the two vertical solid black lines; modes
in the corner of $(k_x,k_y)$ Fourier space represent perpendicular
wavenumbers out to $k_\perp \rho_i = 5.25 \sqrt{2}\simeq 7.42$,
denoted by the vertical dashed black line. Therefore, there is very
little difference in the linear wave frequency over the perpendicular
range of the simulation between $m_i/m_e=36$ and $m_i/m_e=1836$.

The noticeable difference between the two cases arises in the electron
collisionless damping rate $\gamma_e/\omega$ in
\figref{fig:fpac22_gk_disp_mucomp}(b). The ion damping is slightly
smaller for the realistic mass ratio relative to the reduced mass
ratio, but the electron damping drops by nearly an order of magnitude.
Other than the amplitude changes, however, the individual species
damping rates $\gamma_s/\omega$ on a log-log plot have the same the
functional form, only a different relative weighting.  The reduced
mass ratio case has nearly a factor of ten larger relative
contribution to the collisionless damping than the realistic mass
ratio. Note that significant collisionless damping of a wave occurs
when $\gamma/\omega \gtrsim 0.1$ (marked by a horizontal dashed line),
so total collisionless damping is relatively weak over the
perpendicular range of the simulation for a realistic mass ratio
$m_i/m_e=1836$, whereas the damping is very strong with
$\gamma_{tot}/\omega \sim 1.0$ at the smallest perpendicular scales
for the reduced mass ratio $m_i/m_e=36$. This enables collisionless
damping to remove energy from the turbulent fluctuations completely
over the resolve range of scales, avoiding any problematic bottlenecks
at the smallest scales.

Note that reducing the mass ratio to values $m_i/m_e < 32$ (not shown)
leads to a significant qualitative change from the behavior of the
collisionless damping shown here.  At these very low mass ratio
values, the ion collisionless damping does not drop off at $k_\perp
\rho_i \gg 1$, potentially leading to qualitatively incorrect results
about the relative ion and electron damping \citep{Klein:2017b}.

\section{Particle Energization by Component and Species}
\label{app:je}

In the Vlasov-Maxwell system of equations, the rate of change of
particle energy density at a given position in space is given by the
rate of electromagnetic work, $\V{j} \cdot \V{E}$ \citep{Klein:2017b},
confirming the familiar concept the only the electric field can change
the energy of charged particles. In the low-frequency limit of kinetic
plasma theory, one may average the Vlasov-Maxwell equations over the
gyrophase $\theta$ in cylindrical velocity space $(v_\parallel,
v_\perp,\theta)$ to obtain the reduced system of gyrokinetics
\citep{Frieman:1982,Howes:2006}. The benefit of this procedure is the
reduction of velocity space to two dimensions $(v_\parallel,v_\perp)$
at the expense of discarding the physics at cyclotron frequencies and
higher; effectively, the cyclotron resonances and fast magnetosonic
waves are eliminated, while retaining finite Larmor radius effects and
collisionless damping via the Landau resonance.

In addition to this elimination of the cyclotron resonances, a
component of the perpendicular electromagnetic work, $\V{j}_\perp
\cdot \V{E}_\perp$ is alternatively expressed in terms of the magnetic
mirror force, $\V{F}_{mir} = - \mu \nabla_\parallel |\V{B}|$, where the
magnetic moment of a particle is given by $\mu_s = m_s
v_\perp^2/2|B|$.  In the anisotropic limit $k_\parallel \ll k_\perp$
of the gyrokinetic approximation, the change in the magnetic field
magnitude is dominated by the variation in the parallel component of
the field, $\delta |\V{B}| = \delta B_\parallel + \mathcal{O}(|\delta
\V{B}|^2)$. For electromagnetic waves with a fluctuation in the
magnetic field strength, the magnetic mirror force $- \mu
\nabla_\parallel \delta B_\parallel$ \, acting on the magnetic moment
$\mu$ of the particle gyromotion, leads to collisionless damping of
the wave via the Landau resonance, a process denoted by the term
transit-time damping, or alternatively called Barnes damping
\citep{Barnes:1966,Quataert:1998}.

The bottom line is in gyrokinetics there are two separate mechanisms
that can lead to resonant collisionless particle energization: Landau
damping mediated by the parallel electric field $E_\parallel$ and
transit-time damping mediated by gradients in the parallel magnetic
field perturbation $\nabla_\parallel \delta B_\parallel$. In this
paper, we focus solely on the particle energization by Landau damping
through the parallel electric field $E_\parallel$, leaving a detailed
analysis of transit-time damping to future work.

\begin{figure}
  \centerline{\resizebox{2.6in}{!}{\includegraphics{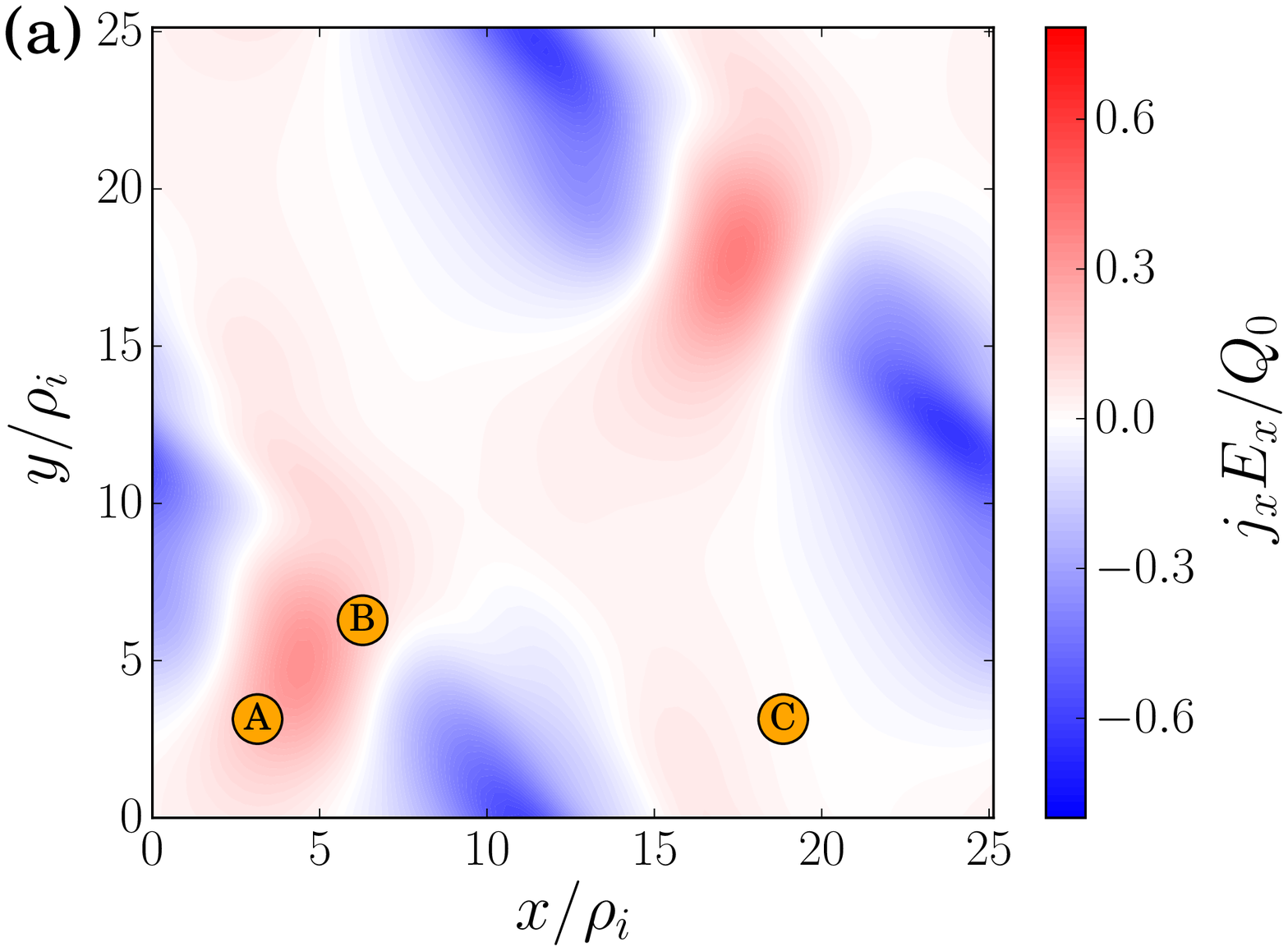}}
\resizebox{2.6in}{!}{\includegraphics{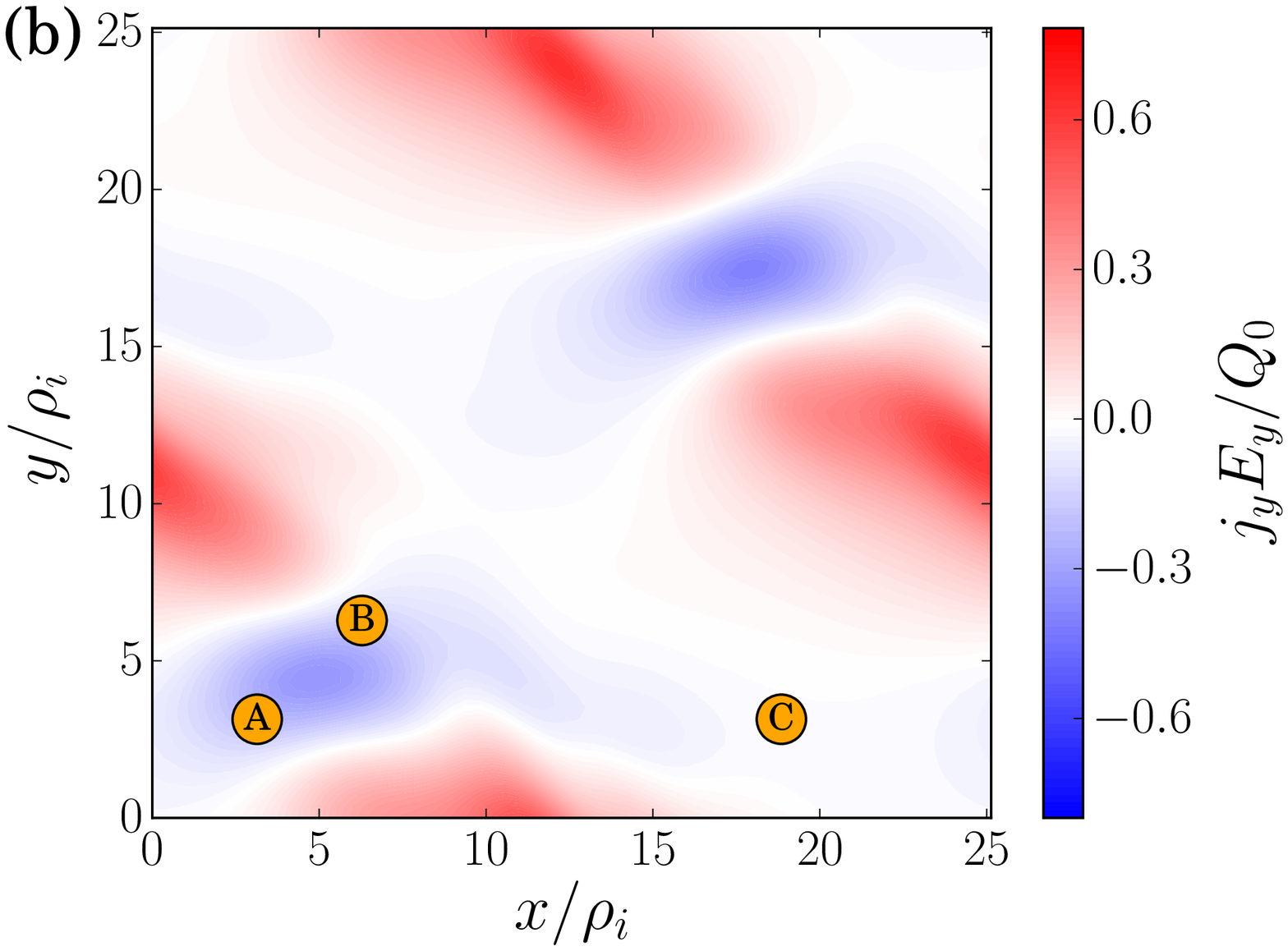} }   }
  \centerline{\resizebox{2.6in}{!}{\includegraphics{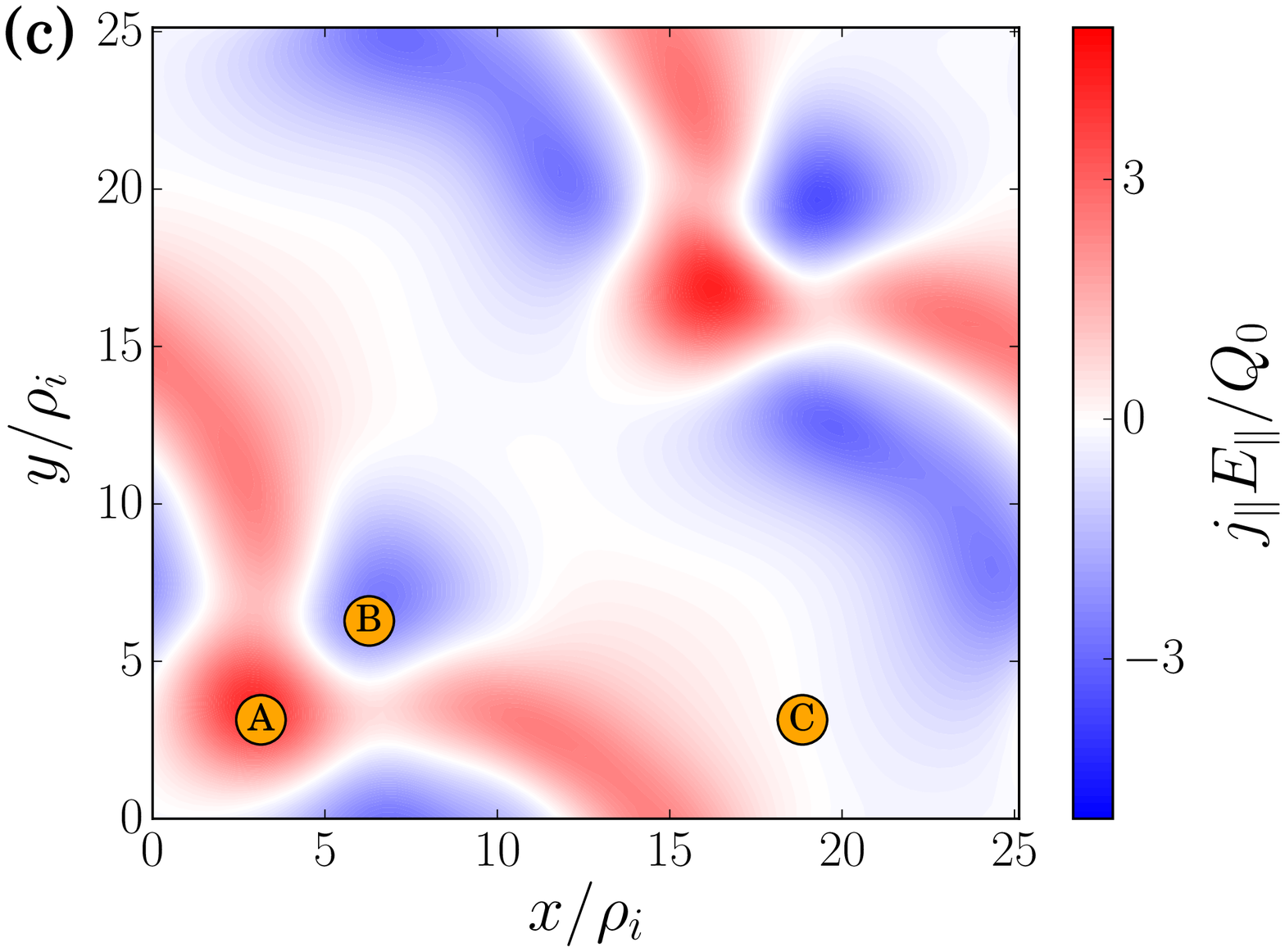}}
\resizebox{2.6in}{!}{\includegraphics{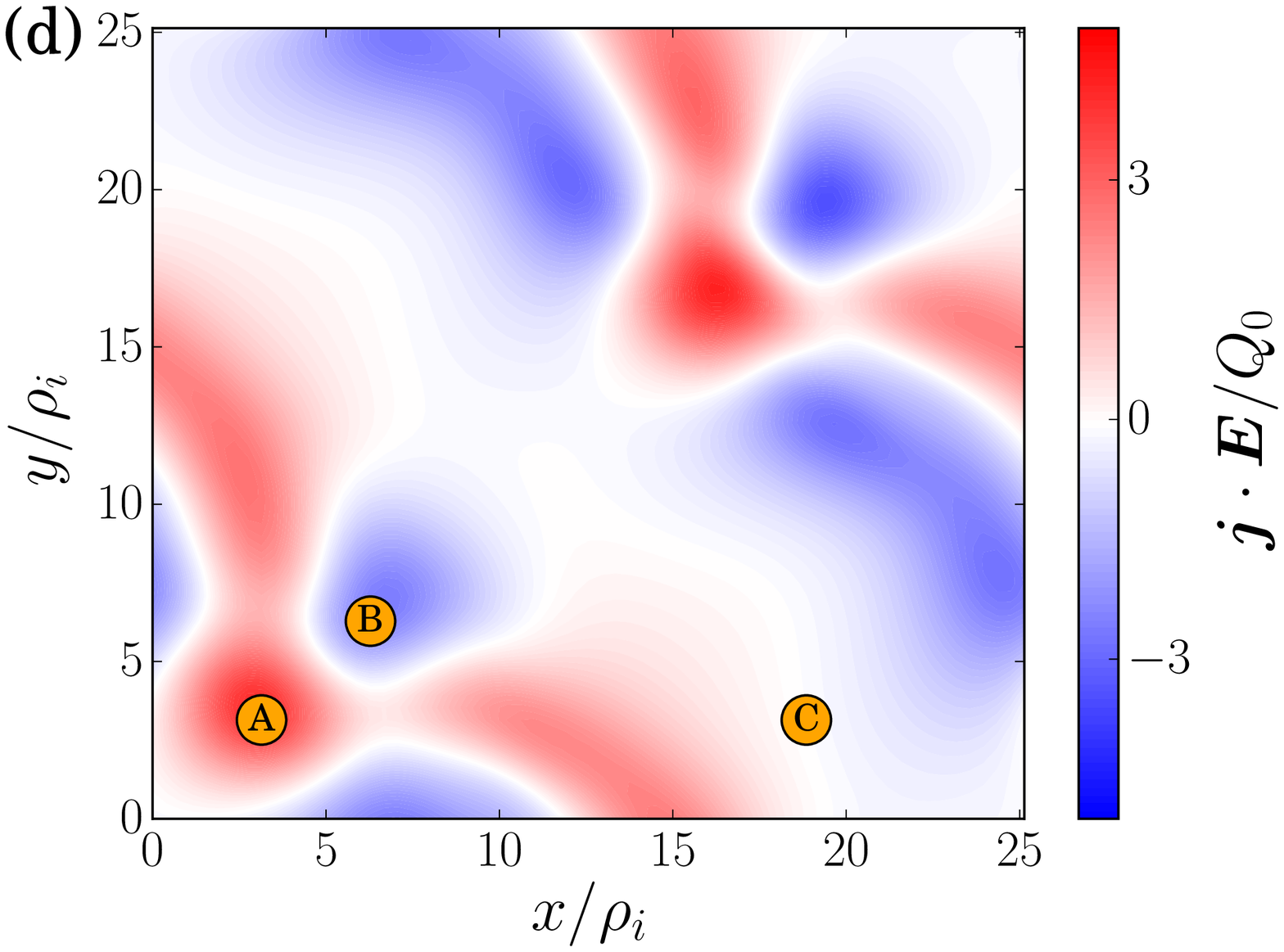} }   }
  \caption{Plots of the different components of the electromagnetic
    work (a) $\langle j_x E_x\rangle_\tau$, (b) $\langle j_y
    E_y\rangle_\tau$, and (c) $\langle j_\parallel
    E_\parallel\rangle_\tau$, as well as the total work (d) $\langle
    \V{j} \cdot \V{E}\rangle_\tau$ averaged over a single wave period
    $\tau=0.992 T_0$ centered at time $t/T_0= 1.86$.}
\label{fig:fpac22_jxyzfull}
\end{figure}

Although gyrokinetics eliminates cyclotron resonant heating, it still
describes the electromagnetic work done by all three components of
$\V{j} \cdot \V{E}$.  The primary focus of this paper is resonant
heating by Landau damping through the parallel electric field
$E_\parallel$, so plots in the body of paper focus only on the
parallel contribution $j_\parallel E_\parallel$.  In
\figref{fig:fpac22_jxyzfull}, we present here for completeness the
three components of the electromagnetic work (a) $\langle j_x
E_x\rangle_\tau$, (b) $\langle j_y E_y\rangle_\tau$, and (c) $\langle
j_\parallel E_\parallel\rangle_\tau$ averaged over single wave period
$\tau=0.992 T_0$ centered at time $t/T_0= 1.86$. The components $ j_x
E_x $ and $j_y E_y$ dominantly represent the energy transfer between
fields and particles associated with undamped wave motion, for example
representing magnetic tension as the restoring force for the \Alfven
wave.  Therefore, $j_x E_x$ and $j_y E_y$ represent oscillatory energy
transfer, and averaged over one wave period there is very little net
energy change.  By comparison, the single-wave period averaged
$\langle j_\parallel E_\parallel \rangle_\tau$ represents the secular
energy transfer associated with collisionless damping,is much larger
than that for $j_x E_x$ or $j_y E_y$.  Note that (d) the total
single-wave period averaged total electromagnetic work $\langle \V{j}
\cdot \V{E} \rangle_\tau$ is dominated by the parallel component
$\langle j_\parallel E_\parallel \rangle_\tau$.  This comparison
motivates our focus in the body of this paper on the parallel
contribution to the electromagnetic work, $j_\parallel E_\parallel$.

\begin{figure}
  \centerline{\resizebox{2.6in}{!}{\includegraphics{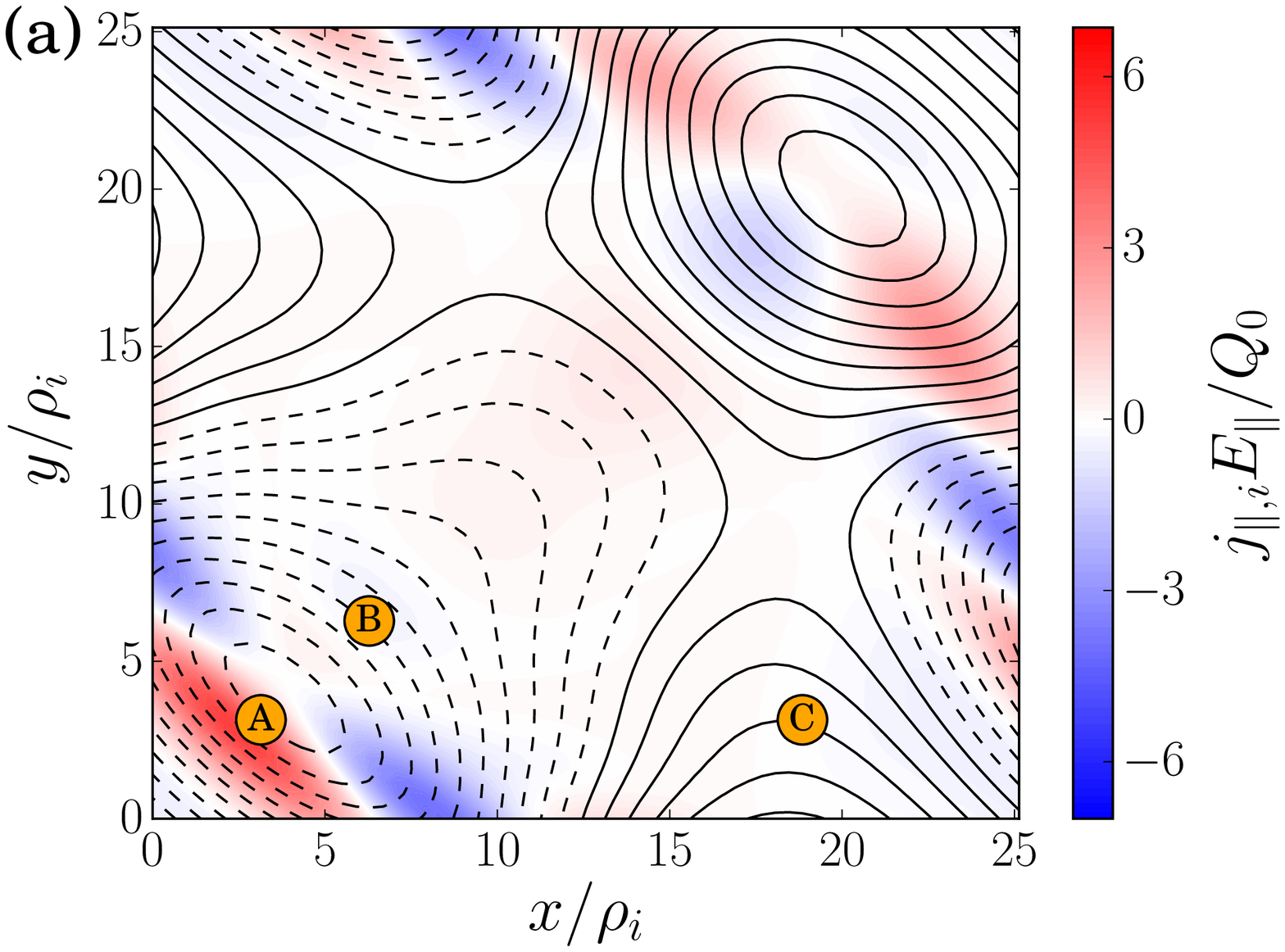}}
\resizebox{2.6in}{!}{\includegraphics{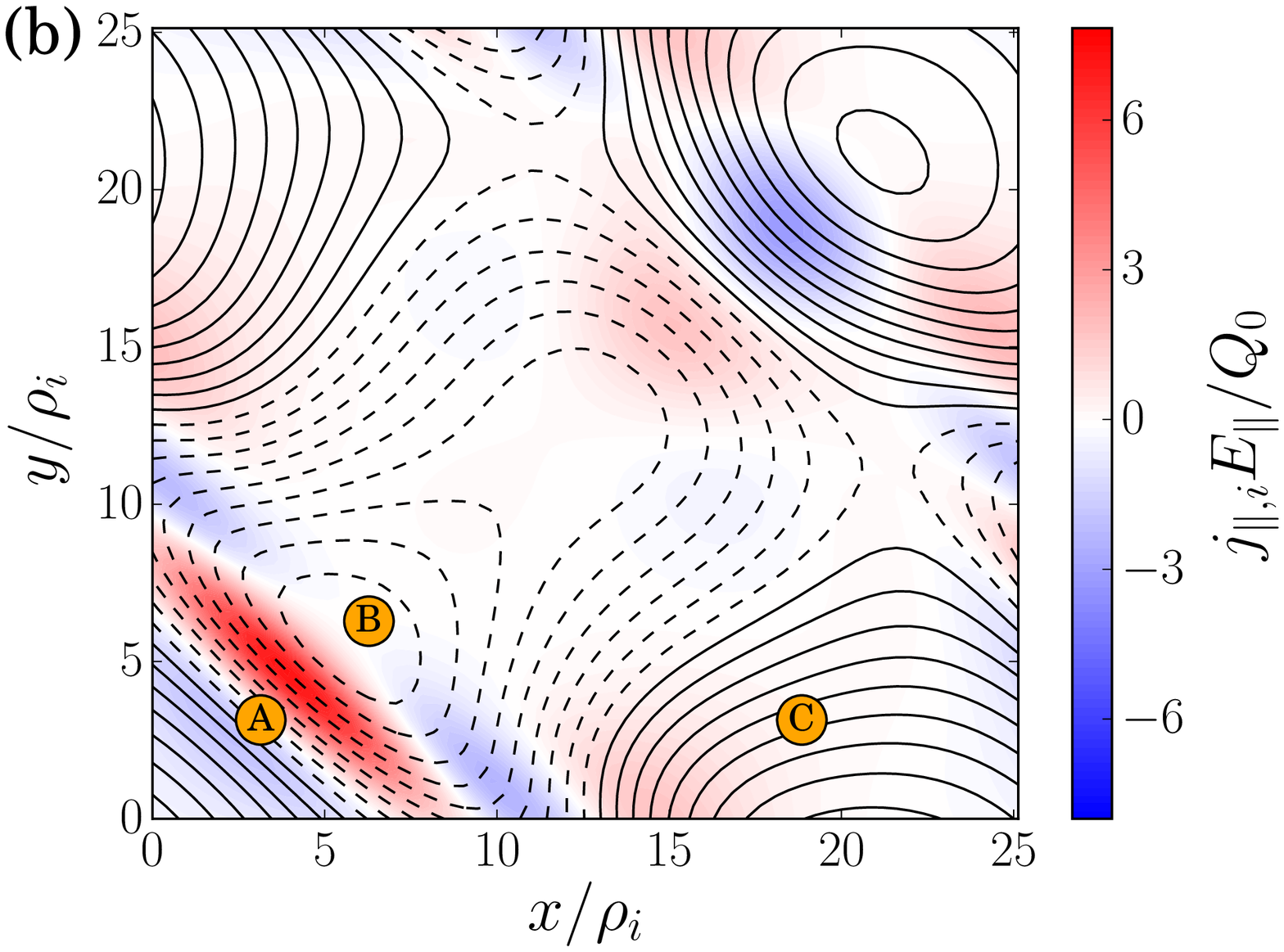} }   }
  \centerline{\resizebox{2.6in}{!}{\includegraphics{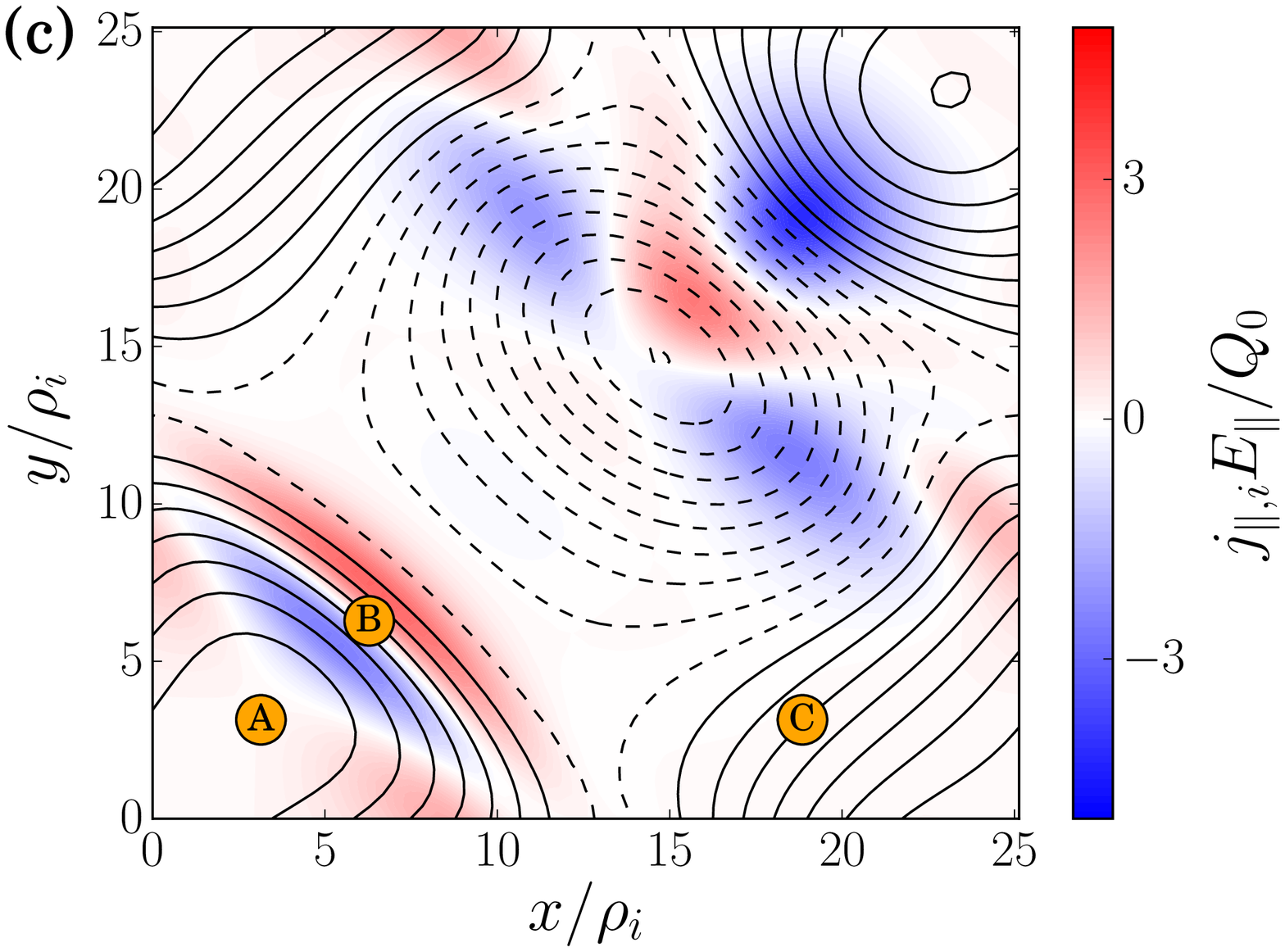}}
\resizebox{2.6in}{!}{\includegraphics{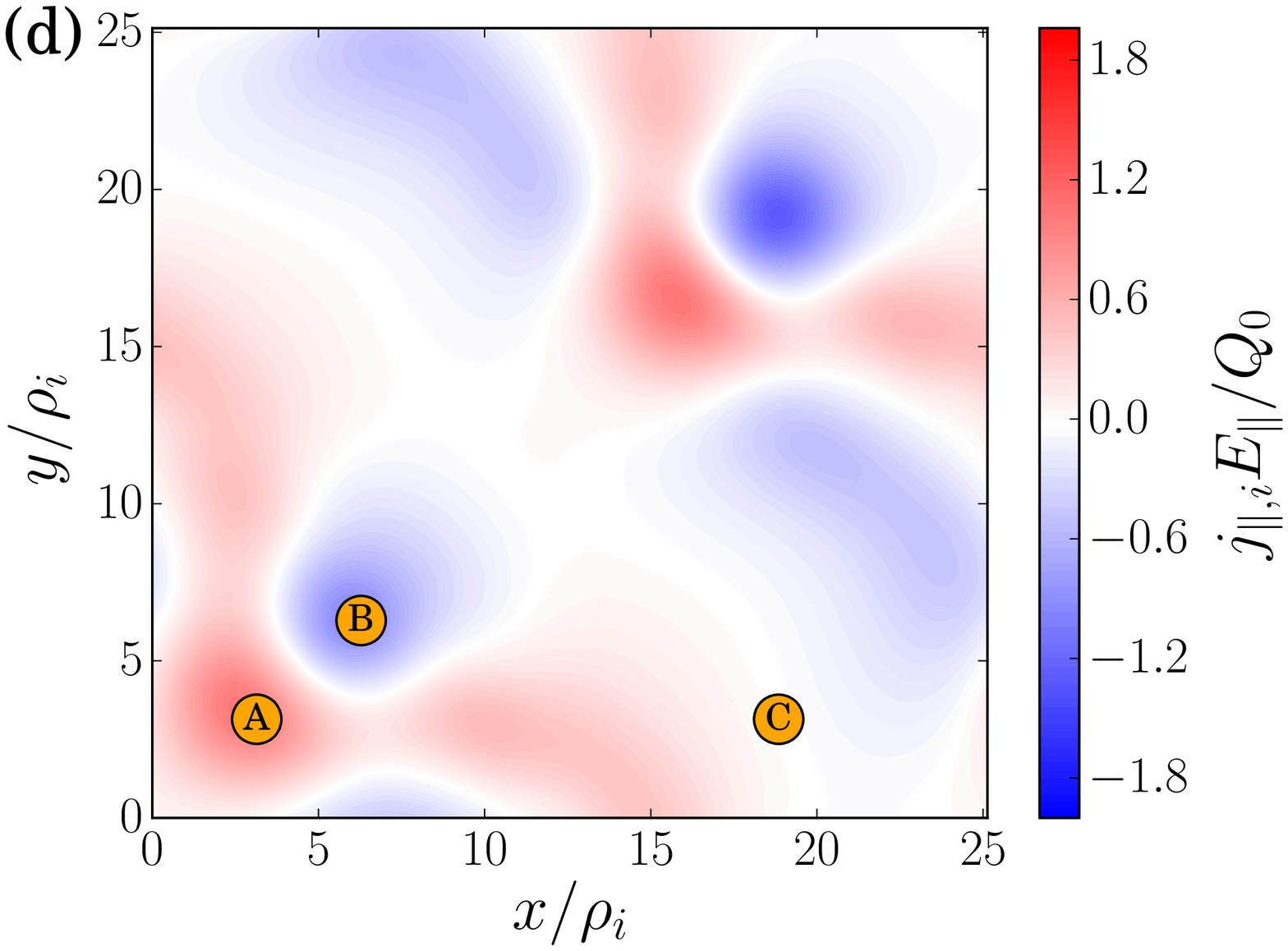} }   }
\caption{Plots of the instantaneous rate of parallel electromagnetic
  work on the ions $j_{\parallel,i} E_\parallel$  and contours of
  the parallel vector potential $A_\parallel$ (contours, positive
  solid, negative dashed) at times $t/T_0=$ (a)
  1.75, (b) 1.86, and (c) 2.03, as well as (d) $\langle j_{\parallel,i}
  E_\parallel \rangle_\tau$ averaged over one full wave period
  $\tau=0.992 T_0$ centered at time $t/T_0= 1.86$.}
\label{fig:fpac22_jzez_s1}
\end{figure}

\begin{figure}
  \centerline{\resizebox{2.6in}{!}{\includegraphics{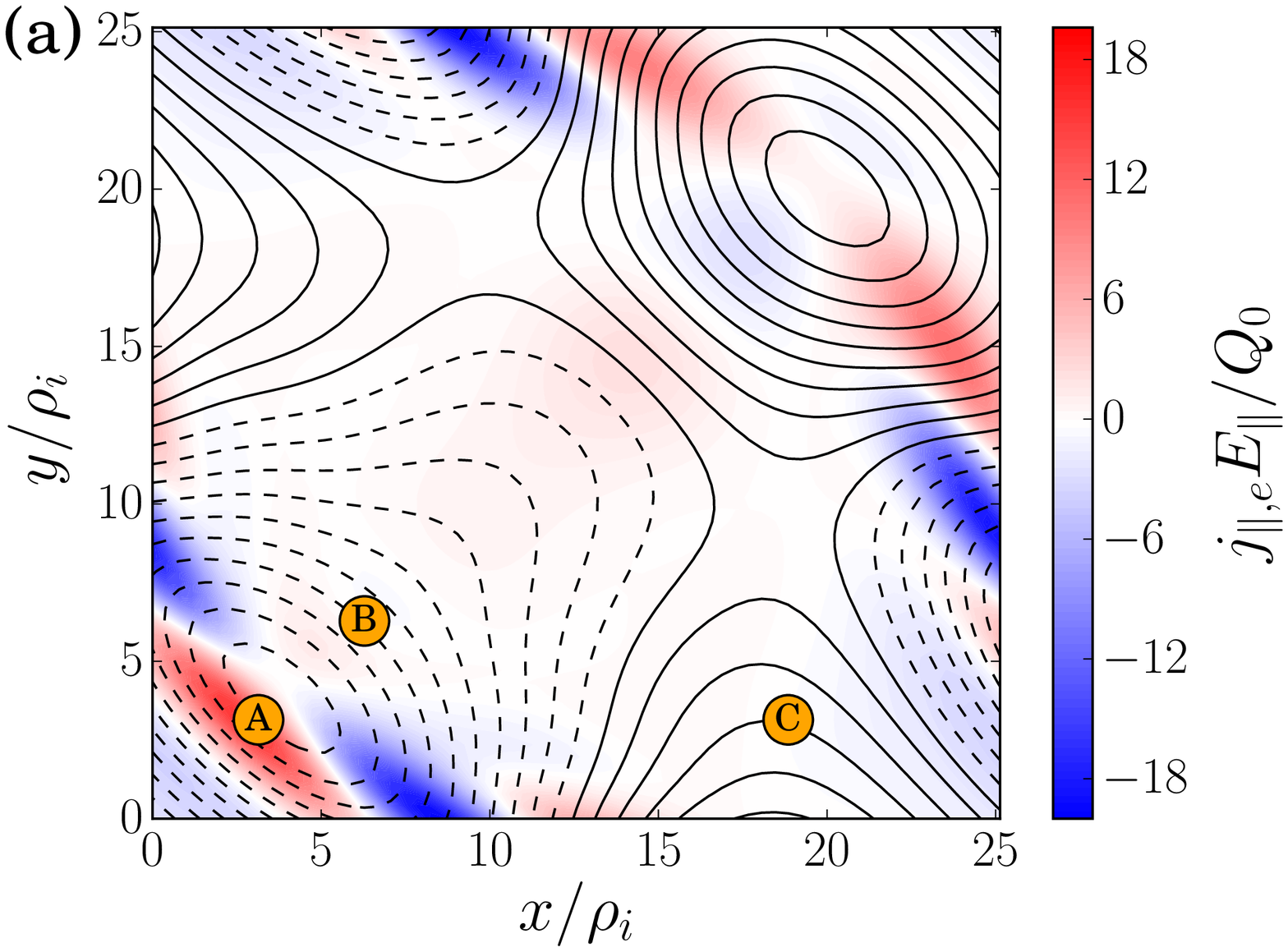}}
\resizebox{2.6in}{!}{\includegraphics{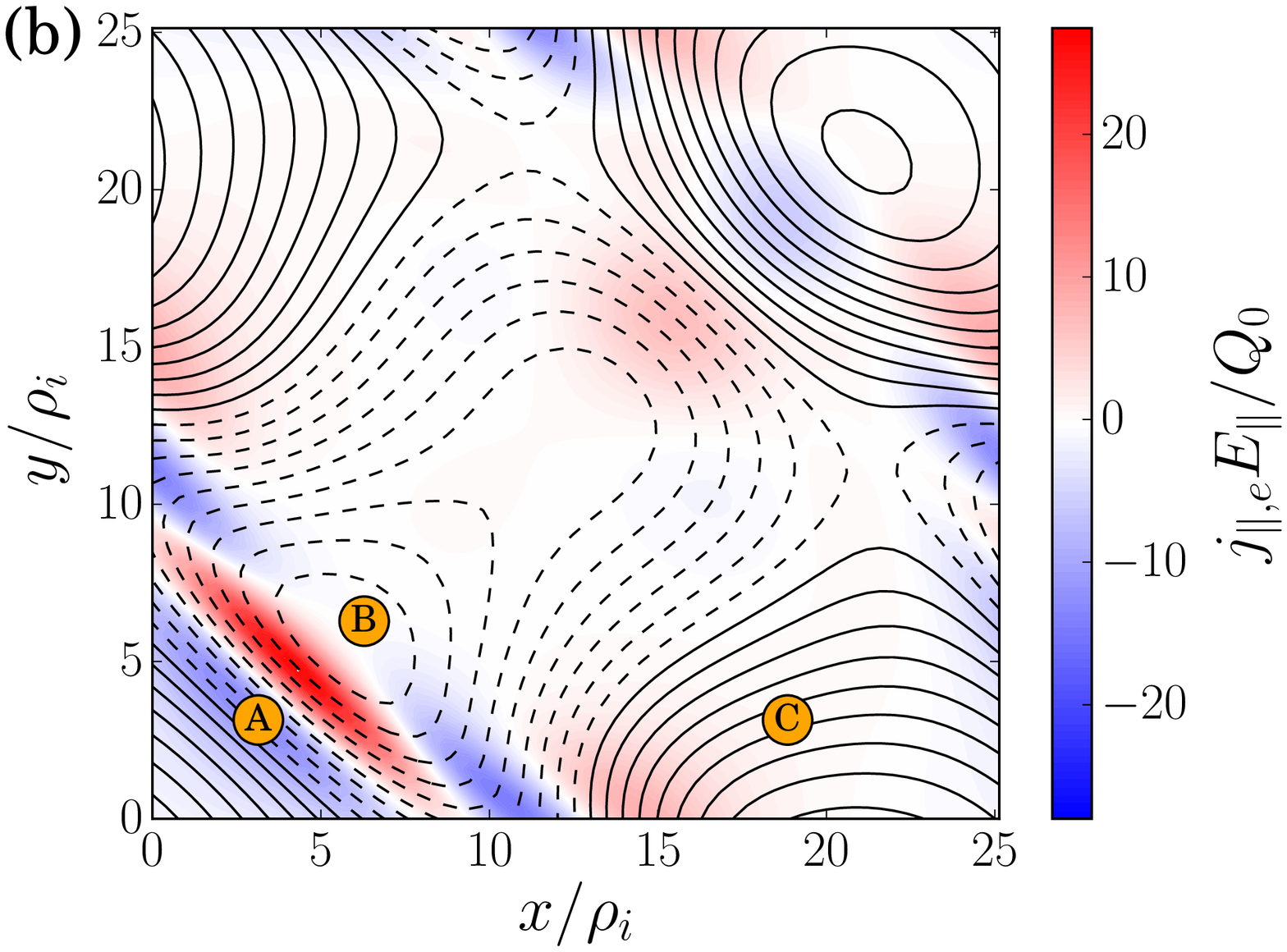} }   }
  \centerline{\resizebox{2.6in}{!}{\includegraphics{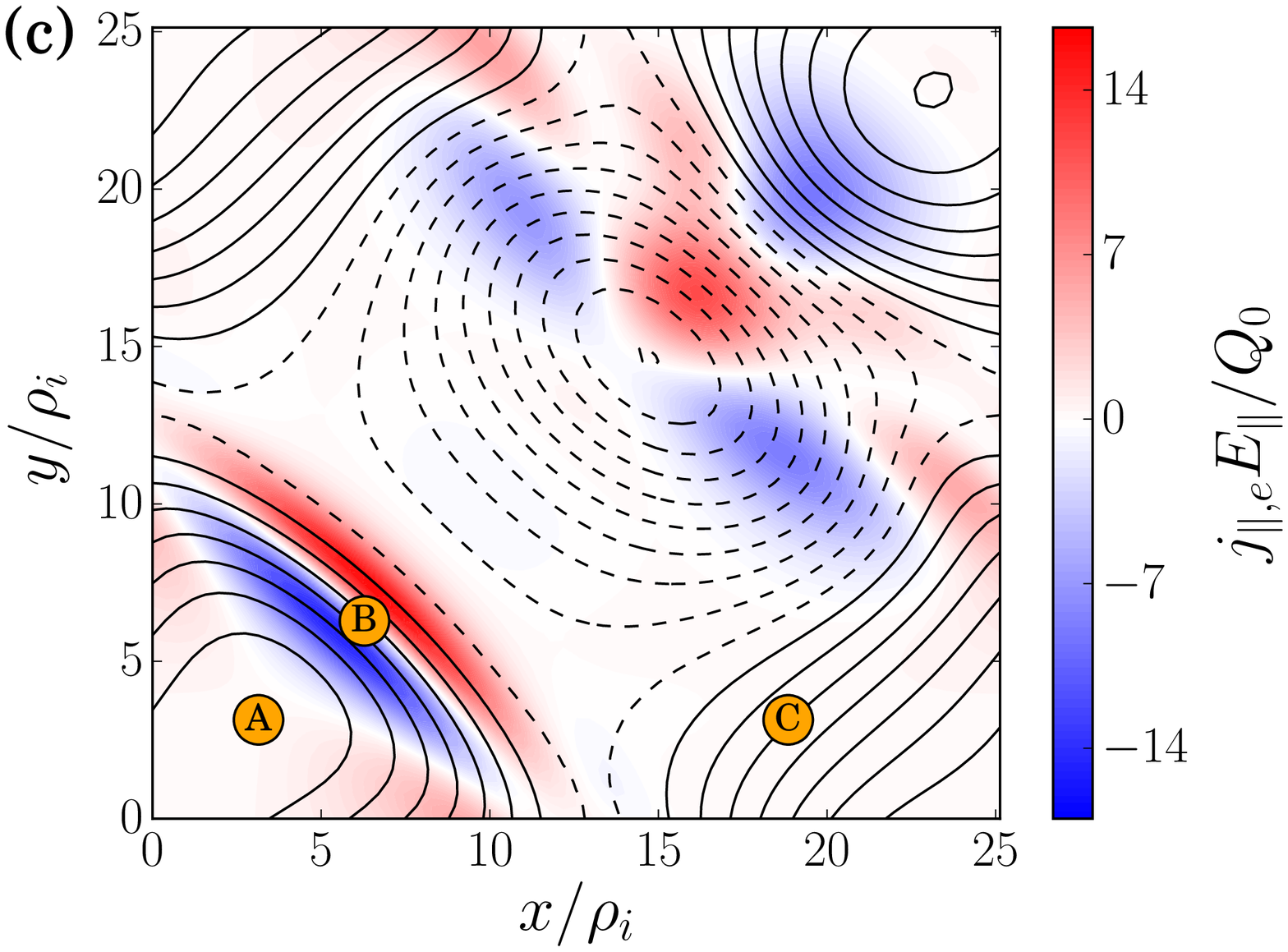}}
\resizebox{2.6in}{!}{\includegraphics{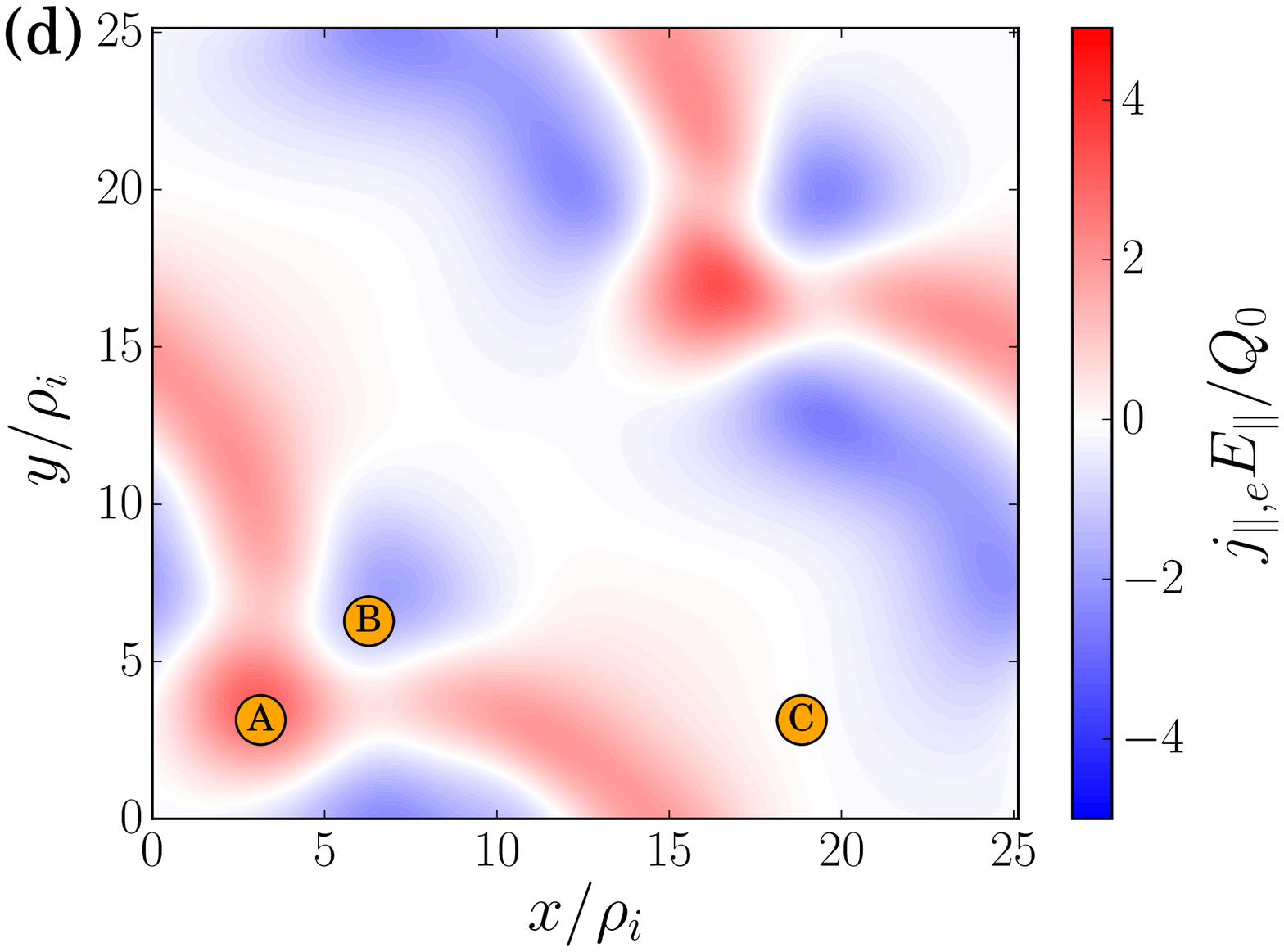} }   }
\caption{Plots of the instantaneous rate of parallel electromagnetic
  work on the electrons $j_{\parallel,e} E_\parallel$  and contours of
  the parallel vector potential $A_\parallel$ (contours, positive
  solid, negative dashed) at times $t/T_0=$ (a)
  1.75, (b) 1.86, and (c) 2.03, as well as (d) $\langle j_{\parallel,e}
  E_\parallel \rangle_\tau$ averaged over one full wave period
  $\tau=0.992 T_0$ centered at time $t/T_0= 1.86$.}
\label{fig:fpac22_jzez_s2}
\end{figure}

We also plot separately the parallel electromagnetic work on the ions
$j_{\parallel,i} E_\parallel$ in \figref{fig:fpac22_jzez_s1} and on
the electrons $j_{\parallel,e} E_\parallel$ in
\figref{fig:fpac22_jzez_s2}.  The spatial patterns of the instantaneous
rate of work at $t/T_0=$ (a) 1.75, (b) 1.86, and (c) 2.03, as well as
(d) the $\langle j_{\parallel,s} E_\parallel \rangle_\tau$ averaged
over one full wave period $\tau=0.992 T_0$ centered at time $t/T_0=
1.86$, are similar for both species, but the rate of electron
energization in this plane is about twice the magnitude of that for
the ions. Since the ion and electron linear damping rates are similar
for $k_\perp \rho_i \lesssim 1$, this may suggest significant energy
removal at higher $k_\perp \rho_i > 1$ where electrons are expected to
receive a greater share of the removed turbulent energy. A future
examination of the ion and electron energization will investigate the
typical length scale at which particles are energized in more detail.

\bibliographystyle{jpp}


\end{document}